\documentclass[%
preprint,
showpacs,
amsmath,amssymb,
aps,
pra,
]{revtex4-1}
\usepackage{graphicx}
\usepackage{dcolumn}
\usepackage{bm}
\usepackage{color}
\usepackage{comment}
\usepackage{diagbox}
\usepackage{verbatim}
\usepackage{multirow}

\begin{document}

\preprint{APS/123-QED}

\title{Correspondence between excited energy eigenstates and local minima of energy landscape in quantum spin systems}

\author{Yang Wei Koh}

\affiliation{Institute of Innovative Research, Tokyo Institute of Technology, Nagatsuta-cho, Midori-ku, Yokohama 226-8503, Japan}

\date{\today}

\begin{abstract}

The quantum-classical correspondence between local minima on the classical energy landscape and excited eigenstates in the energy spectrum is studied within the context of many-body quantum spin systems. In mean-field approximations of a quantum problem, one usually focuses on attaining the global minimum of the resulting energy function, while other minimum solutions are usually ignored. For frustrated systems, a strict distinction between global and local minimum is often not tenable since first-order type transitions can interchange the roles played by two different minima. This begs the question of whether there is any physical interpretation for the local minima encountered in mean-field approximations of quantum systems. We look at the problem from the perspective of quantum spin systems. Two models are studied, a frustrated model with quenched disorder, and a pure system without frustration. Accurate classical energies of the minima are compared with the full spectrum of energy levels, allowing us to search for signs of correspondence between them. It is found that the local minima can generally be interpreted as excited energy eigenstates. Instances of spurious minima are also reported.

\end{abstract}
\maketitle


%
%

\section{Introduction}
\label{sec.introduction}

The study of excited energy eigenstates is recently receiving some interest in the context of many-body quantum spin systems \cite{Zhang21,Seki21}. In quantum optimization algorithms, for instance, excited states appear as suboptimal solutions when the system falls out of equilibrium after undergoing diabatic or thermal transition \cite{Albash18}. Apart from practical issues associated with applications, there are also some basic questions regarding the identity or interpretation of excited states in quantum spin systems, particularly in the presence of frustration. On one hand, from a classical perspective, the defining feature of frustration is a rugged energy landscape exhibiting many minima, with metastable states residing in the local minima. On the other hand, from a purely quantum mechanical point of view, energies are eigenvalues of the Hamiltonian operator, with frustration manifested through the statistics of level spacings or eigenvector amplitudes \cite{Tekur18,Beugeling18}. The question that arises naturally is whether there is a correspondence between the quantum and classical excited energy states. In other words, is there a physical interpretation for the local minima frequently encountered in semiclassical treatments of quantum spin Hamiltonians \cite{Dusuel05}, or are these minima artefacts of the mathematical treatment.

Let us be more precise about what we mean by the term classical energy. The Hartree-Fock wave function $|0\rangle$ is defined as the direct product of single-spin states \cite{Dusuel05,Koh16}
\begin{equation}
|0\rangle
=
\prod_{i=1}^N
{\alpha_i \choose \beta_i}
\label{eq.HF wavefunction.definition}
\end{equation}
where $\alpha_i$ and $\beta_i$ are the spinor components of the $i$th spin in the basis where its $z$-directional Pauli matrix $\sigma^z_i$ is diagonal, and $N$ is the total number of spins. Normalization of $|0\rangle$ requires that $\alpha_i^2+\beta_i^2=1$ for all $i$. By the classical energy landscape of a Hamiltonian operator $H$, we mean the expectation value $\langle 0|H|0\rangle$, which is a $N$-dimensional function of the spin variables $\alpha_i$ \cite{footnote.on meanfield vs classical}. Traditionally, Hartree-Fock approximation gives an estimation of the ground state energy of $H$ by minimizing $\langle 0|H|0\rangle$ with respect to $\alpha_i$
\begin{equation}
E_0^{\mathrm{HF}}(H)
=
\min_{\{\alpha_i\}}
\,
\langle 0|
H
|0\rangle
\label{eq.E0HF energy.definition}
\end{equation}
An implicit assumption is that there is only one minimum on the classical energy surface, so we can unambiguously identify the Hartree-Fock energy $E_0^{\mathrm{HF}}$ with the ground state energy $E_0$. When there is quenched disorder in $H$, however, there can be many local minima on the energy surface. This poses the question of whether evaluating the Hartree-Fock energy at the local minima is meaningful as well, and if these local minima energies correspond to actual energy levels in the quantum spectrum.

There are many previous studies on local minima in disordered quantum spin systems. The nature of the transition between the paramagnetic and spin glass phases has been investigated in a variety of models ranging from the Heisenberg model \cite{Bray80}, the Sherrington-Kirkpatrick model \cite{Yamamoto87,Thirumalai89,Ray89,Buttner90,Goldschmidt90,Usadel91}, rotor models \cite{Kopec94,Read95}, to $p$-spin models \cite{Dobrosavljevic90,Goldschmidt90R,Cesare96,Obuchi07}. A wide range of techniques such as perturbative expansion \cite{Yamamoto87}, replica method \cite{Thirumalai89,Buttner90,Goldschmidt90,Cesare96,Obuchi07}, quantum Monte Carlo simulations \cite{Ray89}, Landau effective-action functional \cite{Read95}, and $1/p$ expansion \cite{Dobrosavljevic90} have also been employed to study these transitions. For a summary of some of these methods and their results see ref. \cite{Usadel91}. The so-called `dynamical effects' induced by quantum fluctuations on the spin glass phase have also been studied in some models at both finite and zero temperatures using a variety of techniques \cite{Miller93,Grempel98,Rozenberg98,Arrachea01}. In most of these earlier works, one usually performs an average over the quenched disorder using the replica method. On the other hand, in the approach via the quantum Thouless-Anderson-Palmer (TAP) equations \cite{Ishii85,Cesare92,Biroli01}, one can choose to study the system under a specific realization of disorder. Compared to the replica approach, this viewpoint offers a more intuitive picture of the glassy phase in the form of a rugged free energy landscape, with the exponentially-many TAP solutions as the local minima. The general formalism of the quantum TAP approach has been developed in ref. \cite{Biroli01}, and applied to the spherical $p$-spin model to obtain  the generic phase diagram for quantum spin glasses. The Hartree-Fock method we adopted above have some similarities with the quantum TAP approach, but also some differences. Like the TAP approach, the Hartree-Fock method works with specific realizations of quenched disorder, and the scenario of a rugged energy landscape emerges as well. In the case of some simple mean-field models (e.g. the ferromagnetic model in Appendix \ref{app.ferromagnetic model.CCSD}), their mean-field equations are also similar. On the other hand, while the quantum TAP approach is applicable at finite temperatures, the Hartree-Fock method is usually only used for zero temperature calculations. The kinds of approximation which are involved in the two methods are also quite different. The TAP equations become exact in the thermodynamic and infinite-dimension limit, whereas for the Hartree-Fock approximation, corrections in the form of correlation energy (discussed below) must be incorporated regardless of system size or dimension.

Another way of studying disordered quantum spin systems is from the perspective of quantum chaos \cite{Tekur18,Beugeling18,Brown08,Zakrzewski23,Gubin12,Georgeot98,Bera22,Winer22,Georgeot00a,Georgeot00b}. In this approach, one mainly focuses on purely quantum mechanical attributes such as level statistics \cite{Gubin12,Tekur18,Brown08,Georgeot00a,Georgeot00b}, level dynamics \cite{Zakrzewski23}, eigenstate amplitudes distribution \cite{Gubin12,Beugeling18}, and entanglement \cite{Brown08}, rather than on the local minima. Also, many works in the literature are devoted to disordered spin chains \cite{Tekur18,Beugeling18,Brown08,Gubin12} or two-dimensional models \cite{Brown08,Georgeot00a,Georgeot00b}, in which the spin glass phase is usually considered to be absent. An exception is an early paper by Georgeot and Shepelyansky which showed how the level spacing distribution changes between Poisson and Wigner-Dyson as one tunes the strength of the transverse field in the quantum Sherrington-Kirkpatrick model \cite{Georgeot98}. Interestingly in this work, the relation between the statistics of the level spacings and the spin glass phase is not very straightforward, because the distribution reverts back to Poisson (i.e. non-chaotic) when deep within the glassy regime. Recently, there is growing interest in the relationship between the spin glass phase and quantum chaos, for instance from the viewpoints of Lyapunov exponent \cite{Bera22} and spectral form factor \cite{Winer22}.

A slightly different angle to view and perhaps bring together these two perspectives is offered by recent discussions on many-body localization-delocalization (MBLD) in quantum spin glasses \cite{Laumann14,Baldwin17,Mukherjee18}, where extensive studies on the energy levels and eigenstates of such systems revealed a transition between localized and delocalized phases reminiscent of many-body localizations observed in spin chains \cite{Imbrie16,Kjall14,Lee17,Biroli20,Tarzia20,Herviou21}. However, as noted by Baldwin et al. in ref. \cite{Baldwin17}, the mechanisms underlying the transitions are quite different between mean-field and chain models, with large energy barriers playing the dominant role in MBLD. The authors also pointed out a possible relation between the mobility edge in MBLD and the dynamical transition ushering a system into the rugged free energy phase (c.f. the quantum TAP approach ref. \cite{Biroli01}). As in studies on quantum chaos, however, in MBLD one focuses mainly on quantum attributes such as level spacings or entanglement entropy, so the intuitive role played by local minima in the localization mechanism is, nevertheless, not very transparent.

This work is in part an effort to bridge and find a relationship between these two different points of view, namely the classical description in terms of local minima, versus the quantum picture based on energy levels and eigenstates. In particular, we are focusing on a specific aspect, namely, to see if a correspondence can be established between the energy levels in the quantum spectrum and the energies of the classical local minima. In many of the works reviewed above, one adopts a statistical stance and looks at the general properties of a system via `bulk' quantities such as the average free energy or the level statistics. Here, on the other hand, our approach is to match pairs of quantum and classical states together individually. Overall, we were able to find evidences of a quantum-classical correspondence between the two. More specifically, the correspondence yields a picture of highly-excited eigenstates which are largely confined around the classical local minima. We stress that it is not our aim here to address the challenging issues reviewed above (i.e. quantum chaos and MBLD). Nonetheless, we feel that our work might still contribute some insights into those questions. 

Apart from theoretical considerations, such a correspondence relation may also be useful in practical scenarios. In quantum annealing and optimization \cite{Albash18}, for instance, a phenomenon known as `freezing' where the system gets stuck in a suboptimal Boltzmann distribution, thereby failing to attain the global minimum solution, has been observed both in numerical simulations \cite{Amin15} and experimentally on the D-wave machine \cite{Boixo16}. Due to the high dimensionality of the Hilbert space involved, quantum annealing processes are usually interpreted in terms of a classical trajectory on an approximate mean-field energy landscape. In refs. \cite{Amin15} and \cite{Boixo16}, freezing has been attributed to trappings by local minima on such energy surfaces. Interestingly, the situation here is somewhat the opposite of quantum chaos and MBLD discussed above. Whether and how the local minima---based on a mean-field picture---represent the actual quantum states of the system is not very clear. A closely related issue concerns the role played by tunneling in quantum annealing, particularly in the D-wave machine \cite{Boixo16,Denchev16}. In these works, tunneling is analyzed with the assumption that the system makes a transition from one local minimum to another. On the other hand, as will be demonstrated in Sec. \ref{sec. Hopfield model section label}, an excited energy eigenstate is sometimes formed by superposing several local minima together. Hence, a more accurate analysis of tunneling in frustrated systems might need to account for such multiple-minima effects. These considerations may play an important role when, say, improving the efficiencies of quantum optimization algorithms. 

A second contribution of this work concerns the numerical calculation of classical energy. In the preceding discussion, we have interpreted the classical energy to be the Hartree-Fock energy. However, Hartree-Fock approximation is sometimes not precise enough to enable one to perform the quantum-classical correspondence mentioned above. In particular, in the high-energy regime where the local minima reside, the spectral lines are quite closely spaced and so the classical energy needs to be relatively close to the appropriate energy eigenvalue in order for one to see their correspondence. The way to improve upon Hartree-Fock approximation is to incorporate the correlation energy via Coupled-Cluster theory \cite{Jensen07,Bartlett07}. In standard Coupled-Cluster theory, one performs a Hausdorff expansion of the transformed Hamiltonian
\begin{equation}
e^{-T} H e^{T}= H + [H,T] + \frac{1}{2}[[H,T],T] + \frac{1}{3!}[[[H,T],T],T] + \cdots
\label{eq.sec01.similarity transform and Hausdorff expansion}
\end{equation}
where $T$ is the excitation operator generating multi-spin excitations when acting on the Hartree-Fock wave function \cite{footnote.two sense of the word excitation}.  The combination of Hartree-Fock approximation and Coupled-Cluster theory has traditionally been used for electronic structure calculations in quantum chemistry \cite{Jensen07,Bartlett07}, and subsequently also in condensed matter physics to treat systems such as the Hubbard model \cite{LeBlanc15}. In these fermionic systems, the models are expressed in terms of creation and annihilation operators $c^{\dagger}$ and $c$, and the expansion Eq. (\ref{eq.sec01.similarity transform and Hausdorff expansion}) terminates after a finite number of terms due to the anticommutator relation $\{c,c^{\dagger}\}=1$. On the other hand, in spin systems the Pauli matrices $\sigma^{\alpha}$ ($\alpha=x,y,z$) obey the relation $[\sigma^{\alpha},\sigma^{\beta}]=2i\epsilon_{\alpha\beta\gamma}\sigma^{\gamma}$ ($\epsilon_{\alpha\beta\gamma}:$ Levi-Civita symbol) where a commutator between two matrices produces a third one, unlike the case of fermions. 
A naive evaluation of Eq. (\ref{eq.sec01.similarity transform and Hausdorff expansion}) for spin systems is therefore not guaranteed to terminate. The formulation of a finite (i.e. terminating) Hausdorff expansion in terms of three operators $\sigma^{\pm}$ and $\sigma^z$ was pioneered by Roger and Hetherington \cite{Roger90}, and Coupled-Cluster theory has since been applied on a variety of spin systems \cite{Roger90, Bishop, Miguel96,Farnell01,Kruger06,Darradi08,Richter10,Gotze11,Gotze15}. However, for an arbitrary Hamiltonian with quenched disorder, the evaluation of this expansion is in general still quite involved, requiring the use of a pattern matching computer-algebra program just for setting up the Coupled-Cluster equations alone \cite{Farnell01,Kruger06,Hirata03}.

In this work, we propose a different way of formulating Coupled-Custer theory for spin systems that circumvents the Hausdorff expansion altogether. For certain types of mean-field systems, our approach simplifies the process of deriving the Coupled-Cluster energy and equations. Consider operators of the form     
\begin{equation}
J_{\alpha}(\bm{a})=\sum_{i=1}^{N} a_i\sigma_i^{\alpha}
\label{eq.definition of generalized operator J(a)}
\end{equation}
where $\sigma_i^{\alpha}$ is the $\alpha$-directional Pauli matrix of the $i$th spin, and $a_i$ is the $i$th component of the $N$-dimensional vector $\bm{a}$. Let us restrict ourselves to models whose $H$ and $T$ can be expressed in terms of polynomials of operators like $J_{\alpha}(\bm{a})$. The transformation $e^{-T} H e^{T}$ can then be evaluated without performing the expansion Eq. (\ref{eq.sec01.similarity transform and Hausdorff expansion}) explicitly. To illustrate, consider a two-body operator $T=\frac{1}{2}[J_{\alpha}(\bm{a})]^2$. We generalize the Hubbard-Stratonovich transformation to operators  
\begin{equation}
e^{\frac{1}{2}[J_{\alpha}(\bm{a})]^2}=\int_{-\infty}^{\infty}\,\frac{dm}{\sqrt{2\pi}} \exp\left[-\frac{1}{2}m^2 +  m J_{\alpha}(\bm{a})\right]
\label{eq.sec01.Hubbard Stratonovich for operators}
\end{equation}
It is seen that the quadratic operator $[J_{\alpha}(\bm{a})]^2$ has been `linearized' on the right side of Eq. (\ref{eq.sec01.Hubbard Stratonovich for operators}) by introducing the random Gaussian field $m$. It is then straightforward to evaluate individual terms in $e^{-T} H e^{T}$ by applying the formula
\begin{equation}
e^{-\lambda J_{\beta}(\bm{b})} \,\, J_{\alpha}(\bm{a})\,\,\, e^{\lambda J_{\beta}(\bm{b})}= J_{\alpha}(\bm{c}) + i \, \epsilon_{\alpha\beta\gamma} \, J_{\gamma}(\bm{d})
\label{eq.sec01.general rotation formula}
\end{equation}
where $c_i=a_i\cosh(2\lambda b_i)$, $d_i=a_i\sinh(2\lambda b_i)$, $\lambda$ is a parameter, and $\epsilon_{\alpha\beta\gamma}$ is the Levi-Civita symbol. Equations (\ref{eq.sec01.Hubbard Stratonovich for operators}) and (\ref{eq.sec01.general rotation formula}) in effect achieve the same purpose as the Hausdorff expansion Eq. (\ref{eq.sec01.similarity transform and Hausdorff expansion}). After they have been applied, the setting up of the Coupled-Cluster energy and equations is straightforward and will be detailed in Sec. \ref{app.random field formulation of CCSD.revised}. The above approach simplifies some of the mathematical derivations encountered in Coupled-Cluster theory. 

As mentioned, Coupled-Clustered theory has been used to study a variety of spin systems. Its performance in spin chains \cite{Roger90,Bishop,Miguel96} and various two-dimensional lattices \cite{Roger90,Bishop,Kruger06,Richter10,Gotze15,Farnell01,Miguel96} has been benchmarked and examined in considerable detail. These studies have shown that Coupled-Cluster theory is an accurate and efficient numerical method for calculating the ground state properties of low-dimensional lattice spin models. On the other hand, its performance in fully-connected models, especially in the presence of quenched disorder, has received lesser attention. It is the secondary objective of this work to assess the feasibility of Coupled-Cluster theory for treating mean-field systems. We study models whose Hamiltonian matrix is amenable to exact diagonalization, thereby allowing us to benchmark the results of Coupled-Cluster calculations. We also apply Coupled-Cluster theory to local minima on the Hartree-Fock energy surface to see if the energies of excited states are obtainable from local minima. We found that at large system sizes, Coupled-Cluster theory is effective and gives significant improvement upon Hartree-Fock approximation. However, at small system size, and particularly when quenched disorder is present, we encountered situations whereby there are no physically valid solutions to the Coupled-Cluster equations. Overall, our assessment of the effectiveness of the Coupled-Cluster method for mean-field models remains inconclusive. In particular, the question of its performance in disordered systems when the number of spins is large remains open and requires further study.

In this paper, we address the questions outlined above by case studying two models. The rest of the paper is organized as follows. Section \ref{app.random field formulation of CCSD.revised} presents the random field formulation of Coupled-Cluster theory at the level of singles doubles approximation (CCSD). Section \ref{sec.cubic model section label} is on a model without frustration, the cubic model under transverse field and antiferromagnetic interactions. This is a simple system, chosen because its energy surface exhibits just one local minimum. After defining the model and reviewing its phase diagram and ground state energy, we focus our attention on the local minimum. The main result here is that in the quantum spectrum, a series of consecutive energy levels form an envelope of avoided crossings, and the energy of the local minimum lies very closely on this envelope. Hence, the quantum-classical correspondence is established between the local minimum and this `secondary' structure, rather than with any particular energy level. We then examine the quantum and classical wave functions along the envelope to ascertain the correspondence. Section \ref{sec. Hopfield model section label} is on the Hopfield model in transverse field. This is a frustrated system exhibiting multiple local minima on its energy surface. After a brief review of the model, we study the rugged structure of the energy landscape in some detail. This is followed by examination of the energy eigenvalues and eigenfunctions, where we found quantum states being localized in the energy basins of local minima. Lastly, Section \ref{sec. summary discussion section label} summarizes and concludes the paper.

To aid in the flow of our presentation, some technical materials are deferred to the appendices. Appendix \ref{app.random field formulation of CCSD} gives a list of CCSD formulas used in this paper. Appendix \ref{app.ferromagnetic model.CCSD} applies CCSD to a simple model, the ferromagnetic model in transverse field, illustrating in some detail the implementation of the method. Appendix \ref{app.wavefunctions} presents the wave functions of the classical approximations, which will be employed when ascertaining quantum-classical correspondence. The reader is referred to these supplements for technical details.

%
%

\section{Random field formulation of Coupled-Cluster theory}
\label{app.random field formulation of CCSD.revised}

In this section, we present the random field formulation of Coupled-Cluster theory. We illustrate our method by outlining the main steps in the derivation of $\langle 0|e^{-T}J_x \,e^{T}|0\rangle$, which is one of the terms (the transverse field) in the Coupled-Cluster energy. We then state the results for the other terms in Appendix \ref{app.random field formulation of CCSD}. The notation $J_{\alpha}$ denotes $J_{\alpha}(\bm{a})$ with all components of $\bm{a}$ set to unity. The Coupled-Cluster energy and equations for the models studied in this paper are obtained by combining terms listed there. For general aspects of Coupled-Cluster theory, the reader is referred to Refs. \cite{Jensen07,Bartlett07}.

In this paper, we employed the singles doubles approximation in the Coupled-Cluster expansion (CCSD). The excitation operator $T$ in Eq. (\ref{eq.sec01.similarity transform and Hausdorff expansion}) is truncated at the second-order term, and the Coupled-Cluster energy is no longer exact. Previous works have shown CCSD to be an economic approximation to try when approaching any new model. As we shall demonstrate in later sections, it leads to very accurate results when when the system size is large. 

For the models we considered, $T$ can be written as 
\begin{equation}
T^{\mathrm{SD}}=y\,\sum_{i=1}^N f_i + \frac{1}{2}\left(\sqrt{w}\,\sum_{i=1}^N f_i\right)^2
\label{eq.app.TSD.definition}
\end{equation}
The operator $f_i=-i\sigma_i^y$ flips the $i$th spinor ${\alpha \choose \beta}$ ($\alpha,\beta$ real). The variables $y$ and $w\,(>0)$ are, respectively, the amplitudes for singles and doubles excitations, and they are determined by solving the CCSD equations. Applying Eq. (\ref{eq.sec01.Hubbard Stratonovich for operators}) followed by Eq. (\ref{eq.sec01.general rotation formula}), one has  
\begin{equation}
e^{-T^{\mathrm{SD}}}
\,
J_x
\,
e^{T^{\mathrm{SD}}}
=
\int \mathcal{D}
\,\,
e^{(i\bar{m} + m)\sqrt{w}\sum_i f_i} 
\bigg(
J_x
\cos2z
+
J_z
\sin2z
\bigg)
\label{eq.app.eJxe.introduce random field}
\end{equation}
where $\int \mathcal{D}=\int_{-\infty}^{\infty}\int_{-\infty}^{\infty} \frac{dm d\bar{m}}{2\pi}\,e^{-\frac{1}{2}(m^2+\bar{m}^2)}$ and $z=y+m\sqrt{w}$. Before taking the expectation value of Eq. (\ref{eq.app.eJxe.introduce random field}) with respect to the Hartree-Fock wave function $|0\rangle={\alpha \choose \beta}^N$, it is convenient to define
\begin{equation}
b_1=\cosh\sqrt{w}(im-\bar{m})\,\,,\,\,b_2=\sinh\sqrt{w}(im-\bar{m})
\label{}
\end{equation}
and
\begin{equation}
P_1=2\alpha\beta + (\alpha^2- \beta^2)(i b_2/b_1)
\,\,,\,\,
P_2=2\alpha\beta(i b_2/b_1)-(\alpha^2- \beta^2)
\label{}
\end{equation}
We then have
\begin{equation}
\langle 0| e^{-T^{\mathrm{SD}}}\, J_x \, e^{T^{\mathrm{SD}}} |0\rangle
=
N \int \mathcal{D}
\,\,
b_1^N
\bigg( P_1 \cos2z - P_2 \sin2z \bigg)
\label{}
\end{equation}
To integrate out the random fields $m$ and $\bar{m}$, we use the identities
\begin{equation}
\int \mathcal{D} \,\, b_1^N \,\,
\left(\frac{ib_2}{b_1}\right)^n \cos\mu z
=
i^n \, e^{-\frac{\mu^2w}{2}} \cosh^N(\mu w)\tanh^n(\mu w)
\left[
\frac{(-1)^n e^{i\mu y}+e^{-i\mu y}}{2}
\right]
\label{}
\end{equation}
\begin{equation}
\int \mathcal{D} \,\, b_1^N \,\,
\left(\frac{ib_2}{b_1}\right)^n \sin\mu z
=
i^n \, e^{-\frac{\mu^2w}{2}} \cosh^N(\mu w)\tanh^n(\mu w)
\left[
\frac{(-1)^n e^{i\mu y}-e^{-i\mu y}}{2i}
\right]
\label{}
\end{equation}
where $n$ and $\mu$ are positive integers. We finally obtain
\begin{equation}
\langle 0| e^{-T^{\mathrm{SD}}}\, J_x \, e^{T^{\mathrm{SD}}} |0\rangle
=
N e^{-2w} \cosh^N(2w)
(1+\tanh2w)
\left[
2\alpha\beta \cos 2y + (\alpha^2-\beta^2)\sin 2y
\right]
\label{eq.app.list.ccsd.energy.Jx}
\end{equation}
The derivation of other terms in the CCSD energy is similar and the results are summarized in Appendix \ref{app.subsec.summary CCSD energy}. For the singles and doubles equations, instead of acting on $e^{-T^{\mathrm{}}}\, H \, e^{T^{\mathrm{}}} |0\rangle$ from the left with $\langle 0|$, one applies $\langle 1|=\langle 0|f_1$ and $\langle 12|=\langle 0|f_1f_2$, respectively. Appendices \ref{app.subsec.summary 1 flip} and \ref{app.subsec.summary 2 flip} summarize the terms used in the CCSD equations. 

%
%

\section{Local minimum as envelope of avoided crossings}
\label{sec.cubic model section label}

\subsection{Model: Cubic potential in transverse field and antiferromagnetic interactions}
\label{}

As our first example, consider the Hamiltonian
\begin{equation}
H_{\mathrm{cubic}}=
-
\frac{s\lambda}{N^2} \left( J_z \right)^3 
+
\frac{s(1-\lambda)}{N}
\left(J_x\right)^2
-
(1-s)
J_x
\label{eq.cubic w antiferro.hamiltonian definition}
\end{equation}
where the parameters $s,\lambda\in[0,1]$ control the relative strengths among the three operators. This model was proposed in the context of quantum annealing \cite{Seki12}, where one seeks to optimize the cubic objective function $(J_z)^3$ via careful adjustments of quantum fluctuations in the form of the transverse field $J_x$ and antiferromagnetic interactions $(J_x)^2$. In the absence of the antiferromagnetic term, quantum annealing is highly inefficient due to a first-order phase transition. Antiferromagnetic fluctuations can, however, soften the transition to a second-order one, leading to a quantum speedup. 

We have two reasons for studying this model. Firstly, it is simple, exhibiting a single local minimum on the classical energy landscape. The second motivation is partly related to the first and concerns the identity and eventual fate of a local minimum in the process of annealing. The computational bottleneck of quantum annealing lies near the Landau-Zener crossing between the ground and first excited states. In the mean-field picture, these two states are embodied by the global and local minima, respectively. While this correspondence holds true near the avoided crossing, when away from the crossing, whether the local minimum still remains as the first excited state, becomes other eigenstates, or if the mean-field description breaks down altogether, is not very clear. In this respect, the simple energy landscape of this model allows us to follow the sole local minimum closely and study this issue without complications from other minima. As we shall show---at least for this model---the local minimum eventually evolves into a very high energy eigenstate, implying bad results for quantum annealing if the system gets trapped by the minimum at the avoided crossing bottleneck.

\subsection{Classical energy landscape and phase diagram}
\label{}

\begin{figure}[h]
\begin{center}
\includegraphics[scale=0.7]{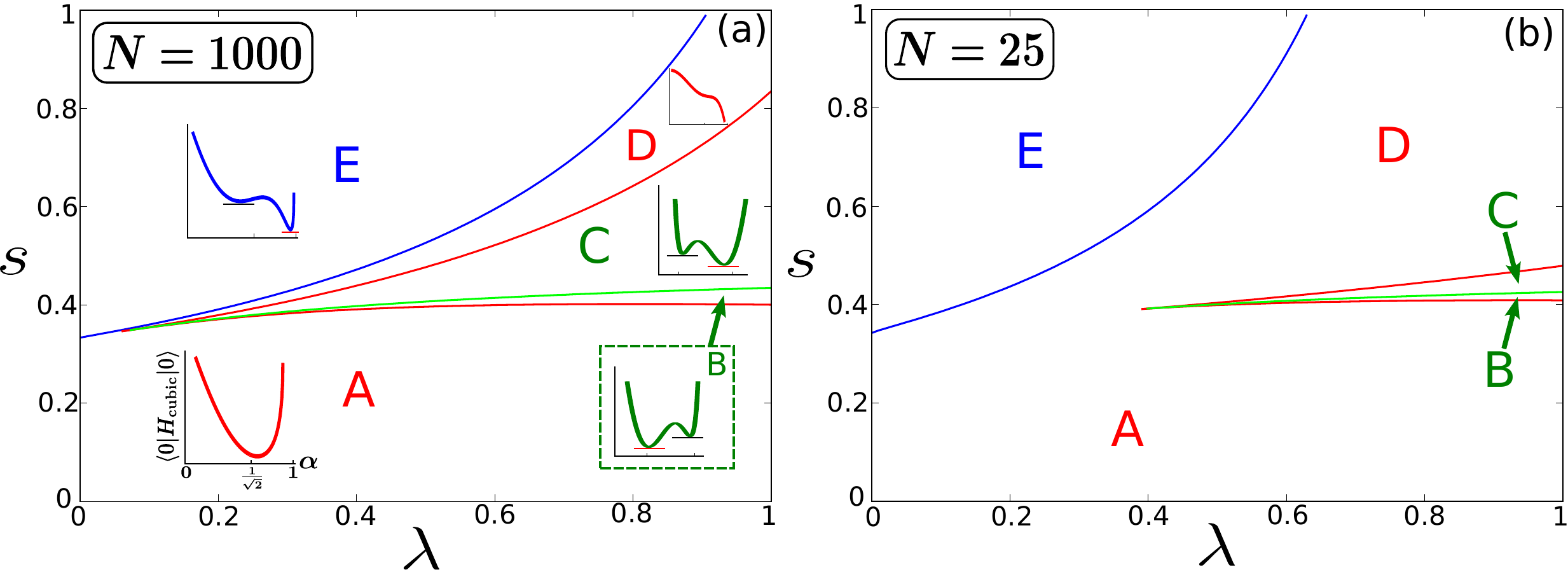}
\caption{Phase diagram of the cubic model $H_{\rm cubic}$ showing its classical energy landscape [Eq. (\ref{eq.cubic model.HF approximation})] in the $s$-$\lambda$ plane. (a) $N=1000$. (b) $N=25$. The plane can be divided into five regions, labeled A to E. At finite $N$, one can go from region A to D without crossing a boundary, as seen from panel (b). In panel (a), a schematic of the energy landscape within a region is shown next to that region's label.}
\label{fig.phase diagram.Hcubic.N1000 and N25}
\end{center}
\end{figure}

Due to permutation symmetry among all spins in $H_{\rm cubic}$, the Hartree-Fock wave function takes the simpler form $|0\rangle={\alpha\choose \beta}^N$, and one has 
\begin{align}
\langle 0|H_{\rm cubic}| 0 \rangle =
&
-s\lambda (N-1)(1-2N^{-1})(2\alpha^2-1)^3
-2N(1-s)\alpha\sqrt{1-\alpha^2}
\nonumber\\
& 
+
4s(1-\lambda)(N-1)\alpha^2(1-\alpha^2)
-
s\lambda(3-2N^{-1})(2\alpha^2-1)
+
s(1-\lambda)
\label{eq.cubic model.HF approximation}
\end{align}
One can summarize the overall behavior of the system in the form of a phase diagram in $s$-$\lambda$ parameter space. Panel (a) of Fig. \ref{fig.phase diagram.Hcubic.N1000 and N25} shows the case for $N=1000$. The phase space consists of five regions, labeled A to E. A schematic of the classical energy landscape within a region is shown next to that region's label. Boundaries between regions are obtained numerically. The energy landscape has one minimum in regions A and D, and two minima in B, C, and E. Our interest will be in the three latter regions, where the energy surface exhibits a local minimum. When going from region B to C, the system undergoes a first-order phase transition. As will be discussed below, only the local minimum in regions B and C exhibits quantum-classical correspondence, while the one in region E seems to be an unphysical minimum.

Panel (a) is representative of the phase diagram for general $N$. As $N$ decreases, the most salient change is that region C becomes smaller,  as seen in panel (b) for the case of $N=25$. When $N$ is finite, regions A and D are not separated by a boundary. This is seen clearly in panel (b) where at small $N$ the effect is more pronounced. As $N$ increases, region D becomes smaller, with regions C and E merging in the thermodynamic limit. This narrowing of region D is, however, quite slow with respect to $N$ as we were still able to observe the region numerically at $N=40\,000$.

\subsection{Ground state energy}
\label{}

\subsubsection{Hartree-Fock approximation}
\label{}

The Hamiltonian $H_{\rm cubic}$ commutes with the total angular momentum operator $J_x^2+J_y^2+J_z^2$, so the ground state energy $E_0(H_{\rm cubic})$ can be obtained numerically by diagonalizing $H_{\rm cubic}$ in the sector with total angular momentum $N/2$. The Hartree-Fock energy $E^{\rm HF}_0(H_{\rm cubic})$ is evaluated at the global minimum of Eq. (\ref{eq.cubic model.HF approximation}). Figure \ref{fig.Hcubic.groundstate energy.N1000.compare E0,EHF,Eccsd}(a) shows the difference between $E_0(H_{\rm cubic})$ and $E^{\rm HF}_0(H_{\rm cubic})$ when $N=1000$. To accentuate small differences, the fractional error $\frac{|E^{\rm HF}_0-E_0|}{|E_0|}$ is plotted on logarithmic scale in the vertical axis. Each curve shows the error as a function of $s$ at constant $\lambda$, representing a vertical slice through the phase diagram [cf. Fig. \ref{fig.phase diagram.Hcubic.N1000 and N25}(a)]. The three curves convey an overall picture of the system's behavior. Generally speaking, although $E^{\rm HF}_0(H_{\rm cubic})$ approximates $E_0(H_{\rm cubic})$ quite accurately when $s\lesssim 0.4$, the error is significant when $s\gtrsim 0.4$. 

\begin{figure}[h]
\begin{center}
\includegraphics[scale=0.7]{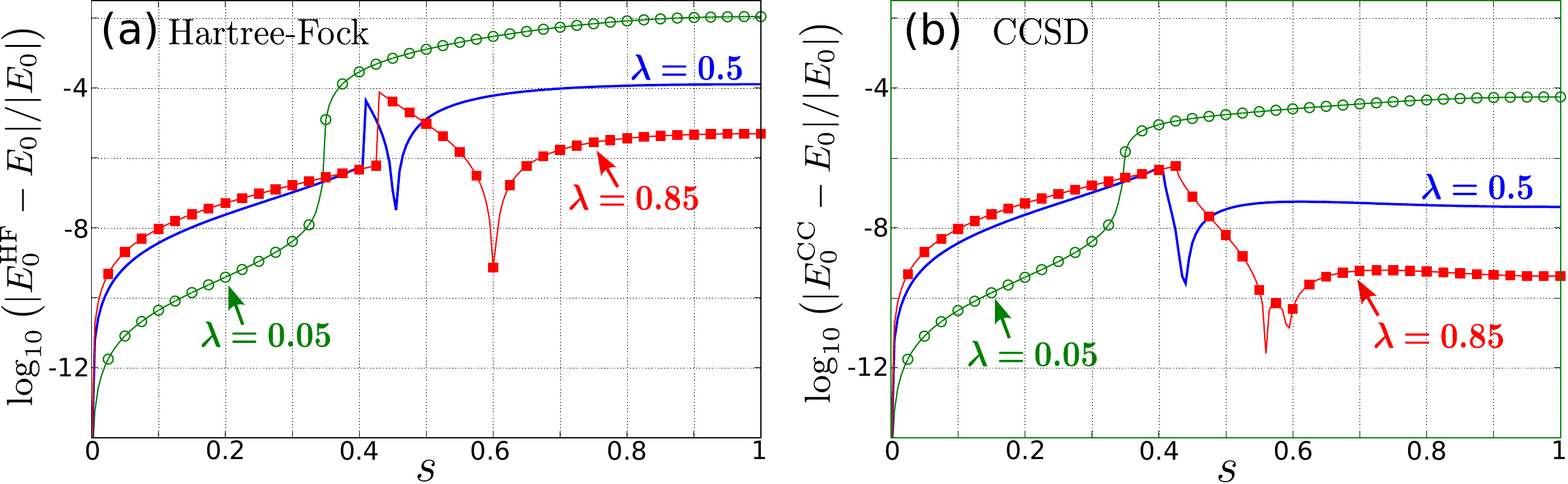}
\caption{Comparing the ground state energy of the cubic model $E_0(H_{\rm cubic})$ with (a) Hartree-Fock energy $E^{\rm HF}_0(H_{\rm cubic})$ and (b) CCSD energy $E^{\rm CC}_0(H_{\rm cubic})$, for $N=1000$. Each curve shows the logarithm of the fractional error as a function of $s$ at constant $\lambda$. It is seen that $E^{\rm CC}_0(H_{\rm cubic})$ gives an accurate approximation of $E_0(H_{\rm cubic})$ when $N$ is large.}
\label{fig.Hcubic.groundstate energy.N1000.compare E0,EHF,Eccsd}
\end{center}
\end{figure}

\subsubsection{Coupled-Cluster singles doubles (CCSD) approximation}
\label{subsubsec.CCSD of Hcubic.E0CC}

We now improve upon Hartree-Fock energy via CCSD approximation. Details are given in Appendices \ref{app.random field formulation of CCSD} and \ref{app.ferromagnetic model.CCSD}. Here we simply outline the procedure. 

The CCSD excitation operator $T^{\rm SD}$ is given by Eq. (\ref{eq.app.TSD.definition}). The singles equation is obtained by substituting Eqs. (\ref{eq.app.list.ccsd.1flip.Jx}), (\ref{eq.app.list.ccsd.1flip.Jx2}), and (\ref{eq.app.list.ccsd.1flip.Jz3}) into $\langle 1|e^{-T^{\mathrm{SD}}} H_{\mathrm{cubic}}e^{T^{\mathrm{SD}}} |0\rangle=0$, and the doubles equation by substituting Eqs. (\ref{eq.app.list.ccsd.2flip.Jx}), (\ref{eq.app.list.ccsd.2flip.Jx2}), and (\ref{eq.app.list.ccsd.2flip.Jz3}) into $\langle 12|e^{-T^{\mathrm{SD}}} H_{\mathrm{cubic}}e^{T^{\mathrm{SD}}} |0\rangle=0$. The equations are solved numerically to obtain the CCSD parameters $y$ and $w$ in Eq. (\ref{eq.app.TSD.definition}), which are then substituted into $\langle 0|e^{-T^{\mathrm{SD}}} H_{\mathrm{cubic}}e^{T^{\mathrm{SD}}} |0\rangle$ via Eqs. (\ref{eq.app.list.ccsd.energy.Jx}), (\ref{eq.app.list.ccsd.energy.Jx2}), and (\ref{eq.app.list.ccsd.energy.Jz3}), yielding the CCSD energy $E_0^{\rm CC}(H_{\rm cubic})$. Figure \ref{fig.Hcubic.groundstate energy.N1000.compare E0,EHF,Eccsd}(b) shows the fractional error between $E_0(H_{\rm cubic})$ and $E_0^{\rm CC}(H_{\rm cubic})$. It is seen that CCSD leads to significant improvement in accuracy in the region $s\gtrsim 0.4$. This close agreement between $E_0(H_{\rm cubic})$ and $E_0^{\rm CC}(H_{\rm cubic})$ improves even further as $N$ increases. To summarize, when $N$ is large CCSD energy is accurate enough to enable one to correlate it with an energy eigenvalue.

\subsection{Local minimum in regions B and C}
\label{subsec.Hcubic.local min.regions B and C}

\begin{figure}[h]
\begin{center}
\includegraphics[scale=0.7]{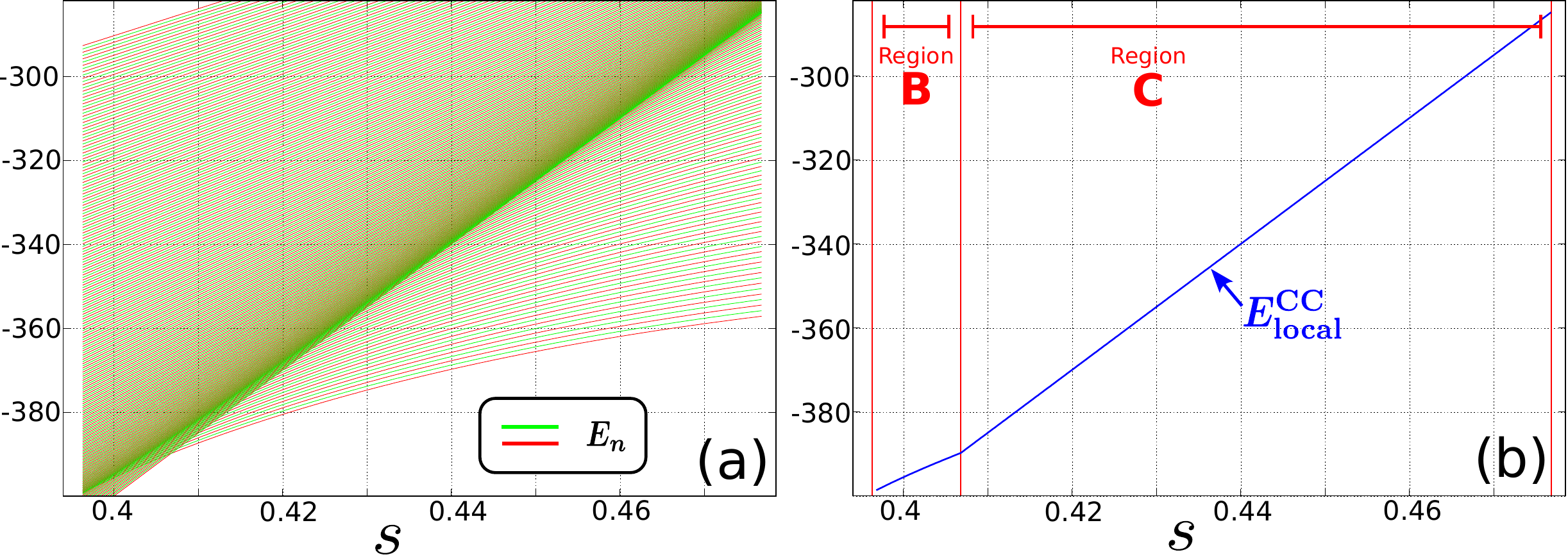}
\caption{Correspondence between quantum spectrum and classical energy of local minimum for the cubic model $H_{\rm cubic}$ ($\lambda=0.5$, $N=1000$). (a) The full energy spectrum as a function of $s$ (sector of $H_{\rm cubic}$ with total angular momentum $N/2$). Alternate energy levels are plotted with the same color. The densely colored band spanning across the diagonal is a region with many avoided crossings among energy levels. (b) The CCSD energy of the local minimum $E_{\rm local}^{\rm CC}(H_{\rm cubic})$ in the same range of $s$. The boundaries of regions B and C [cf. Fig. \ref{fig.phase diagram.Hcubic.N1000 and N25}(a)] are also shown. Comparing the two panels, one sees that the curve (blue) in panel (b) coincides with the band (dense green) in panel (a).}
\label{fig.Hcubic.full En and Eccsd.global view}
\end{center}
\end{figure}

We now discuss the local minimum, here focusing on the one in regions B and C. Its CCSD energy, denoted as $E^{\rm CC}_{\rm local}(H_{\rm cubic})$, is calculated in the same way as $E^{\rm CC}_0(H_{\rm cubic})$, with the Hartree-Fock wave function $|0\rangle$ evaluated at the local minimum of Eq. (\ref{eq.cubic model.HF approximation}). Figure \ref{fig.Hcubic.full En and Eccsd.global view}(b) shows $E^{\rm CC}_{\rm local}(H_{\rm cubic})$ as a function of $s$ ($\lambda=0.5$, $N=1000$). To track $E^{\rm CC}_{\rm local}(H_{\rm cubic})$ as it evolves amidst the energy eigenvalues, one needs to look at the entire energy spectrum. Figure \ref{fig.Hcubic.full En and Eccsd.global view}(a) shows the spectrum of $H_{\rm cubic}$ in the sector with total angular momentum $N/2$. We have plotted alternate levels using the same color to help the reader distinguish between individual eigenvalues. Each curve in panel (a) shows an energy level $E_n(H_{\rm cubic})$ ($n=0,\,\cdots,N$) as a function of $s$. One discerns a densely colored band (green) spanning across the diagonal, which is a region where many energy levels undergo avoided crossings. Notice that the curve of $E^{\rm CC}_{\rm local}(H_{\rm cubic})$ coincides with this band, which can be seen by comparing the two panels visually.

\begin{figure}[h]
\begin{center}
\includegraphics[scale=0.7]{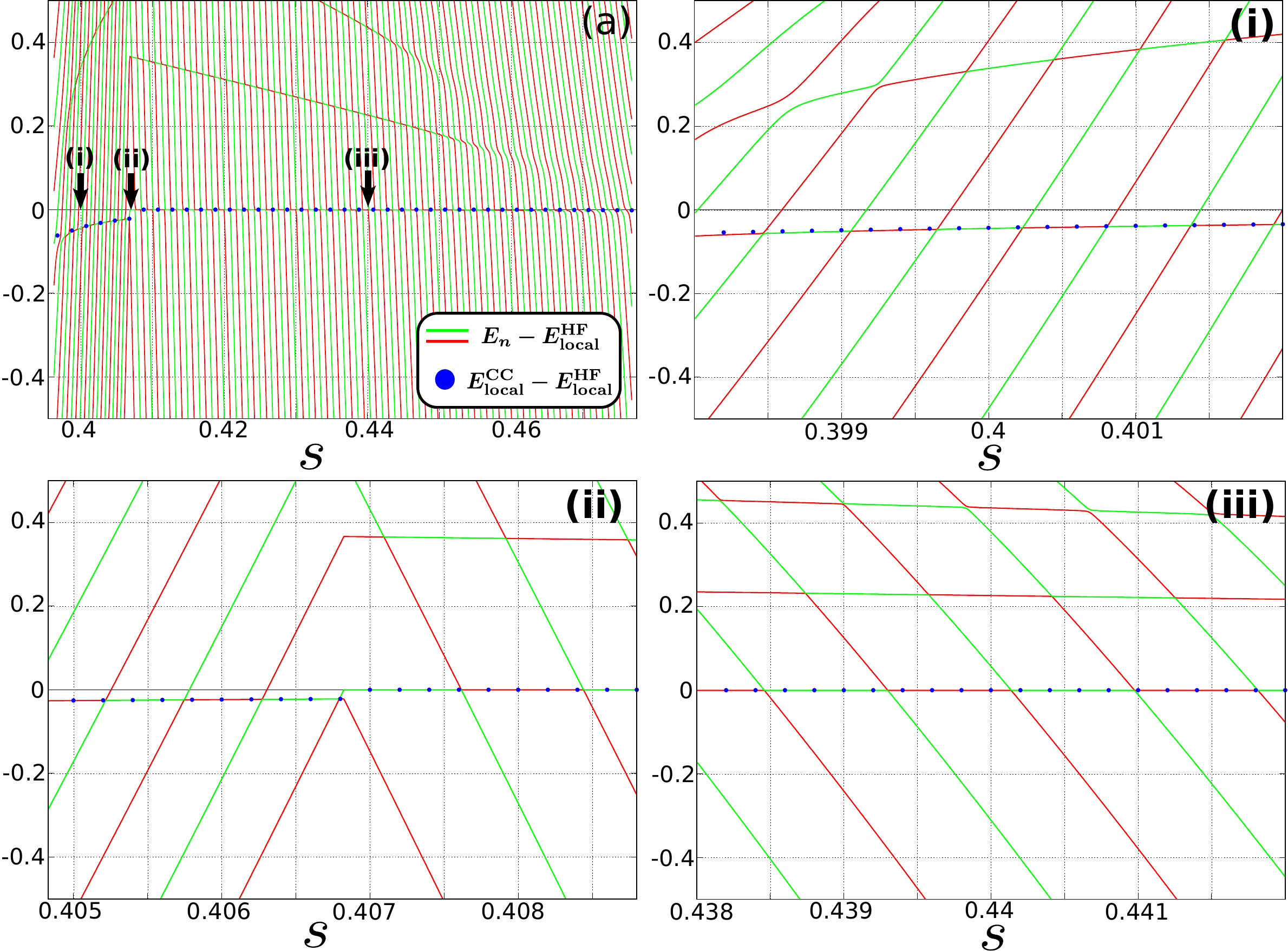}
\caption{Closed-up view of Fig. \ref{fig.Hcubic.full En and Eccsd.global view}, with the curve of $E_{\rm local}^{\rm CC}(H_{\rm cubic})$ of Fig. \ref{fig.Hcubic.full En and Eccsd.global view}(b) superimposed on the full spectrum of Fig. \ref{fig.Hcubic.full En and Eccsd.global view}(a). The vertical axes here have been adjusted by subtracting the Hartree-Fock energy of the local minimum $E_{\rm local}^{\rm HF}(H_{\rm cubic})$ from $E_{\rm local}^{\rm CC}$ and all the energy levels of the spectrum. The legend shown in panel (a) applies to all panels. (a) Close-up view of the horizontal axis (i.e. around the Hartree-Fock energy). (i) Close-up view of the region around $s=0.4$, indicated in panel (a). (ii) Around $s=0.407$, where the first-order phase transition occurs. (iii) Around $s=0.44$. Overall, one sees that the CCSD energy of the local minimum coincides with the lowest envelope.}
\label{fig.Hcubic.closeup of local min. 4 panels}
\end{center}
\end{figure}

Figure \ref{fig.Hcubic.closeup of local min. 4 panels} zooms in on the region where the CCSD energy coincides with the band. To facilitate visual inspection, the curve of $E^{\rm CC}_{\rm local}(H_{\rm cubic})$ is superimposed on the full spectrum $E_n(H_{\rm cubic})$, with the Hartree-Fock energy $E^{\rm HF}_{\rm local}(H_{\rm cubic})$ subtracted from both energies. Panel (a) shows a close-up view of 
a narrow range of energy around the horizontal axis. One sees that an energy level sometimes exhibits a `z'-shaped kink in its curve, which is due to a succession of two avoided crossings with neighboring levels. Interestingly, these kinks chain together to form an envelope of avoided crossings, and several of these are seen in panel (a). The lowest one lie along the horizontal axis, which incidentally is also the Hartree-Fock energy. After CCSD correction, we see that the energy of the local minimum (blue circles) agrees almost exactly with the lowest envelope. Panels (i) to (iii) zoom in on three locations, indicated in panel (a), for closer inspection.

\begin{figure}[h]
\begin{center}
\includegraphics[scale=0.75]{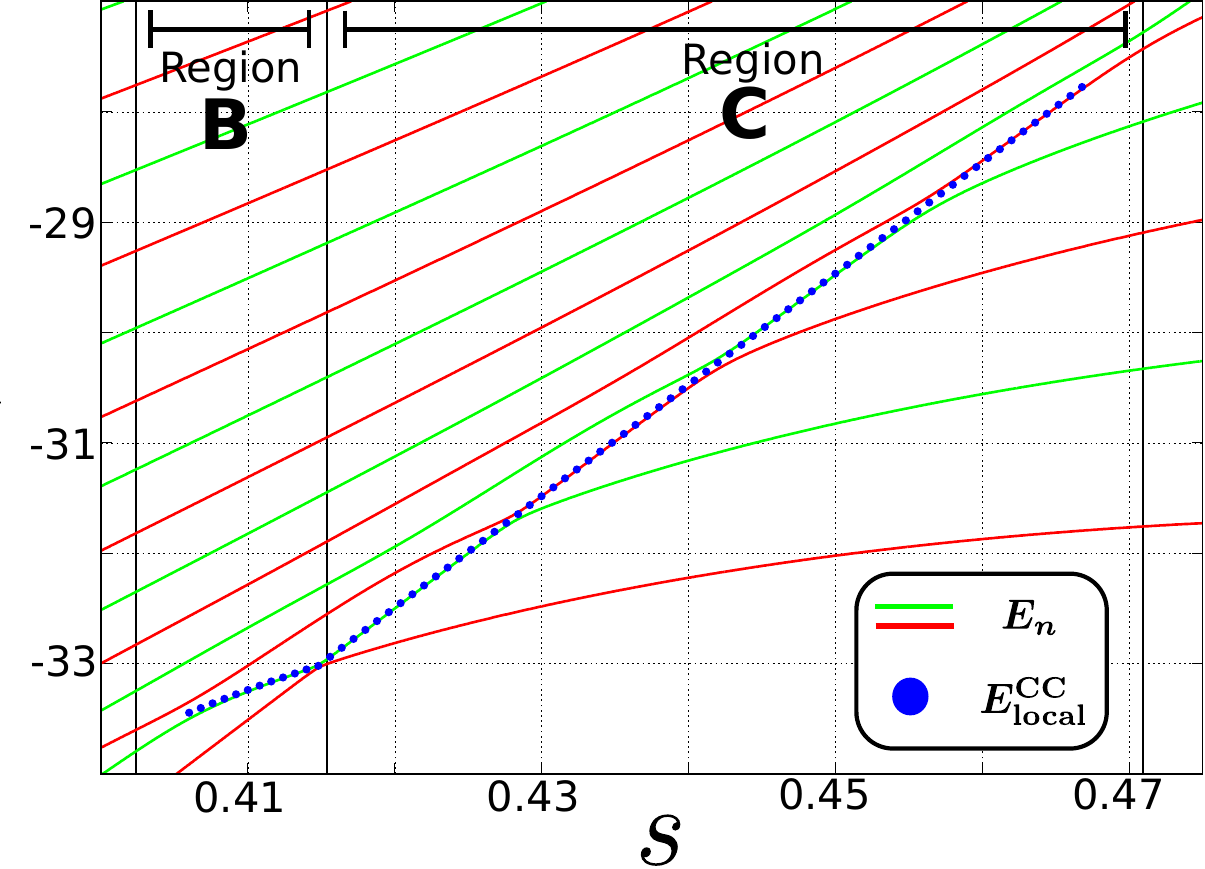}
\caption{Correspondence between the envelope of avoided crossings and the CCSD energy of local minimum $E^{\rm CC}_{\rm local}(H_{\rm cubic})$ when the system size is small ($N=75$, $\lambda=0.65$). The figure is organized similar to Fig. \ref{fig.Hcubic.full En and Eccsd.global view}, this time with $E^{\rm CC}_{\rm local}$ superimposed on the spectrum. The quantum-classical correspondence is observable even at this smaller system size.}
\label{fig.Hcubic.juxtapose Eccsd on En.N 75}
\end{center}
\end{figure}

Hence, there appears to be a quantum-classical correspondence between the local minimum and the energy spectrum. Usually, one associates a classical energy state with a single energy level, as in the case of the ground state. Here, the local minimum is associated instead with an envelope of avoided crossings formed by concatenation of consecutive energy levels. This quantum-classical correspondence between the envelope and the local minimum improves when $\lambda$ is closer to unity (constant $N$), and also as $N$ increases. Nevertheless, the correspondence is already apparent at smaller $N$. Figure \ref{fig.Hcubic.juxtapose Eccsd on En.N 75} shows the CCSD energy superimposed on the energy spectrum at a smaller system size ($N=75$, $\lambda=0.65$). Again, one sees that the CCSD energy coincides with the envelope of avoided crossings.

\subsection{Wave functions along the envelope of avoided crossings}
\label{}

Let us examine the classical and quantum wave functions along the envelope. The energy eigenfunctions $| E_n(H_{\rm cubic})\rangle$ are obtained from diagonalization and we represent them using the eigenfunctions of the $J_z$ operator $|m_z\rangle$. The Hartree-Fock wave function of the local minimum $|E^{\rm HF}_{\rm local}(H_{\rm cubic})\rangle$ is also represented using the same basis. Details on $|m_z\rangle$ and the calculation of $\langle m_z|E^{\rm HF}_{\rm local}\rangle$ are given in Appendix \ref{app.wavefunctions}.

\begin{figure}[h]
\begin{center}
\includegraphics[scale=0.9]{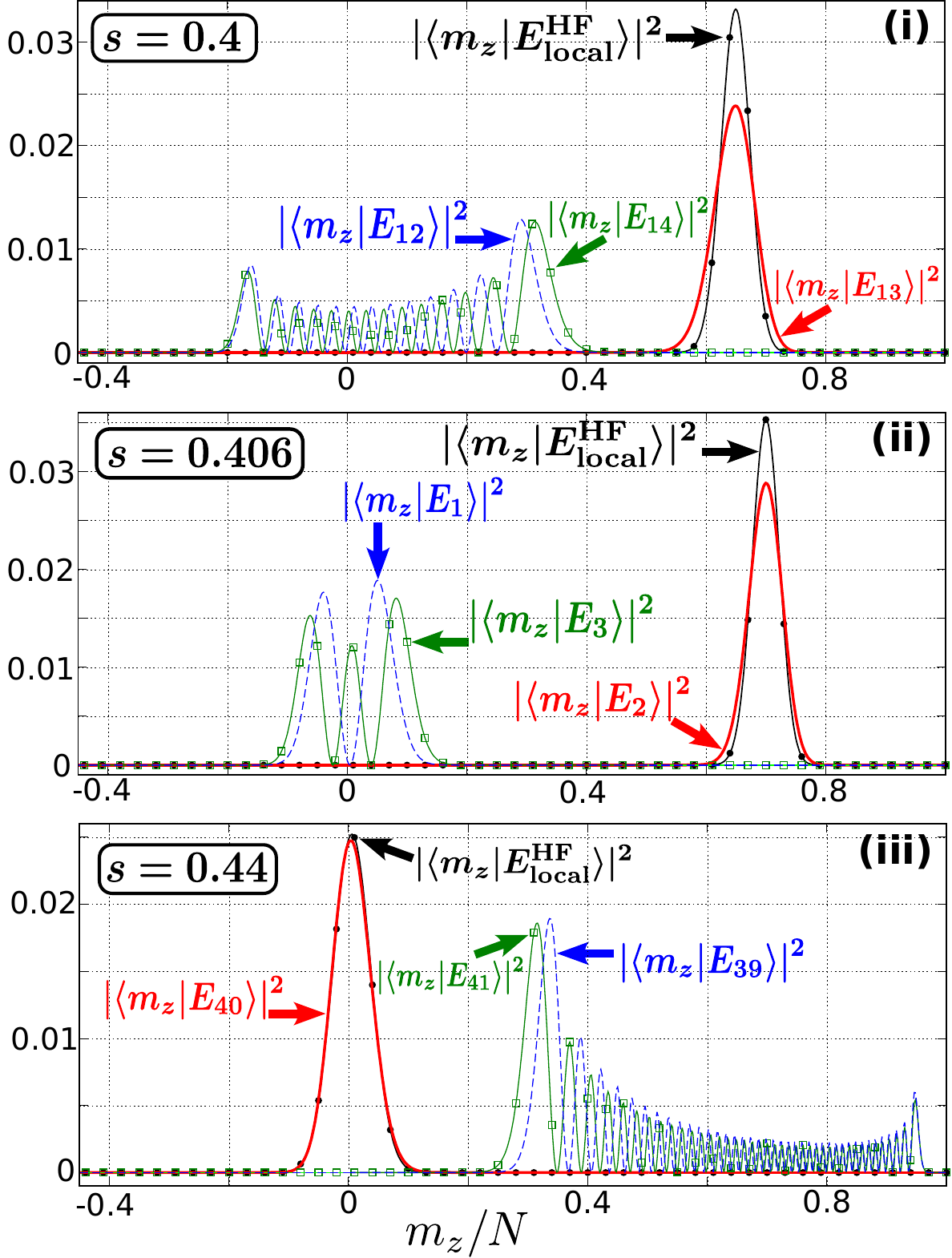}
\caption{Wave functions at three values of $s$ along the envelope shown in Fig. \ref{fig.Hcubic.closeup of local min. 4 panels}(a). (i) $s=0.4$. (ii) $s=0.406$. (iii) $s=0.44$. In each panel, the probability densities of the Hartree-Fock wave function $\langle m_z|E_{\rm local}^{\rm HF}\rangle$ (circle) and the three nearest energy eigenfunctions $\langle m_z | E_{\tilde{n}}\rangle$ (solid) and $\langle m_z | E_{\tilde{n}\pm 1}\rangle$ (dashed and square) are plotted. One sees that the classical $\langle m_z|E_{\rm local}^{\rm HF}\rangle$ has a quantum counterpart in $\langle m_z | E_{\tilde{n}}\rangle$.}
\label{fig.Hcubic.compare psis at 3 s.region B and C}
\end{center}
\end{figure}

To evince the correspondence between the wave functions, we define $\tilde{n}$ as the level label such that the energy difference $|E^{\rm HF}_{\rm local}(H_{\rm cubic})-E_{\tilde{n}}(H_{\rm cubic})|$ (fixed $s$ and $\lambda$) is the smallest among all the energy eigenvalues; in other words, the eigenvalue closest to $E^{\rm HF}_{\rm local}$ is $E_{\tilde{n}}$. Figure \ref{fig.Hcubic.compare psis at 3 s.region B and C} compares the probability densities $|\langle m_z|E^{\rm HF}_{\rm local}\rangle|^2$ (black circles) and $|\langle m_z|E_{\tilde{n}}\rangle|^2$ (red solid) at three locations of the envelope in Fig. \ref{fig.Hcubic.closeup of local min. 4 panels}(a); specifically, panels (i) to (iii) show for $s=0.4$, 0.406, and 0.44, respectively. The densities of the two neighboring energy levels $|\langle m_z|E_{\tilde{n}\pm 1}\rangle|^2$ (blue dashed, green square) are also shown, for comparison. One sees that there is close correspondence between the classical state at the local minimum and the nearest energy eigenstate \cite{footnote.concerning using HF psi over CCSD psi in Hcubic}.

\clearpage

\begin{figure}[h]
\begin{center}
\includegraphics[scale=0.9]{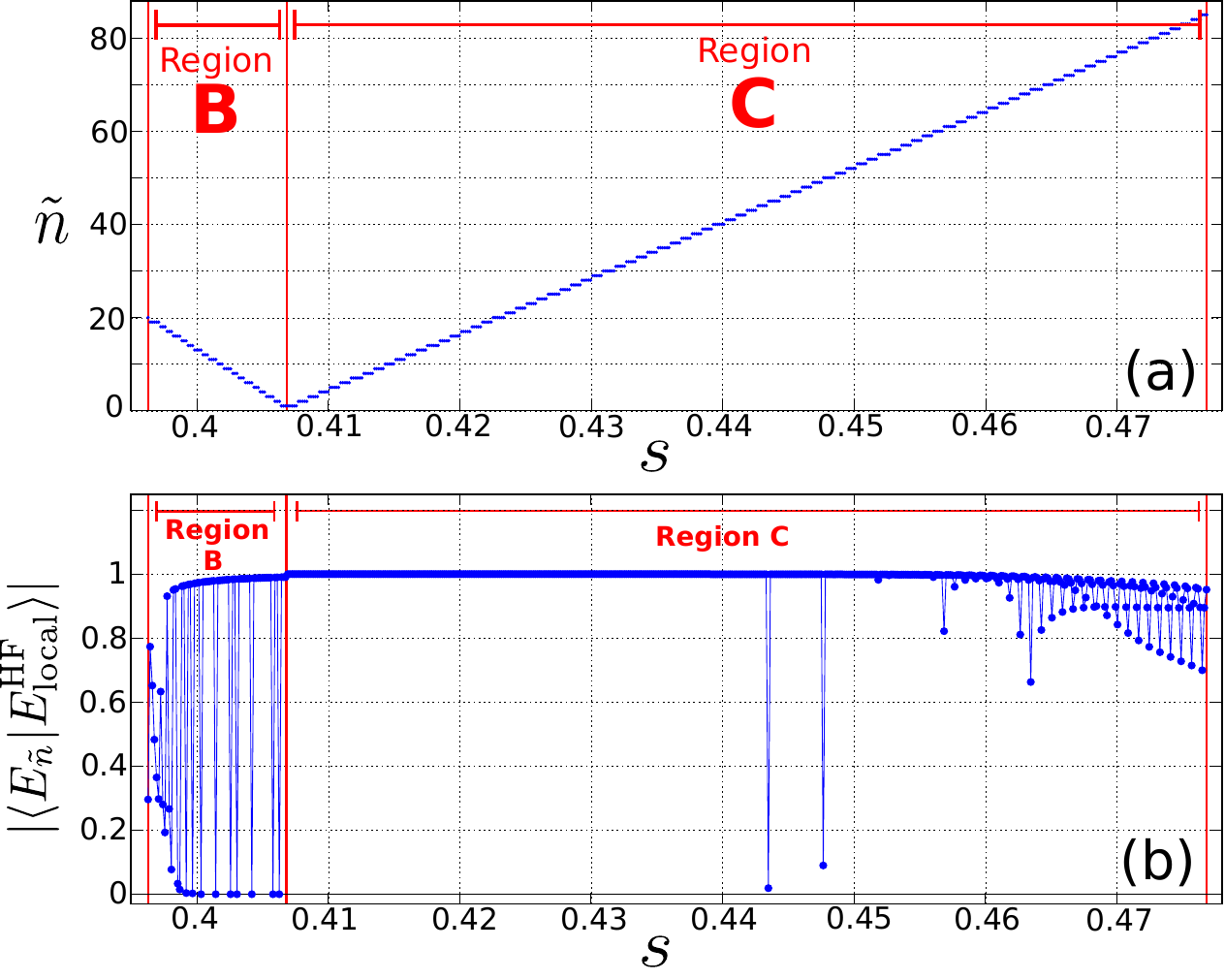}
\caption{(a) Level label $\tilde{n}$ along the envelope in Fig. \ref{fig.Hcubic.closeup of local min. 4 panels}(a). (b) Overlap $\left\langle E_{\tilde{n}}| E_{\rm local}^{\rm HF} \right\rangle$ between the Hartree Fock wave function at the local minimum and the nearest energy eigenfunction, along the envelope. Lines connecting data points (circles) are to guide the eye only. The overlap is close to unity along the entire envelope.}
\label{fig.Hcubic.n tilde and overlap}
\end{center}
\end{figure}

A more comprehensive picture can be attained by monitoring the overlap between the wave functions. Panel (a) of Fig. \ref{fig.Hcubic.n tilde and overlap} shows the level label $\tilde{n}$ along the envelope in Fig. \ref{fig.Hcubic.closeup of local min. 4 panels}(a), and panel (b) shows the overlap $\left\langle E_{\tilde{n}}| E_{\rm local}^{\rm HF}\right\rangle$. We see that the overlap is very close to unity along the entire envelope. (Sudden deteriorations at isolated points are due to avoided crossings, where $E_{\tilde{n}}$ and $E_{\tilde{n}-1}$ are almost degenerate.)

We close this section with a short comment. In this model, we were able to evince the quantum-classical correspondence partly because the Hamiltonian commutes with the total angular momentum. This had allowed us to focus on the energy spectrum of one sector (with $N+1$ levels) instead of the entire spectrum (with $2^N$ levels), greatly simplifying our visual inspection of the classical and quantum energies. In general, however, when the Hamiltonian does not exhibit any symmetry, such correspondences---if present---will be difficult to discover due to the much higher density of levels in the spectrum.

\subsection{Unphysical local minimum in region E}
\label{}

\begin{figure}[h]
\begin{center}
\includegraphics[scale=0.75]{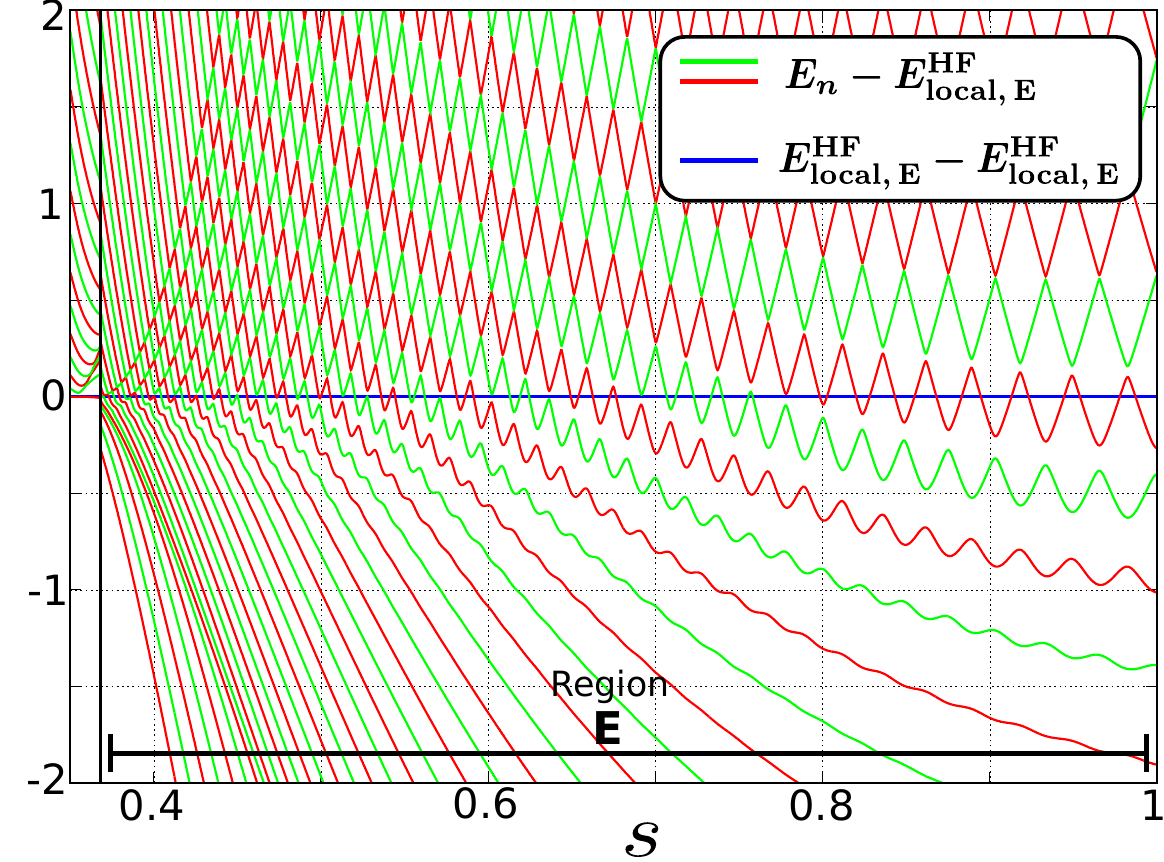}
\caption{No quantum-classical correspondence for the local minimum in region E ($N=100, \lambda=0.1$). The Hartree-Fock energy of the local minimum, denoted $E^{\rm HF}_{\rm local,\,E}(H_{\rm cubic})$ (blue), is superimposed on the energy spectrum (red and green). The vertical axis has been shifted by $E^{\rm HF}_{\rm local,\,E}$ , as in Fig. \ref{fig.Hcubic.closeup of local min. 4 panels}. Visually, there appears to be no correspondence between the classical and quantum energies.} 
\label{fig.regionE.En and EHFunphys}
\end{center}
\end{figure}

One might ask whether all the local minima on the classical energy landscape has correspondences in the quantum spectrum. Figure \ref{fig.regionE.En and EHFunphys} compares the Hartree-Fock energy of the local minimum in region E, denoted $E_{\rm local,\,E}^{\rm HF}(H_{\rm cubic})$, with the energy spectrum (for $N=100, \lambda=0.1$). The figure is organized similar to Fig. \ref{fig.Hcubic.closeup of local min. 4 panels}, with the Hartree-Fock energy subtracted along the vertical axis. One sees that the energy of the local minimum (horizontal axis, blue) cuts across many levels, and visually there does not appear to be any correspondence between the classical and quantum energies. We tried solving the CCSD equations at this local minimum, but were unable to obtain any solutions (perhaps another indication that the minimum is not physically meaningful). In analogy to Fig. \ref{fig.Hcubic.n tilde and overlap}(b), we examined the overlap of the Hartree-Fock wave function with the nearest energy eigenfunction, and found that the overlap is small (between 0 and 0.6). Hence, it is also possible that a local minimum is unphysical.

%
%


\section{Quantum-classical correspondence in a frustrated system}
\label{sec. Hopfield model section label}

\subsection{Hopfield model in transverse field}
\label{}

The cubic model is not frustrated and its energy surface is relatively simple. Let us now consider a frustrated one, the Hopfield model in transverse field \cite{Ma.and.Gong.93,Ma.etal.93,Nishimori96}
\begin{equation}
H_{\mathrm{hop}}=-\frac{1}{2N}\sum_{\mu=1}^{p}\left[J_z(\bm{\xi^{\mu}})\right]^2 - \Gamma J_x
\label{eq.Hhop.hamiltonian.definition}
\end{equation}
where $J_z(\bm{\xi^{\mu}})$ is defined by Eq. (\ref{eq.definition of generalized operator J(a)}), and $\Gamma$ is the strength of the transverse field. The original Hopfield model, without the transverse field, was proposed in a classical context within the field of associative memory \cite{AmitBOOK}. $\bm{\xi^{\mu}}$ is an $N$-dimensional binary vector with components $+1$ or $-1$. A total of $p$ such vectors are embedded as memory patterns, which are to be retrieved from the temporal dynamics of the system. The original model was subsequently generalized to Eq. (\ref{eq.Hhop.hamiltonian.definition}) by various authors \cite{Ma.and.Gong.93,Ma.etal.93,Nishimori96}. Recently, this model is receiving renewed attention in diverse fields such as quantum machine learning \cite{Rebentrost18}, image recognition \cite{Liu20}, and quantum computing \cite{Miller21}. 

Apart from its wide range of applications, we are motivated to study this model mainly because it is very flexible in the sense that the complexity of its energy landscape is tunable via the embedded patterns $\bm{\xi^{\mu}}$. For finite loading where the ratio $\frac{p}{N}$ is non-zero in the thermodynamic limit, the Hopfield model crosses over into the Sherrington-Kirkpatrick model and exhibits spin glass behavior. However, as the dimension of the Hilbert space is very large, it is not feasible to diagonalize the Hamiltonian matrix numerically in this regime. On the other hand, in the low loading regime where $p\ll N$ at finite $N$, the energy landscape still retains some vestige of ruggedness where local minima are located near to the patterns $\bm{\xi^{\mu}}$ in configuration space. It is in this regime that we shall study the model since we have some control over the energy surface via our choice of patterns, and also because we can study a small system size whose Hamiltonian matrix is amenable to exact numerical diagonalization. The case of uncorrelated patterns (each $\xi^{\mu}_i=+1$ or $-1$ with equal probability) in the low loading regime has been studied analytically using mean-field approximation by Nishimori and Nonomura \cite{Nishimori96}. Using our Hartree-Fock framework, discussed below, we were able to verify their predictions on the structure of the free energy landscape. On the other hand, the case of correlated patterns is harder to treat analytically and has received lesser attention by comparison. We shall therefore focus on such patterns in this work.

It is not feasible to perform an exhaustive study of all possible pattern combinations. In the following we shall focus instead on just one set of correlated patterns \cite{footnote.the patterns being used in Hhop}. Table \ref{tab.correlation amg patts and HF solns} shows the correlations among the patterns as measured by the inner product $\bm{\xi^{\mu} \cdot \xi^{\nu}}=N^{-1}\sum_{i=1}^N \xi^{\mu}_i \xi^{\nu}_i$. It is seen that there is significant correlation among the embedded patterns. This instantiation has parameters $N=16$ and $p=4$. We are limited to these small values because it is computationally expensive both to obtain the energy eigenvalues and eigenstates, and to perform a thorough sampling of the energy landscape.

\begin{table}[h]
\begin{center}
\begin{tabular}{c | ccccccc c |c}
\hline
\hline
 & \multicolumn{7}{c}{$\bm{a \cdot b}$} & &\\
\cline{1-9}  
\diagbox[innerwidth=2.25cm, height=1.0cm,innerleftsep=0.25cm, innerrightsep=0.35cm]{$\bm{a}$}{$\bm{b}$} & $\bm{\xi^1}$ &\,\,\,\,\,& $\bm{\xi^2}$ &\,\,\,\,\,& $\bm{\xi^3}$ &\,\,\,\,\,& $\bm{\xi^4}$ & & \,\,\, Energy \,\,\, \\ 
\cline{1-9} \cline{10-10}
$\bm{\xi^1}, \bm{\vec{\alpha}_5}$  & 1 && 0 && 0 && 0 && -8\\
$\bm{\xi^2}, \bm{\vec{\alpha}_4}$  & 0 && 1  && -0.25  && 0 && -8.5 \\
$\bm{\xi^3}, \bm{\vec{\alpha}_2}$  & 0 && -0.25  && 1 && 0.25 && -9\\
$\bm{\xi^4}, \bm{\vec{\alpha}_3}$  & 0 && 0 && 0.25 && 1 && -8.5\\
$\bm{\vec{\alpha}_1}$  & 0.25 && 0.5 && -0.75 && -0.5 && -9 \\
$\bm{\vec{\alpha}_6}$  & 0.5 && -0.5 && 0 && 0.5      && -6\\
\hline
\hline
\end{tabular}
\caption{Correlations among the embedded patterns $\bm{\xi^{\mu}}$ (c.f. \cite{footnote.the patterns being used in Hhop}) and Hartree-Fock solutions $\bm{\vec{\alpha}_n}$ (at $\Gamma=0$) for the instance of $H_{\rm hop}$ being studied ($N=16$, $p=4$). Correlation between $\bm{a}$ and $\bm{b}$ is measured by $\bm{a\cdot b}=N^{-1}\sum_{i=1}^Na_ib_i$.  The last column lists the energy Eq. (\ref{eq.HFenergy.Hhop}) at $\bm{a}$ with $\Gamma=0$.}
\label{tab.correlation amg patts and HF solns}
\end{center}
\end{table}

\subsection{Hartree-Fock approximation}
\label{}

As $\xi_i^{\mu}$ is binary, there are $2^p$ possible combinations of the tuple $(\xi_i^{1},\cdots,\xi_i^p)$, and each spin label $i$ belongs to one of the combinations. Let a combination be labeled $\gamma$. The $N$ spins can then be divided into $2^p$ groups where all the spins in each group belongs to the same combination. The Hamiltonian Eq. (\ref{eq.Hhop.hamiltonian.definition}) can be rewritten as
\begin{equation}
H_{\mathrm{hop}}
=
-\frac{1}{2N}
\sum_{\mu=1}^{p}
\left[
\sum_{\gamma=1}^{2^p}
\xi^{\mu}_{\gamma}
J_z^{\gamma}
\right]^2 
- 
\Gamma 
\sum_{\gamma=1}^{2^p}
J^{\gamma}_x
\label{eq.Hhop.hamiltonian.regroup into gamma groups}
\end{equation}
where $\xi_{\gamma}^{\mu}$ denotes the component of the $\mu$th pattern in the tuple $\gamma$, and modifying Eq. (\ref{eq.definition of generalized operator J(a)}) slightly we define
\begin{equation}
J_{\alpha}^{\gamma}=\sum_{i_{\gamma}=1}^{N_{\gamma}}\sigma_{i_{\gamma}}^{\alpha}
\label{}
\end{equation}
where $i_{\gamma}$ labels a spin in the group of $\gamma$, and $N_{\gamma}$ is the total number of spins in the group. Equation (\ref{eq.Hhop.hamiltonian.regroup into gamma groups}) simply makes the grouping of the spins manifest, and is equivalent to Eq. (\ref{eq.Hhop.hamiltonian.definition}). The point is that one can work with a Hartree-Fock wave function with reduced dimensionality 
\begin{equation}
|0\rangle=\prod_{\gamma=1}^{2^{p}}
\left[
\prod_{i_{\gamma}=1}^{N_{\gamma}}
{\alpha_{i_{\gamma}}\choose \beta_{i_{\gamma}}}
\right]
=
\prod_{\gamma=1}^{2^{p}}
{\alpha_{\gamma} \choose \beta_{\gamma}}^{N_{\gamma}}
\label{}
\end{equation}
instead of Eq. (\ref{eq.HF wavefunction.definition}). The second equality emphasizes that all the spins within a group exhibits permutation symmetry. The classical energy surface is then
\begin{align}
\langle 0| H_{\rm hop} |0 \rangle
=
&
-
\frac{1}{2N}
\sum_{\mu=1}^{p}
\left[
\sum_{\gamma}^{} N_{\gamma} \xi_{\gamma}^{\mu}(2\alpha_{\gamma}^2-1)
\right]^2
+
\frac{p}{2N}
\sum_{\gamma}N_{\gamma}(2\alpha_{\gamma}^2-1)^2
\nonumber \\
&
-
\frac{p}{2}
-
2\Gamma\sum_{\gamma}^{}N_{\gamma}\alpha_{\gamma}\sqrt{1-\alpha_{\gamma}^2}
\label{eq.HFenergy.Hhop}
\end{align}
Equation (\ref{eq.HFenergy.Hhop}) is valid for arbitrary patterns. Substituting the instantiation of patterns mentioned above leads to a seven-dimensional function for $\langle 0 |H_{\rm hop}|0\rangle$.

\subsection{Visualization of high-dimensional energy landscape}
\label{}

In this section, we examine the classical energy Eq. (\ref{eq.HFenergy.Hhop}) in detail.

\subsubsection{$\Gamma=0$}

Let us first consider the case of $\Gamma=0$ (i.e. no transverse field). We obtained all the energy minima numerically. An initial condition is randomly chosen, and its energy minimized using zero temperature Metropolis algorithm. This process is repeated, each time starting from a different initial condition, until all minima are found. Six solutions, and their spin-flipped counterparts, were obtained. Each solution is then converted back to an $N$-dimensional spin-labeled vector $\bm{\vec{\alpha}}$ where the component of the $i_{\gamma}$th spin $(\bm{\vec{\alpha}})_{i_{\gamma}}$ is related to its Hartree-Fock solution $\alpha_{\gamma}$ by 
\begin{equation}
(\bm{\vec{\alpha}})_{i_{\gamma}}=2\alpha_{\gamma}^2-1
\label{}
\end{equation}
Four of the six solutions coincide with the embedded patterns. The correlations of the remaining two solutions, denoted $\bm{\vec{\alpha}_1}$ and $\bm{\vec{\alpha}_6}$, are shown in Table \ref{tab.correlation amg patts and HF solns}. These so-called spurious states correspond to incorrect recall of the memorized patterns. The last column of Table \ref{tab.correlation amg patts and HF solns} gives the Hartree-Fock energies of the six solutions. It is seen that the spurious state $\bm{\vec{\alpha}_1}$ is one of the ground states.

\begin{figure}[h]
\begin{center}
\includegraphics[scale=0.9]{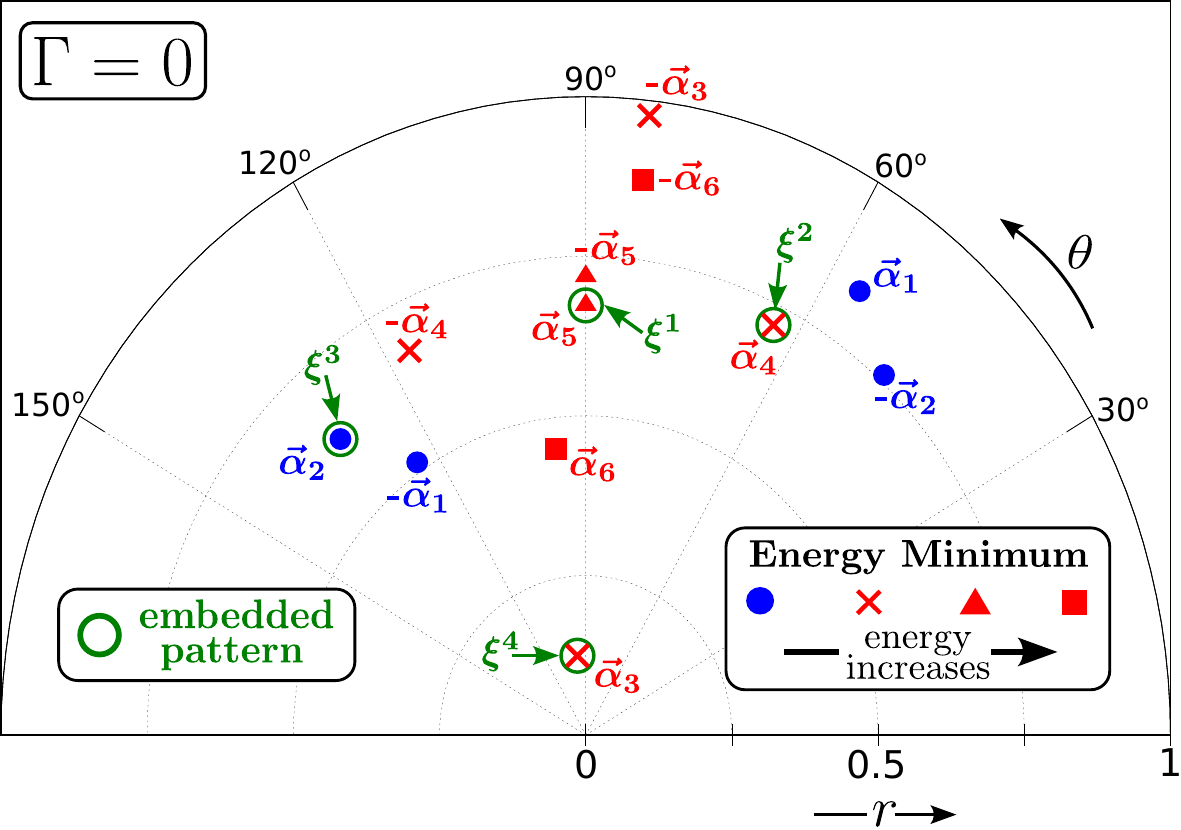}
\caption{Polar plot showing the relative positions of all the energy minima of Eq. (\ref{eq.HFenergy.Hhop}) when $\Gamma=0$, instantiated by the patterns \cite{footnote.the patterns being used in Hhop}. The coordinates $(r,\theta)$ of a Hartree-Fock solution $\bm{\vec{\alpha}_n}$ are plotted as the radial and angular displacements, with $\bm{\omega}$ and $\bm{p}$ given by \cite{footnote.omega and pole vectors}. Spin-flipped solutions are denoted -$\bm{\vec{\alpha}_n}$. Global minima are plotted with solid circles (blue), local minima in red, and the embedded patterns with open circles (green).} 
\label{fig.polar plot.HF solns and embedded patts}
\end{center}
\end{figure}

The vectors $\bm{\vec{\alpha}_n}$ are high-dimensional. Let us introduce a plotting system to visualize them. An $N$-dimensional vector $\bm{a}$ can be projected onto the polar coordinates $(r,\theta)$, where $r$ is the normalized Euclidean distance of $\bm{a}$ from the \emph{origin} vector $\bm{\omega}$ 
\begin{equation}
r^2=\frac{1}{4N}\sum_{i=1}^N (a_i-\omega_i)^2
\label{eq.r.definition}
\end{equation}
and $\theta$ is the angle between $\bm{a}$ and the \emph{pole} vector $\bm{p}$
\begin{equation}
\theta=\cos^{-1}\left[\frac{\bm{a\cdot p}}{(\bm{a\cdot a})(\bm{p\cdot p})}\right]
\label{eq.theta.definition}
\end{equation}
with the inner product $\bm{a\cdot p}$ defined in the caption of Table \ref{tab.correlation amg patts and HF solns}. Figure \ref{fig.polar plot.HF solns and embedded patts} shows the plot of the polar coordinates of all the energy minima $\bm{\vec{\alpha}_n}$, where $r$ and $\theta$ are plotted as the radial and angular displacements, respectively. We have chosen $\bm{\omega}$ and $\bm{p}$ such that all the $\bm{\vec{\alpha}_n}$ have distinct coordinates \cite{footnote.omega and pole vectors}. The spin-flipped solutions (denoted -$\bm{\vec{\alpha}_n}$) and the embedded patterns (green open circles) are also shown. The global minima are indicated by solid circles (blue), while local minima are plotted in red. 

\begin{figure}[h]
\begin{center}
\includegraphics[scale=0.9]{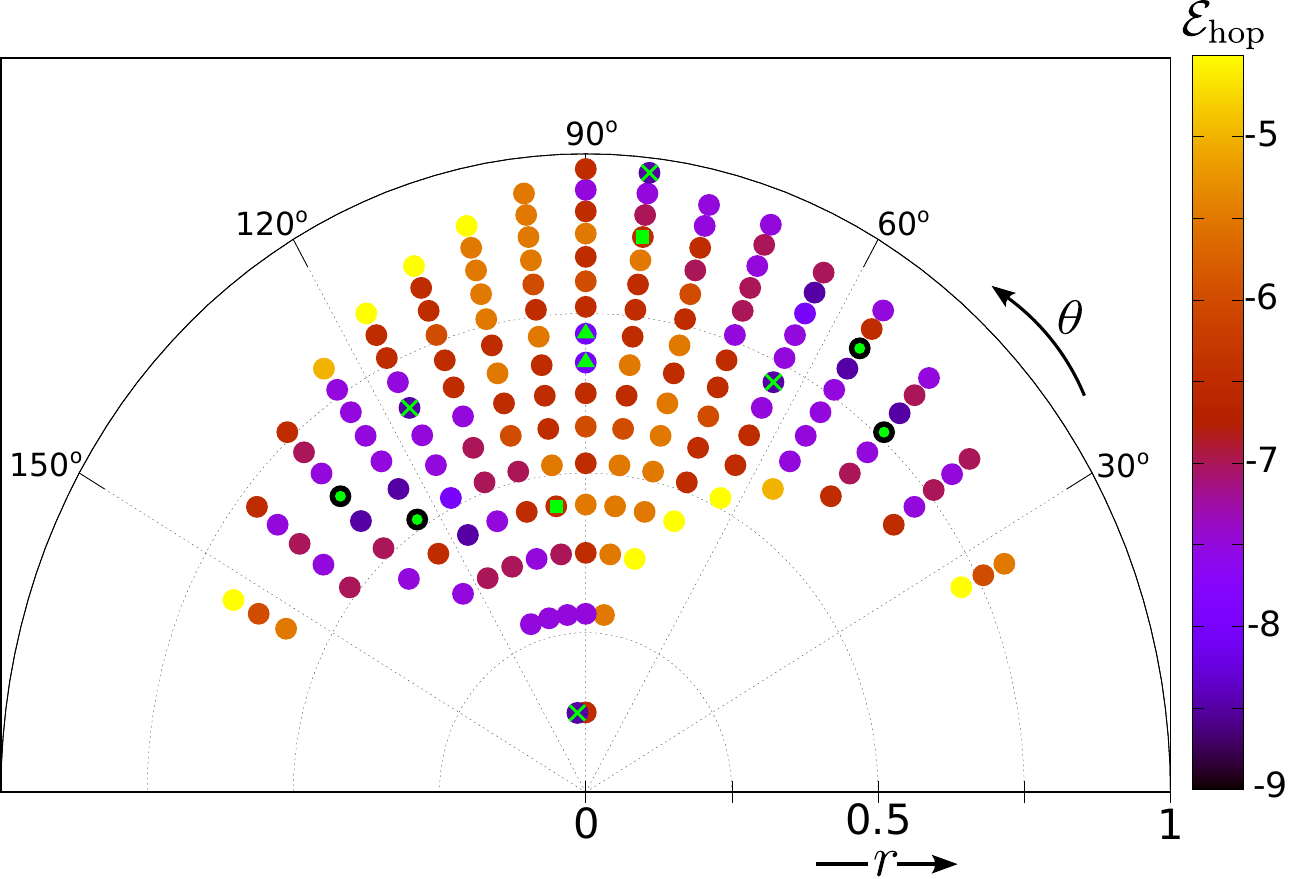}
\caption{Visualizing the landscape of the energy $\mathcal{E}_{\rm hop}$. The energy minima in Fig. \ref{fig.polar plot.HF solns and embedded patts} are reproduced using the same symbols (green). Each solid circle represents the polar coordinates of a point in configuration space, with its energy $\mathcal{E}_{\rm hop}$ indicated by color. Higher-energy regions are indicated by warmer colors. The topography of the high-dimensional energy surface can be discerned from this polar plot.} 
\label{fig.energy landscape.Ehop.original Hopfield energy}
\end{center}
\end{figure}

This method of plotting provides a way to visualize high-dimensional energy landscapes. As a first example, consider the energy of the original Hopfield model, which is defined on the vertices of an $N$-dimensional cube. Let us denote $\mathcal{E}_{\rm hop}$ as the energy of $H_{\rm hop}$  evaluated at the $2^N$ classical spin configurations [i.e. Eq. (\ref{eq.Hhop.hamiltonian.definition}) with $\Gamma=0$]. The definition of $\mathcal{E}_{\rm hop}$ differs slightly from Eq. (\ref{eq.HFenergy.Hhop}) in that the latter is a continuous function and takes into account spin permutation symmetry. Figure \ref{fig.energy landscape.Ehop.original Hopfield energy} shows the energy landscape of $\mathcal{E}_{\rm hop}$ based on the plotting system used for displaying the energy minima. The polar coordinates of a spin configuration is calculated in the same way as for the minima, and the $\mathcal{E}_{\rm hop}$ of the configuration is indicated by color. If more than one configuration is mapped onto the same polar coordinates, the lowest $\mathcal{E}_{\rm hop}$ is adopted. The energy minima shown in Fig. \ref{fig.polar plot.HF solns and embedded patts} are also plotted here, using the same symbols (green). In this heat map, high-energy regions are indicated by warm colors. With this system of plotting, one can gain some intuition into the topography of the energy function $\mathcal{E}_{\rm hop}$. For instance, the relative depth among the minima, the energy barriers separating them, as well as possible transition pathways, can all be discerned from the plot.

This method of plotting can also be applied to the classical energy Eq. (\ref{eq.HFenergy.Hhop}). In this case, the configuration space is continuous, so we set up a discrete rectangular grid in the $r$-$\theta$ plane. A randomly sampled point in configuration space is first projected onto polar coordinates, and then assigned to the appropriate bin in the grid. If the energy of the sampled point is lower than the energy already registered in the bin, replace the previous energy with the new one. This sampling process is repeated until there are no further updates in all the bins. Figure \ref{fig.energy landscape.classical energy.3values of Gamma}(a) shows the energy landscape of $\langle 0|H_{\rm hop}|0\rangle$ when $\Gamma=0$, obtained via uniform sampling of configuration space $10^{11}$ times. For presentation purposes, the polar coordinates are plotted in Cartesian format. Energy minima (green) are also indicated, as in Fig. \ref{fig.energy landscape.Ehop.original Hopfield energy}. The heat map allows us to visualize the ruggedness of the energy landscape, which is usually difficult for high-dimensional surfaces. For instance, we can discern topographical features such as corridors connecting energy basins (narrow blue channels), as well as the shape and size of each basin (black patches).

\subsubsection{$\Gamma>0$}

\begin{figure}[h]
\begin{center}
\includegraphics[scale=0.65]{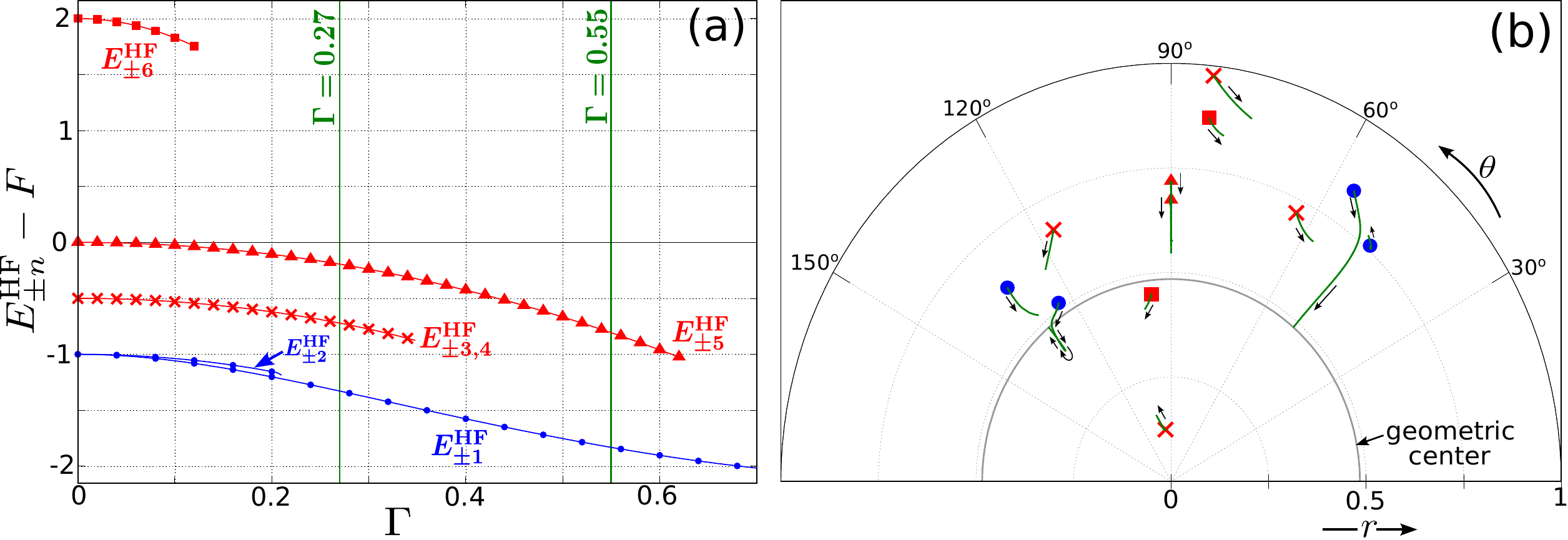}
\caption{Evolutions of the energies and positions of the minima of $\langle 0|H_{\rm hop}|0\rangle$ with $\Gamma$. (a) Hartree-Fock energy $E_n^{\rm HF}(H_{\rm hop})$ of the minimum $\bm{\vec{\alpha}_n}$ as a function of $\Gamma$, adjusted by the free energy $F$. At $\Gamma=0.27$, $\pm\bm{\vec{\alpha}_{2,6}}$ have vanished from the energy surface; at $\Gamma=0.55$, only $\pm\bm{\vec{\alpha}_{1,5}}$ are left. (b) Polar coordinates of the minima. Starting from $\Gamma=0$ (c.f. Fig. \ref{fig.polar plot.HF solns and embedded patts}), each minimum evolves along a trajectory where the accompanying arrows indicate the direction of increasing $\Gamma$. The paramagnetic state, located at the geometric center of configuration space, is also indicated.} 
\label{fig.HFsolns.evolution with Gamma.panels a and b}
\end{center}
\end{figure}

Let us now turn on the transverse field. We tracked the evolution of all the energy minima numerically as $\Gamma$ is increased. Denote the Hartree-Fock energy of the minimum $\pm\bm{\vec{\alpha}_n}$ as $E_{\pm n}^{\rm HF}(H_{\rm hop})$. Figure \ref{fig.HFsolns.evolution with Gamma.panels a and b}(a) shows how the energy of each minimum varies with $\Gamma$. Note that the free energy $F=-(N/2)(1+\Gamma^2)$ \cite{Nishimori96} has been subtracted from the curves. One sees that the local minima disappear from the energy surface at different $\Gamma$. At $\Gamma=0.55$, for example, only the local minima $\pm\bm{\vec{\alpha}_5}$ is remaining on the surface.

\begin{figure}[h]
\begin{center}
\includegraphics[scale=0.65]{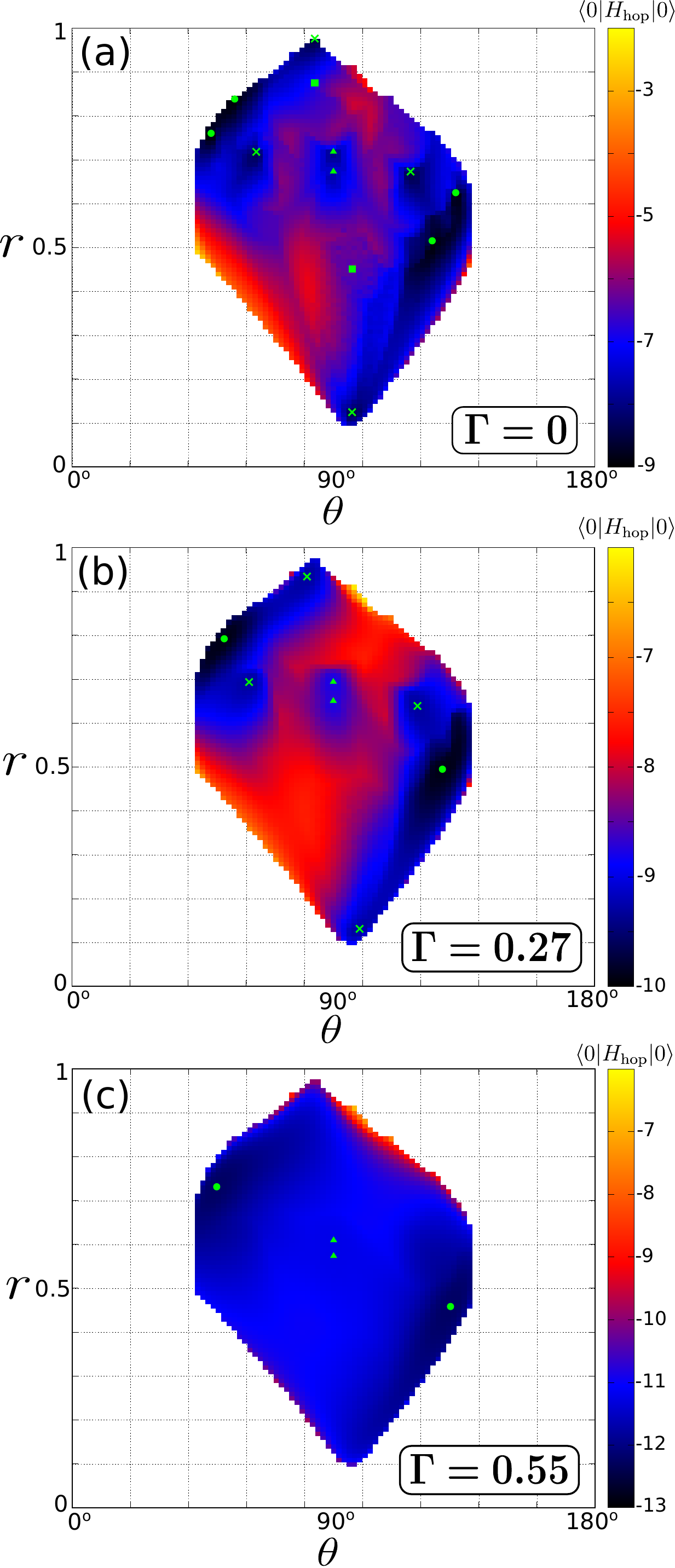}
\caption{Visualizing the landscape of the classical energy $\langle 0|H_{\rm hop}|0 \rangle$ at different $\Gamma$. Polar coordinates are plotted in Cartesian format, with $\theta$ for abscissa and $r$ for ordinate. Energy minima present at each $\Gamma$ are indicated with the same symbols (green) used in Fig. \ref{fig.polar plot.HF solns and embedded patts}. (a) At $\Gamma=0$. (b) At $\Gamma=0.27$. (c) At $\Gamma=0.55$. One discerns the smoothening of the landscape and the vanishing of the minima with increase in $\Gamma$.} 
\label{fig.energy landscape.classical energy.3values of Gamma}
\end{center}
\end{figure}

The procedure for plotting Fig. \ref{fig.energy landscape.classical energy.3values of Gamma}(a) can also be applied to $\Gamma>0$ to visualize the evolution of the energy surface with $\Gamma$. The landscapes at $\Gamma=0.27$ and 0.55---indicated in Fig. \ref{fig.HFsolns.evolution with Gamma.panels a and b}(a) by vertical lines---are shown in panels (b) and (c) of Fig. \ref{fig.energy landscape.classical energy.3values of Gamma}, respectively. In addition to the vanishing of the local minima, one can also see the smoothening of the energy surface with increase in $\Gamma$.

To round up this section on classical energy landscape, let us look at how the positions of the minima change with $\Gamma$. Figure \ref{fig.HFsolns.evolution with Gamma.panels a and b}(b) shows the evolution of the polar coordinates of the minima as $\Gamma$ increases. The paramagnetic state, located at the geometric center of configuration space, is indicated by a semi-circle. It is seen that the global minima $\pm\bm{\vec{\alpha}_1}$ trace out rather complicated paths when evolving towards the paramagnetic state.

\clearpage

\subsection{Localization of excited eigenstates in the energy basins of local minima}
\label{}

We now bring in the quantum aspects of the model. The energy eigenvalues $E_n(H_{\rm hop})$ and the corresponding eigenvectors are obtained using the Lanczos algorithm \cite{footnote.concerning Spectra and convergence}. As the ground state has been discussed in many previous works, in the following we shall focus on the local minima.

Figure \ref{fig.FQ En vs HF EHFm.Hopfieldmodel.for a3,4 and a5}(a) shows the energy levels around the four-fold degenerate Hartree-Fock energy $E^{\rm HF}_{\pm 3,4}$ (blue circles). The levels $E_4$ to $E_7$ (highlighted in red) are closest to it. There is some disagreement between the classical and quantum energies, originating from the smallness of the system size. Nevertheless, convincing evidences of quantum-classical correspondence can still be discerned by examining the wave functions. Table \ref{tab.for EHF pm3,4.wvefns projections} shows the overlaps between the four eigenfunctions $|E_4\rangle$ to $|E_7\rangle$ and the four Hartree-Fock wave functions $\left|E^{\rm HF}_{\pm 3,4}\right\rangle$, taken at $\Gamma=0.1$ and 0.32 [vertical lines in Fig. \ref{fig.FQ En vs HF EHFm.Hopfieldmodel.for a3,4 and a5}(a)]. As the overlap between a $|E_n\rangle$ and each of the $\left|E^{\rm HF}_{m}\right\rangle$ differ by only a sign, for compactness the square of the overlap $\left|\left\langle E^{\rm HF}_{m} |E_n \right\rangle\right|^2$ is shown. At $\Gamma=0.1$, the projections of, say $|E_4\rangle$, onto the four Hartree-Fock states account for $4\times 23=92$ percent of its probability mass, meaning that the eigenfunction is localized mainly in the four local minima $\pm\bm{\vec{\alpha}_{3,4}}$. This behavior is exhibited by the other three eigenfunctions as well. The high degree of overlap between the quantum and classical states persists---dropping only slightly---until $\Gamma=0.32$, just before the minima vanish from the energy surface.

\begin{table}[h]
\begin{center}
\begin{tabular}{c   cccc  }
\hline
\hline
  & \multicolumn{4}{c}{$\left|\left\langle E_{\pm 3,4}^{\rm HF} |E_n \right\rangle\right|^2$}  \\
\cline{2-5}
\hspace{0.4cm} $\Gamma$ \hspace{0.4cm}
& \hspace{0.1cm} $E_{4}$ \hspace{0.1cm} & \hspace{0.1cm} $E_{5}$ \hspace{0.1cm} & \hspace{0.1cm} $E_{6}$ \hspace{0.1cm} & \hspace{0.1cm} $E_{7}$ \hspace{0.1cm} 
\\
\hline
0.1 & 0.23 & 0.23 & 0.23 & 0.23 
\\
0.32  & 0.23  & 0.23 & 0.21 & 0.21 
\\
\hline
\hline
\end{tabular}
\caption{Squares of the overlap, $\left|\left\langle E^{\rm HF}_{\pm 3,4} |E_n \right\rangle\right|^2$, between the Hartree-Fock wave functions at the local minima $\pm\bm{\vec{\alpha}_{3,4}}$ and the eigenfunctions $|E_4\rangle$ to $|E_7\rangle$ of the Hopfield model, evaluated at $\Gamma=0.1$ and 0.32 [c.f. Fig. \ref{fig.FQ En vs HF EHFm.Hopfieldmodel.for a3,4 and a5}(a)]. The overlaps $\left\langle E^{\rm HF}_{\pm 3,4} |E_n \right\rangle$ of each $\left|E_m^{\rm HF}\right\rangle$ with a $|E_n\rangle$ are identical apart from a sign.}
\label{tab.for EHF pm3,4.wvefns projections}
\end{center}
\end{table}

The next minima $\pm\bm{\vec{\alpha}_{5}}$ present some additional features. Figure \ref{fig.FQ En vs HF EHFm.Hopfieldmodel.for a3,4 and a5}(b) shows the energy levels around the two-fold degenerate $E^{\rm HF}_{\pm 5}$ (blue circles). In this case, two groups of levels are involved: $E_{14}, E_{15}$ (red) and $E_{16}, E_{17}$ (green). It is seen that there is an avoided crossing between $E_{14,15}$ and $E_{16,17}$ at $\Gamma \approx 0.26$. Before the crossing, localization is manifested by the eigenfunctions $|E_{16}\rangle$ and $|E_{17}\rangle$. For instance, at $\Gamma=0.25$, projecting $|E_{16}\rangle$ onto the two Hartree-Fock states gives $\left|\left\langle E^{\rm HF}_{5}|E_{16}\right\rangle\right|^2+\left|\left\langle E^{\rm HF}_{-5}|E_{16}\right\rangle\right|^2\approx2\times 0.48=0.96$. After the crossing, the localizations are inherited by $|E_{14}\rangle$ and $|E_{15}\rangle$. For example, at $\Gamma=0.28$ projection of $|E_{14}\rangle$ gives $\left|\left\langle E^{\rm HF}_{5}|E_{14}\right\rangle\right|^2+\left|\left\langle E^{\rm HF}_{-5}|E_{14}\right\rangle\right|^2\approx2\times 0.44=0.88$. The localizations persist until $\Gamma=0.4$, where the projections drop to 0.66 for $|E_{14}\rangle$ and 0.57 for $|E_{15}\rangle$, after which the overlaps between the quantum and classical states decay very rapidly.

\begin{figure}[h]
\begin{center}
\includegraphics[scale=0.7]{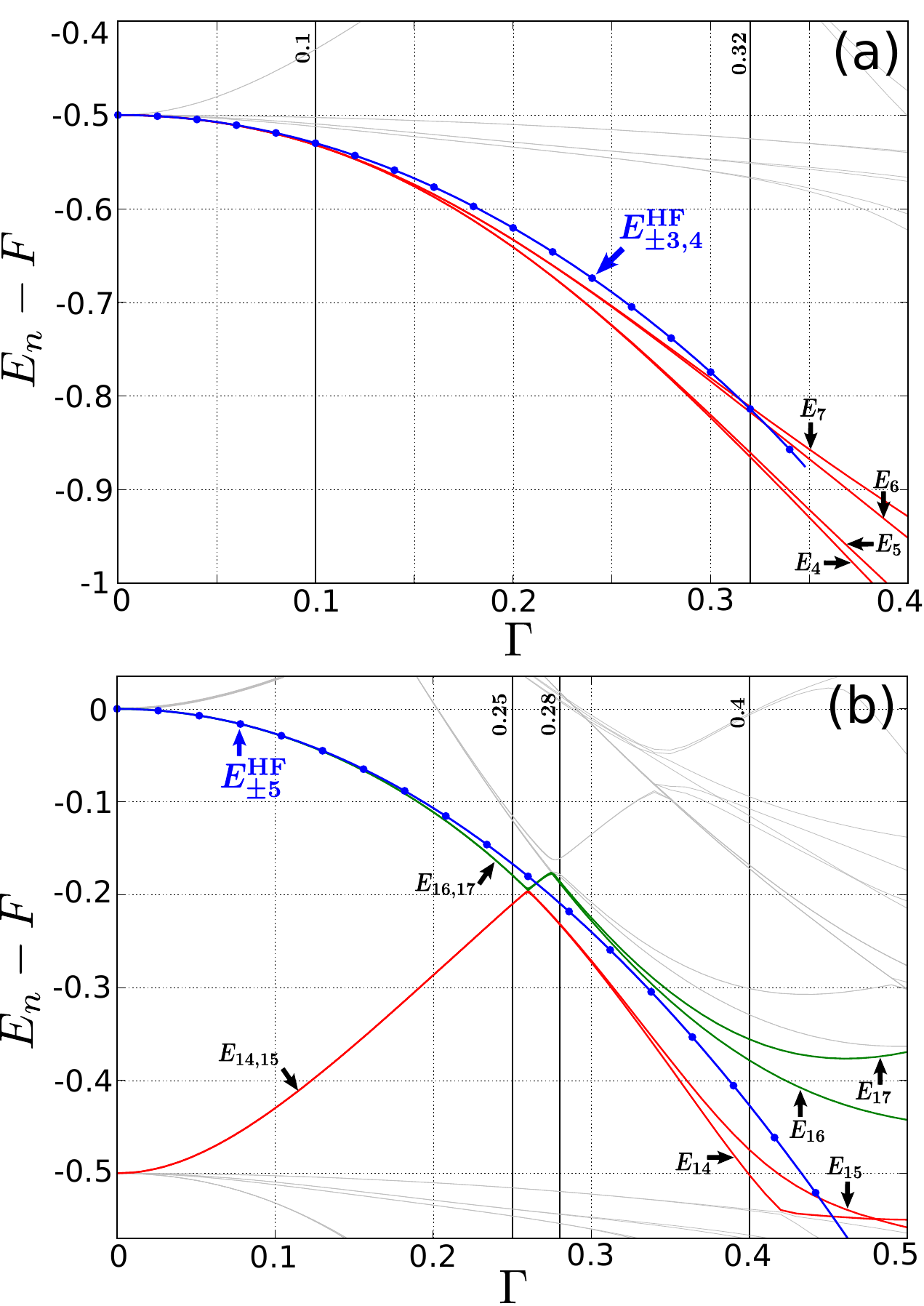}
\caption{Energy eigenvalues $E_n(H_{\rm hop})$ in the vicinity of (a) $E_{\pm 3,4}^{\rm HF}$ and (b) $E_{\pm 5}^{\rm HF}$ [c.f. Fig. \ref{fig.HFsolns.evolution with Gamma.panels a and b}(a)]. The levels $E_n$ are plotted with solid lines. Curves of Hartree-Fock energy are adorned with circles (blue). All energies are shifted by the free energy $F$. Identities of the levels closest to the Hartree-Fock energy are labeled and colored (red, green), while irrelevant levels are plotted in grey. Vertical lines indicate the $\Gamma$ values where the energy eigenfunctions are examined.} 
\label{fig.FQ En vs HF EHFm.Hopfieldmodel.for a3,4 and a5}
\end{center}
\end{figure}

\begin{figure}[h]
\begin{center}
\includegraphics[scale=0.8]{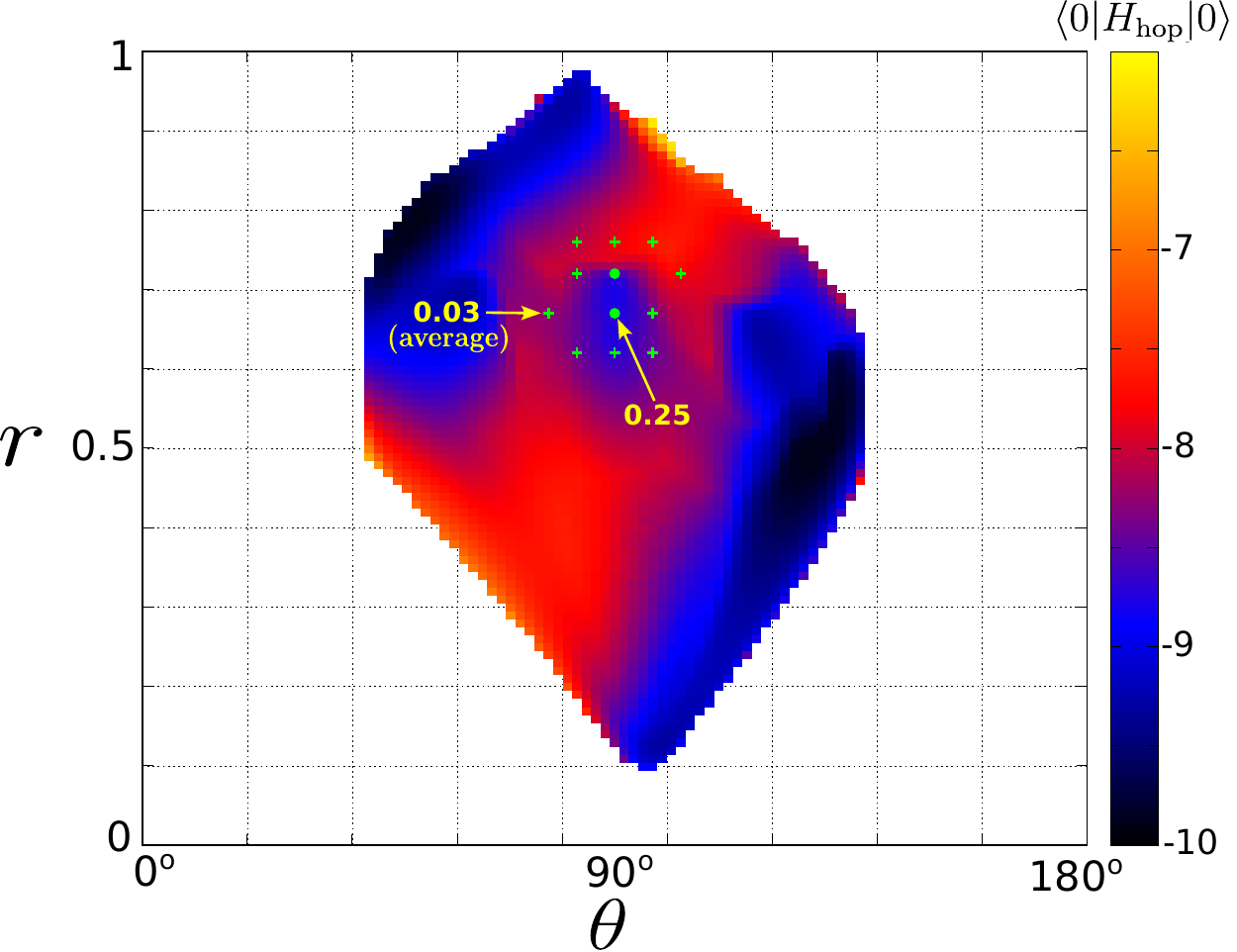}
\caption{Localization of the eigenstate $|E_{14}\rangle$ around the energy basins of the minima $\pm\bm{\vec{\alpha}_5}$ at $\Gamma=0.28$. The heat map shows the energy landscape of $\langle 0|H_{\rm hop}|0\rangle$. The eigenfunction $|E_{14}\rangle$ is projected onto the $r$-$\theta$ plane, and the circle and cross symbols (green) indicate the locations with the highest cumulative probability. Arrows indicate the probability carried by each symbol. It is seen that more than 80 percent of the probability mass of $|E_{14}\rangle$ is localized around $\pm\bm{\vec{\alpha}_5}$.}  
\label{fig.localization E14 psi.HF landscape.gamma 0.28}
\end{center}
\end{figure}

One can visualize the eigenstates' localizations using the method of plotting the energy landscape presented earlier. We performed the Lanczos algorithm in the $\sigma_i^z$ representation, and each energy eigenfunction is returned as a superposition of $2^N$ classical spin configurations. Each basis vector (a spin configuration) is first projected onto the $r$-$\theta$ plane and assigned to the appropriate grid bin. Then, instead of updating the energy, we accumulate the probability (i.e. amplitude squared) of the basis vector onto the histogram at that bin. As an example, Fig. \ref{fig.localization E14 psi.HF landscape.gamma 0.28} shows the localization of the eigenstate $|E_{14}\rangle$ around the minima $\pm\bm{\vec{\alpha}_5}$ at $\Gamma=0.28$ [c.f. Fig. \ref{fig.FQ En vs HF EHFm.Hopfieldmodel.for a3,4 and a5}(b), just after the avoided crossing]. The heat map shows the classical energy landscape where the energy basin of $\pm\bm{\vec{\alpha}_5}$ is the blue oval at the center. The histogram of $|E_{14}\rangle$ is shown as symbols (green) superimposed on the energy landscape. For clarity, only the twelve bins with the highest cumulative probability are shown. The circle symbols each carry a probability of $\approx$ 0.25. The probabilities at each of the cross symbols are actually not equal, but on average each carries a probability of $\approx$ 0.03. It is seen that more than 80 percent of the probability mass of $|E_{14}\rangle$ is localized around the energy basin of the minima $\pm\bm{\vec{\alpha}_5}$.

It is instructive to compare our polar plot (Fig. \ref{fig.localization E14 psi.HF landscape.gamma 0.28}) to the Fock space diagram introduced by Mukherjee et al. for visualizing many-body localization-delocalization \cite{Mukherjee18}. The authors examined how high-energy eigenvectors of the Sherrington-Kirkpatrick model exhibit localization-delocalization as the strength of the transverse field changes. In their work, each of the $2^N$ basis configurations is given a label $i\in[0,2^N-1]$ and mapped onto coordinates $i\rightarrow (i_A,i_B)$; the amplitudes of an eigenvector are then plotted on the two-dimensional $i_A$-$i_B$ plane in the form of a heat map. Their diagram reveals the MBLD transition in a visual way. However, in this system of plotting the label $i$ assigned to a basis configuration is somewhat arbitrary, and the physical significance of the coordinates $(i_A,i_B)$ is not very clear. On the other hand, in our polar plot the coordinates $(r,\theta)$ are always referenced to the classical energy surface, so our diagram is more informative. For instance, changes in the energy landscape with the parameter $\Gamma$ are reflected in the polar plot, so the mechanism underlying the shifts in positions of the eigenfunction amplitudes with $\Gamma$ becomes evident. If one were to visualize this with the Fock space diagram, only the shifts of the amplitudes in $i_A$-$i_B$ space would be visible, with no insights on localization.

\subsection{Unphysical local minima at $\pm\bm{\vec{\alpha}_6}$}
\label{}

\begin{figure}[h]
\begin{center}
\includegraphics[scale=0.75]{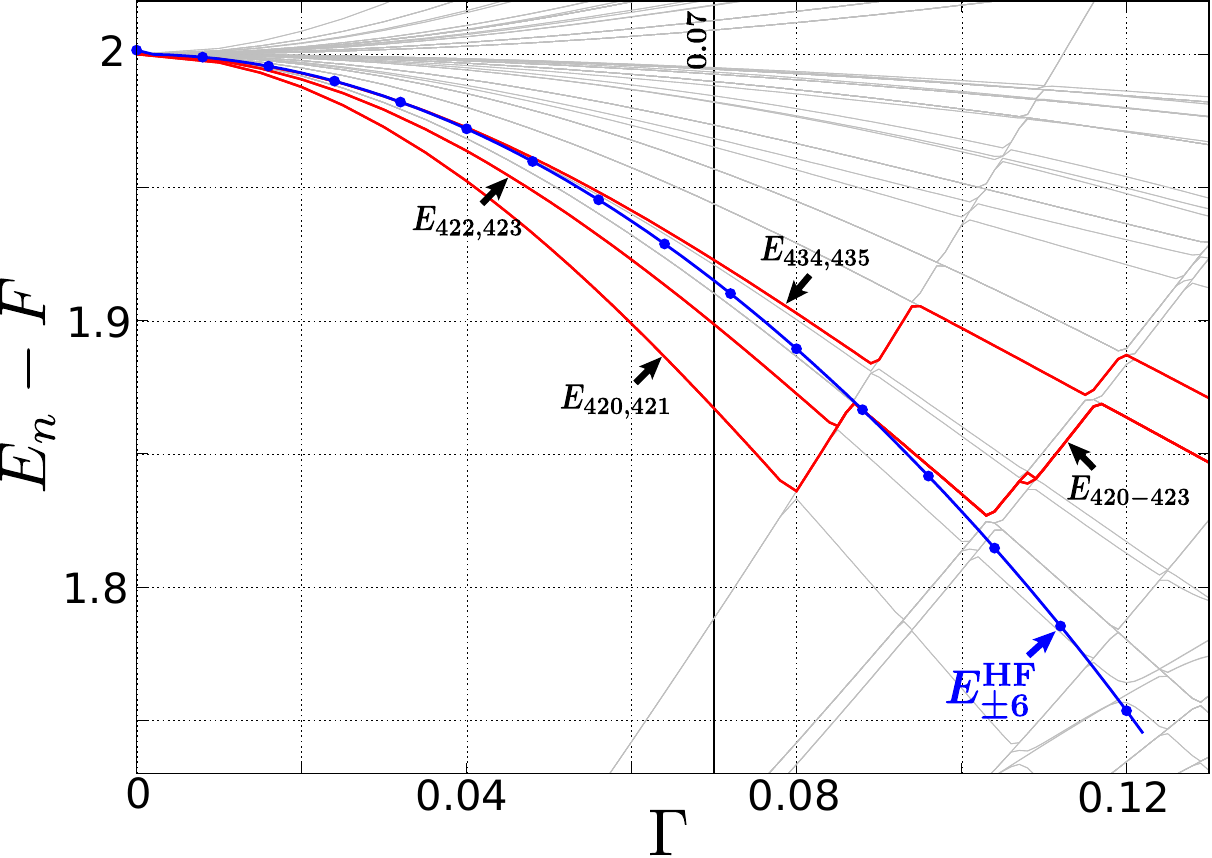}
\caption{Energy eigenvalues $E_n$ of the Hopfield model in the vicinity the highest Hartree-Fock energy $E_{\pm 6}^{\rm HF}$. The figure is organized similar to Fig. \ref{fig.FQ En vs HF EHFm.Hopfieldmodel.for a3,4 and a5}. Discussion in the text focuses on the region before the first avoided crossing at $\Gamma\approx 0.08$. The labeled energy levels (red) are those whose eigenfunctions exhibit the largest overlaps with $\left|E^{\rm HF}_{\pm6}\right\rangle$ before the crossing. The wave function overlaps at $\Gamma=0.07$ (vertical line) are given in Table \ref{tab.for EHF pm6 at Gamma 0.07.wavefunctions projections}.}  
\label{fig.FQ En vs HF EHFm.Hopfieldmodel.for a6}
\end{center}
\end{figure}

\begin{figure}[h]
\begin{center}
\includegraphics[scale=0.8]{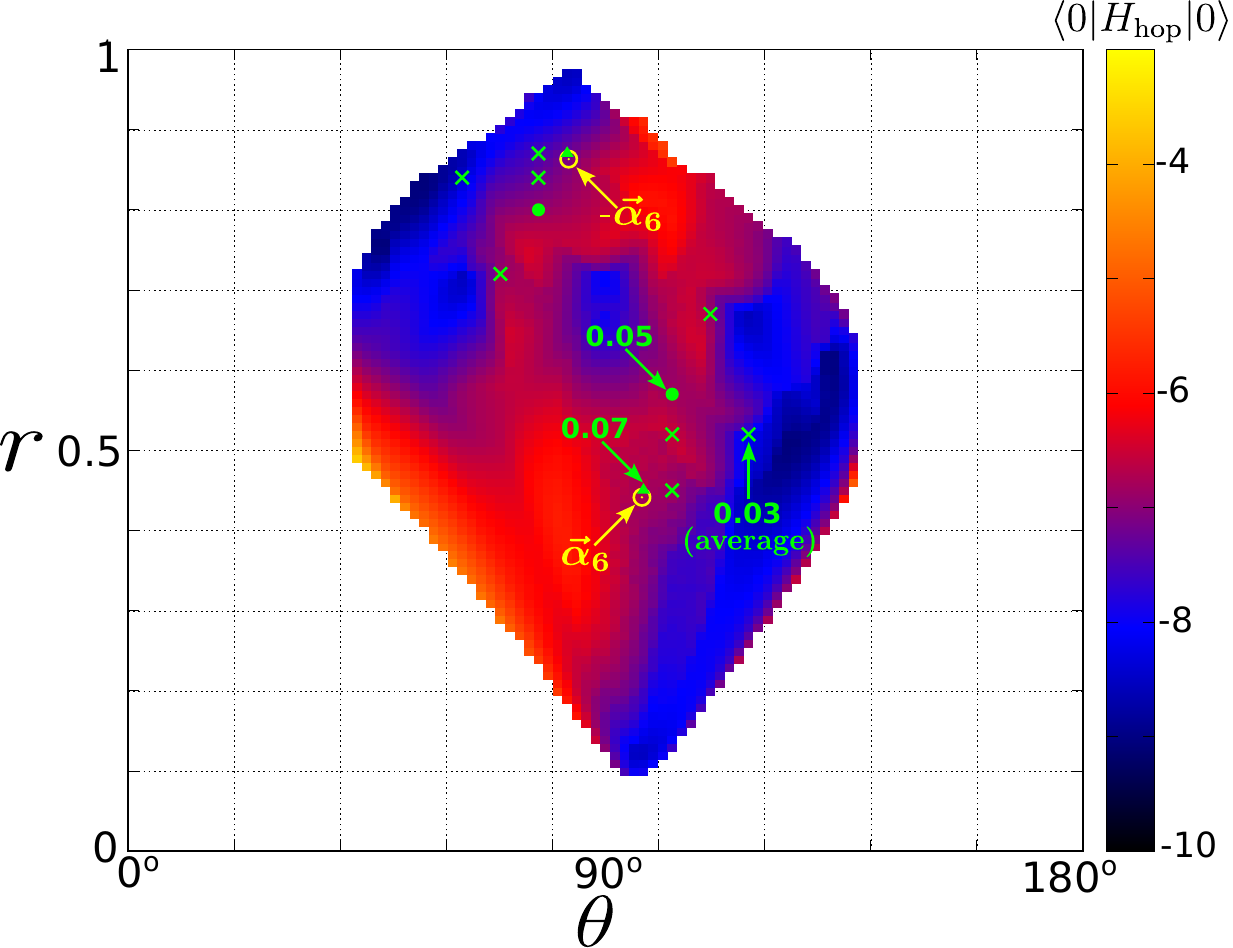}
\caption{Projection of the eigenstate $|E_{420}\rangle$ at $\Gamma=0.07$ onto the $r$-$\theta$ plane. The figure is organized similar to Fig. \ref{fig.localization E14 psi.HF landscape.gamma 0.28}. The heat map shows the energy landscape at $\Gamma=0.07$, with the minima $\pm\bm{\vec{\alpha}_6}$ indicated by empty circles (yellow). The other symbols (green) represent the locations where $|E_{420}\rangle$ has the highest cumulative probability. We see that only about 20 percent of the probability mass of the eigenfunction is close to $\pm\bm{\vec{\alpha}_6}$.}  
\label{fig.E420 psi.HF landscape.gamma 0.07}
\end{center}
\end{figure}

We now turn our attention to the final local minima $\pm\bm{\vec{\alpha}_6}$. These have the highest Hartree-Fock energy among all the minima, and from the landscape diagram Fig. \ref{fig.energy landscape.classical energy.3values of Gamma}(a) one can also see that their energy basins are relatively shallow. Unlike for the other minima, we were unable to establish quantum-classical correspondence for $\pm\bm{\vec{\alpha}_6}$. Figure \ref{fig.FQ En vs HF EHFm.Hopfieldmodel.for a6} shows the energy levels around the two-fold degenerate $E_{\pm 6}^{\rm HF}$ (blue circles). There is an avoided crossing at $\Gamma=0.08$ ($E_n-F\approx$1.83). We first focus our discussion on the $\Gamma$ regime before the crossing. We scanned through the overlaps between $\left|E_{\pm6}^{\rm HF}\right\rangle$ and all the eigenstates $|E_n\rangle$ and found that $\left\langle E^{\rm HF}_{\pm 6} |E_n \right\rangle$ is appreciable only for six of the eigenstates. The energy levels of these six eigenstates are labeled in Fig. \ref{fig.FQ En vs HF EHFm.Hopfieldmodel.for a6} (highlighted in red). Generally speaking, even for these six states the overlaps are quite low. For instance, the squared overlaps $\left|\left\langle E^{\rm HF}_{\pm 6} |E_n \right\rangle\right|^2$ at $\Gamma=0.07$ (vertical line in Fig. \ref{fig.FQ En vs HF EHFm.Hopfieldmodel.for a6}) are listed in Table \ref{tab.for EHF pm6 at Gamma 0.07.wavefunctions projections}. It is seen that for, say $|E_{420}\rangle$, only 
$2\times 11\approx 22$ percent of its probability mass is accounted for by $\left|E_{\pm6}^{\rm HF}\right\rangle$. One can also ascertain this visually by projecting an eigenstate onto the $r$-$\theta$ plane and comparing it with the energy landscape, as we have done for $|E_{14}\rangle$ earlier. Figure \ref{fig.E420 psi.HF landscape.gamma 0.07} shows the probability histogram of $|E_{420}\rangle$ at $\Gamma=0.07$ superimposed on the classical energy landscape. For clarity of presentation, only the twelve largest bins---comprising approximately 50 percent of the total probability mass---are plotted (green symbols). The local minima $\pm\bm{\vec{\alpha}_6}$ are indicated by empty circles (yellow). One sees that only about 20 percent of the probability mass of $|E_{420}\rangle$ is located close to $\pm\bm{\vec{\alpha}_6}$. Overall, it seems that the minima $\pm\bm{\vec{\alpha}_6}$ cannot be associated with any of the energy eigenstates.

\begin{table}[h]
\begin{center}
\begin{tabular}{ccc}
\hline
\hline
 \multicolumn{3}{c}{$\left|\left\langle E_{\pm6}^{\rm HF} |E_n \right\rangle\right|^2$}  \\
\hline
\,$E_{420,421}$ \, & \, $E_{422,423}$ \, & \, $E_{434,435}$ \,\\
\hline
 0.11   & 0.18 & 0.12 \\
\hline
\hline
\end{tabular}
\caption{Squared overlaps $\left|\left\langle E^{\rm HF}_{\pm 6} |E_n \right\rangle\right|^2$ of the Hopfield model for the minima $\pm\bm{\vec{\alpha}_6}$, evaluated at $\Gamma=0.07$. Only six eigenfunctions---$|E_{420}\rangle$ to $|E_{423}\rangle$, $|E_{434}\rangle$, and $|E_{435}\rangle$---exhibit appreciable overlaps with the Hartree-Fock states $\left|E^{\rm HF}_{\pm6}\right\rangle$. As in Table \ref{tab.for EHF pm3,4.wvefns projections}, $\left\langle E^{\rm HF}_{6} |E_n \right\rangle$ and $\left\langle E^{\rm HF}_{-6} |E_n \right\rangle$ differ only by a sign. The small overlaps suggest that $\pm\bm{\vec{\alpha}_6}$ have no correspondence with physical states.}
\label{tab.for EHF pm6 at Gamma 0.07.wavefunctions projections}
\end{center}
\end{table}

After the first avoided crossing at $\Gamma=0.08$, the relationship between $\pm\bm{\vec{\alpha}_6}$ and the energy levels becomes rather complex because of the subsequent crossings. We tracked the overlaps in the manner discussed above until $\Gamma=0.12$, and found that the behavior, nevertheless, remains similar. Hence, like the cubic model, for the Hopfield model one can also encounter unphysical local minima.

\subsection{Incorporation of Coupled-Cluster effects}
\label{}

Hitherto, our discussions have centered around the Hartree-Fock energy. We tried incorporating Coupled-Cluster excitations into the classical energy, but did not obtain improvement in accuracy for the particular realization of Hopfield model being studied here. For singles excitations, the Coupled-Cluster energy and equations can be derived by applying the results of Appendix \ref{app.random field formulation of CCSD} to individual $J_{\alpha}^{\gamma}$ operators in Eq. (\ref{eq.Hhop.hamiltonian.regroup into gamma groups}). At the global minima $\pm\bm{\vec{\alpha}_1}$, we found that the singles energy is higher than the Hartree-Fock energy $E^{\rm HF}_{\pm1}$, and is numerically less accurate than the latter. Furthermore, as we increase $\Gamma$ beyond $\approx 0.1$, the basin of attraction of the singles solution diminishes while spurious solutions appear, making it difficult to identify and attain the correct solution. At the local minima, the accuracy issue is even more acute as the singles energies that we managed to obtain are very far from both the Hartree-Fock and quantum energies. We also implemented a limited version of doubles excitations where the two excited spins both lie within the same group $\gamma$, and so the CCSD energy and equations can again be derived using the results of Appendix \ref{app.random field formulation of CCSD}. The effects of doubles excitations are, however, very small and do not affect our above conclusions. Despite these results, we think that the discouraging performance of CCSD in our case study is largely due to the smallness of our system size ($N=16$). As our detailed analysis of the ferromagnetic model in Appendix \ref{app.ferromagnetic model.CCSD} shows, CCSD becomes very accurate when the system size is large. In that regime, numerical diagonalization of the Hamiltonian is no longer feasible, so our formulas in Appendix \ref{app.random field formulation of CCSD} may come in useful in numerical studies of frustrated quantum spin systems.

%
%

\section{Summary and discussions}
\label{sec. summary discussion section label}

In this paper, we performed a detailed numerical investigation on the relationship between the energy eigenstates and the local minima on the energy landscape of quantum spin systems. We adopted an empirical approach, working with specific models, so the results here might not be generally applicable to every system. Nonetheless, we think that the illustrations here can shed some insights on the quantum-classical correspondence between the two quantities. 

We studied two models, and found that the local minima on the classical energy landscape can be interpreted as excited eigenstates in the energy spectrum. The first model is a non-frustrated one exhibiting a single local minimum. Here, it was found that the minimum corresponds to an envelope of avoided crossings, which is a secondary structure on the energy spectrum. The second model is a frustrated one exhibiting multiple local minima. In this case, we found energy eigenfunctions which are localized in the energy basins of the minima. Apart from these affirmative cases, however, we also found instances of unphysical local minima, so one needs to exercise some caution when applying these observations to other systems.

One possible extension of this work might be in the classification or labeling of the energy levels of frustrated systems. In non-frustrated models such as Eqs. (\ref{eq.cubic w antiferro.hamiltonian definition}) and (\ref{eq.app.ferro model.definition}), the total angular momentum is conserved, so its quantum number can be used to classify the energy levels of the system into different sectors. On the other hand, for frustrated systems such as Eq. (\ref{eq.Hhop.hamiltonian.definition}), it is not \emph{a priori} clear what the conserved quantity---if it exists---should be. For such systems, one might consider assigning labels to levels based on the energy basin (or basins) which they are associated with. It would be interesting to see how the excitations in a basin can be generated within the equation-of-motion framework \cite{Bartlett07,Rowe68}, and if the correspondence reported here continues to be valid. These labels could then serve as generalized `quantum numbers' for frustrated systems.

In our studies of the Hopfield model, we noticed some features of its energy spectrum that might be indicative of a path to quantum chaos. In Figs. \ref{fig.FQ En vs HF EHFm.Hopfieldmodel.for a3,4 and a5}(b) and \ref{fig.FQ En vs HF EHFm.Hopfieldmodel.for a6}, there is a regime of weak transverse field (small $\Gamma$) where the spectrum is regular in the sense that there are no collisions among the energy levels. The appearance of the first avoided crossing then initiates a cascade of complex level collisions in subsequent larger $\Gamma$ values. This transition from regular to `level-collision' phase can be quantitatively characterized by various measures based on level statistics \cite{Gubin12,Zakrzewski23}. Curiously, the disappearance of local minima from the energy surface [c.f. Fig. \ref{fig.HFsolns.evolution with Gamma.panels a and b}(a)] occurs in the vicinity of this onset of chaos. It might be interesting to examine whether there is any relationship between the two in our future studies.  

As mentioned in the Introduction, one of the purposes of this work is to study the feasibility of the Coupled-Cluster method as a numerical technique for quantum spin systems. Perhaps the most important observation we made is that the existence of solutions (to the Coupled-Cluster equations) depends on the system size and also on parameters within the model. Generally speaking, the method tends to breakdown near second-order phase transitions and at small system sizes. Hence, the Coupled-Cluster expansion is not uniformly convergent and does not provide improvement upon the Hartree-Fock approximation unconditionally. On the other hand, for pure systems at large system sizes, we found the method to be quite reliable and accurate. It remains to assess its performance for disordered systems at large system size in our future studies.

In this work, we implemented CCSD using the random field approach. Higher-order excitations can be treated similarly. For instance, quadruples and triples excitations can be written as $X^4=(X^2)^2$ and $4X^3=(X+X^2)^2-(X-X^2)^2$, after which linearization can be achieved by repeated applications of the Hubbard-Stratonovich transform. Cross terms can treated similarly: $4XY=(X+Y)^2-(X-Y)^2$. It might be worth mentioning that the Fourier transform of the Airy function
\begin{equation}
e^{\frac{i}{3}(2\pi X)^3}=\int_{-\infty}^{\infty}\mathrm{Ai}(k)e^{-2\pi i k X} dk
\label{}
\end{equation}
allows us to linearize a cubic term $X^3$ at the price of just a single random field variable $k$. This additional economy might help simplify the derivation of the Coupled-Cluster energy and equations.
 
We briefly comment on our usage of Eqs. (\ref{eq.definition of generalized operator J(a)}) to (\ref{eq.sec01.general rotation formula}). The formulas in Appendix \ref{app.random field formulation of CCSD} are derived by setting the vector $\bm{a}$ in $J_{\alpha}(\bm{a})$ to unity, which did not utilize the full generality of the three relations. For more complicated models such as the Sherrington-Kirkpatrick model, the formulas in the appendix are no longer applicable, and one should employ the more general framework when dealing with such systems.
 
Lastly, let us touch upon the accuracy of Hartree-Fock approximation. In the course of our study, we found that Hartree-Fock approximation is sometimes quite inaccurate as a leading approximation, especially at large system sizes. On the other hand, note that in the quantum TAP approach \cite{Biroli01}, inclusion of the Onsager correction is necessary. This correction removes the effects of self-response in the internal field, and is important when dealing with random systems. Curiously, such a correction term does not arise in the traditional Hartree-Fock framework, and this might account for some of its inadequacies as a mean-field approximation. It would be interesting to see if terms analogous to the Onsager correction can be accommodated within the Hartree-Fock framework so as to yield a more accurate leading approximation.

\begin{acknowledgements}
This work is partly funded by a project commissioned by the New Energy and Industrial Technology Development Organization (NEDO) of Japan. We also thank the anonymous referees for their valuable comments.
\end{acknowledgements}

\appendix

%
%

\section{Summary of formulas for Coupled-Cluster theory}
\label{app.random field formulation of CCSD}

This appendix is a contiuation of the derivation given in Sec. \ref{app.random field formulation of CCSD.revised}. In the following, to simplify notations, we omit the superscript from $T^{\mathrm{SD}}$, and define $e_n\equiv e^{-nw}, c_n\equiv\cosh(nw),  t_n\equiv\tanh(nw), \tilde{c}_n\equiv\cos(ny)$, and $\tilde{s}_n\equiv\sin(ny)$.

\subsection{For CCSD energy $\langle 0| e^{-T^{\mathrm{}}}\, H \, e^{T^{\mathrm{}}} |0\rangle$}
\label{app.subsec.summary CCSD energy}

\begin{eqnarray}
\langle 0| e^{-T^{\mathrm{}}}\, J_z \, e^{T^{\mathrm{}}} |0\rangle 
&=&
N e_2  c_2^N
(1+t_2)
\left[
2\alpha\beta \tilde{s}_2 - (\alpha^2-\beta^2)\tilde{c}_2
\right]
\label{eq.app.list.ccsd.energy.Jz}\\
 \langle 0| e^{-T^{\mathrm{}}}\, J_x^2 \, e^{T^{\mathrm{}}} |0\rangle &=& N + {\textstyle \frac{N(N-1)}{2}}\{ 1 + e_8c_4^N (1+t_4)^2  [ 4\alpha\beta (\alpha^2-\beta^2)\tilde{s}_4 -[(\alpha^2-\beta^2)^2-4\alpha^2\beta^2] \tilde{c}_4]\}
\nonumber\\
&& 
\label{eq.app.list.ccsd.energy.Jx2}\\
\langle 0| e^{-T^{\mathrm{}}}\, J_z^2 \, e^{T^{\mathrm{}}} |0\rangle &=&
 N + {\textstyle \frac{N(N-1)}{2}}\{ 1 - e_8 c^N_4 (1+t_4)^2   
  [ 4\alpha\beta (\alpha^2-\beta^2)\tilde{s}_4 -[(\alpha^2-\beta^2)^2-4\alpha^2\beta^2] \tilde{c}_4]\}
\nonumber\\
&&
\label{eq.app.list.ccsd.energy.Jz2}\\
\langle 0| e^{-T^{\mathrm{}}}\, J_z^3 \, e^{T^{\mathrm{}}} |0\rangle &=
&
-{\textstyle \frac{N(N-1)(N-2)}{4}}e_{18}c^N_6(1+t_6)^3\{2\alpha\beta(3-16\alpha^2\beta^2)\tilde{s}_6
 + (\alpha^2-\beta^2)(16\alpha^2\beta^2-1)\tilde{c}_6\}
\nonumber\\
&& -{\textstyle\frac{3N}{4}}e_2c_2^N(1+t_2)[N(N+1)-(N-1)(N-2)t^2_2]
\,\,[2\alpha\beta\tilde{s}_2 - (\alpha^2-\beta^2)\tilde{c}_2]
\nonumber\\
&& -{\textstyle\frac{N}{2}}e_{18}c_6^N(1-t_6)[2\alpha\beta\tilde{s}_6 +(\alpha^2-\beta^2)\tilde{c}_6 ]
\label{eq.app.list.ccsd.energy.Jz3}
\end{eqnarray}

\subsection{For singles equation $\langle 1| e^{-T^{\mathrm{}}}\, H \, e^{T^{\mathrm{}}} |0\rangle =0$}
\label{app.subsec.summary 1 flip}

\begin{eqnarray}
\langle 1| e^{-T^{\mathrm{}}}\, J_x \, e^{T^{\mathrm{}}} |0\rangle 
&=&
e_2c_2^N[1 + Nt_2 + (N-1)t^2_2]\,[2\alpha\beta\tilde{s}_2 - (\alpha^2-\beta^2)\tilde{c}_2]
\label{eq.app.list.ccsd.1flip.Jx}\\
\langle 1| e^{-T^{\mathrm{}}}\, J_z \, e^{T^{\mathrm{}}} |0\rangle 
&=&
e_2c_2^N[1 + Nt_2 + (N-1)t_2^2]\,[2\alpha\beta\tilde{c}_2 + (\alpha^2-\beta^2)\tilde{s}_2]
\label{eq.app.list.ccsd.1flip.Jz}\\
\langle 1| e^{-T^{\mathrm{}}}\, J_x^2 \, e^{T^{\mathrm{}}} |0\rangle 
&=&
(N-1)e_8c^N_4(1+t_4)^2 (1+{\textstyle\frac{N-2}{2}}t_4)
\{ [4\alpha^2\beta^2 - (\alpha^2-\beta^2)^2] \tilde{s}_4 - 4\alpha\beta(\alpha^2-\beta^2)\tilde{c}_4 \}
\nonumber\\&&
\label{eq.app.list.ccsd.1flip.Jx2}\\
\langle 1| e^{-T^{\mathrm{}}}\, J_z^2 \, e^{T^{\mathrm{}}} |0\rangle 
&=&
-\langle 1| e^{-T^{\mathrm{}}}\, J_x^2 \, e^{T^{\mathrm{}}} |0\rangle
\label{eq.app.list.ccsd.1flip.Jz2}\\
\langle 1| e^{-T^{\mathrm{}}}\, J_z^3 \, e^{T^{\mathrm{}}} |0\rangle 
&=&
e_2c_2^N(1+t_2)[2\alpha\beta\tilde{c}_2+(\alpha^2-\beta^2)\tilde{s}_2]
\nonumber\\
&&
\times
\{
(3N-2)[1+(N-1)t_2]
+
{\textstyle \frac{3(N-1)(N-2)}{4}}(1-t^2_2)[1+(N-3)t_2]
\}
\nonumber\\
&&
+
{\textstyle\frac{(N-1)(N-2)}{4}}e_{18}c^N_6[3+(N-3)t_6]
(1+t_6)^3
\nonumber\\
&&
\times
[\,\,2\alpha\beta(3-16\alpha^2\beta^2)\tilde{c}_6+(\alpha^2-\beta^2)(1-16\alpha^2\beta^2)\tilde{s}_6 \,\, ]
\label{eq.app.list.ccsd.1flip.Jz3}
\end{eqnarray}

\subsection{For doubles equation $\langle 12| e^{-T^{\mathrm{}}}\, H \, e^{T^{\mathrm{}}} |0\rangle =0$}
\label{app.subsec.summary 2 flip}

\begin{align}
\langle 12| e^{-T^{\mathrm{}}}\, J_x \, e^{T^{\mathrm{}}} |0\rangle =
&
-e_2c_2^N t_2 [2+Nt_2+(N-2)t_2^2]\,
[2\alpha\beta \tilde{c}_2 + (\alpha^2-\beta^2)\tilde{s}_2]
\label{eq.app.list.ccsd.2flip.Jx}\\
\langle 12| e^{-T^{\mathrm{}}}\, J_z \, e^{T^{\mathrm{}}} |0\rangle =
&
\quad\,
e_2 c_2^N t_2 [2+Nt_2+(N-2)t_2^2]\,
[2\alpha\beta\tilde{s}_2- (\alpha^2-\beta^2)\tilde{c}_2]
\label{eq.app.list.ccsd.2flip.Jz}\\
\langle 12| e^{-T^{\mathrm{}}}\, J_x^2 \, e^{T^{\mathrm{}}} |0\rangle =
&
1+e_8c_4^N(1+t_4)^2
\{1+2(N-2)t_4+{\textstyle\frac{(N-2)(N-3)}{2}}t_4^2\}
\nonumber\\
&
\times
\{[(\alpha^2-\beta^2)^2-4\alpha^2\beta^2]\tilde{c}_4 - 4\alpha\beta(\alpha^2-\beta^2)\tilde{s}_4\}
\label{eq.app.list.ccsd.2flip.Jx2}\\
\langle 12| e^{-T^{\mathrm{}}}\, J_z^2 \, e^{T^{\mathrm{}}} |0\rangle =
&
1 - e_8c^N_4(1+t_4)^2
\{1+2(N-2)t_4+{\textstyle\frac{(N-2)(N-3)}{2}}t^2_4\}
\nonumber\\
&
\times
\{[(\alpha^2-\beta^2)^2-4\alpha^2\beta^2]\tilde{c}_4 - 4\alpha\beta(\alpha^2-\beta^2)\tilde{s}_4\}
\label{eq.app.list.ccsd.2flip.Jz2}\\
\langle 12| e^{-T^{\mathrm{}}}\, J_z^3 \, e^{T^{\mathrm{}}} |0\rangle =
&
e_2c_2^N[2\alpha\beta\tilde{s}_2-(\alpha^2-\beta^2)\tilde{c}_2]
\{\,\, (3N-2)t_2 [2+Nt_2+(N-2)t_2^2]
\nonumber\\
&
-{\textstyle\frac{3(N-2)}{2}}c_2^{-2}(1+t_2)
[1-(N-3)t_2-{\textstyle\frac{(N-3)(N-4)}{2}}t_2^2]\,\,\}
\nonumber\\
&
+{\textstyle\frac{3(N-2)}{2}}e_{18}c_6^N(1+t_6)^3
\{ 1+(N-3)t_6 + {\textstyle\frac{(N-3)(N-4)}{6}}t_6^2 \}
\nonumber\\
&
\times
[
2\alpha\beta(3-16\alpha^2\beta^2)\tilde{s}_6
+
(\alpha^2-\beta^2)(16\alpha^2\beta^2-1)\tilde{c}_6
]
\label{eq.app.list.ccsd.2flip.Jz3}
\end{align}

%
%

\section{CCSD approximation of ferromagnetic model}
\label{app.ferromagnetic model.CCSD}

In this appendix, we illustrate the implementation of CCSD by applying it to the ferromagnetic model in transverse field
\begin{equation}
H_{\mathrm{ferro}}=-(J/N)\left( J_z \right)^2 -\Gamma J_x
\label{eq.app.ferro model.definition}
\end{equation}
where $J$ and $\Gamma$ are the strengths of the ferromagnetic coupling and the transverse field, respectively. This is perhaps the simplest fully-connected quantum spin model possible. Nevertheless, even the ground state energy is not analytically solvable when $N$ is finite. In the thermodynamic limit $N\rightarrow\infty$, as $\Gamma$ is lowered the ground state undergoes a second-order phase transition from a para to a ferromagnet at the critical point $\Gamma_{\mathrm{c}}=2J$. The ground state energy of this model under Configuration Interaction singles doubles (CISD) approximation was studied previously in ref. \cite{Koh16}. The results here therefore also offer some insights into how CCSD improves upon CISD.

\subsection{Hartree-Fock approximation}
\label{}

The Hartree-Fock energy of $H_{\mathrm{ferro}}$ is obtained by minimizing 
\begin{equation}
\langle 0 |H_{\mathrm{ferro}}|0 \rangle
=
-J(N-1)(2\alpha^2-1)^2
-2N\Gamma\alpha\sqrt{1-\alpha^2}
-J
\label{}
\end{equation}
with respect to $\alpha$. The energy surface exhibits no local minimum, and the solution is $\alpha_{0}=1/\sqrt{2}$ when $\Gamma\ge 2J(1-N^{-1})$ (paramagnet), and
\begin{equation}
\alpha_{\pm}
=
\sqrt{
\frac{1\pm \sqrt{1-\left(\frac{N}{N-1}\frac{\Gamma}{2J}\right)^2}}{2}
}
\label{}
\end{equation}
when $\Gamma<2J(1-N^{-1})$ (ferromagnet), where $\alpha_+$ and $\alpha_-$ are related by a spin flip. The Hartree-Fock energy is
\begin{equation}
E_0^{\mathrm{HF}}(H_{\mathrm{ferro}})
=
\left\{
\begin{array}{ccc}
-N\Gamma-J & \mathrm{when} & \Gamma\ge 2J(1-N^{-1}),
\\
-N\left(\frac{N}{N-1}\frac{\Gamma^2}{4J}+J\right) & \mathrm{when} & \Gamma<2J(1-N^{-1}).
\end{array}
\right.
\label{eq.app.EHF.ferro.two phases}
\end{equation}

\subsection{Solutions of CCSD equations}
\label{}

\begin{figure}[h]
\begin{center}
\includegraphics[scale=0.75]{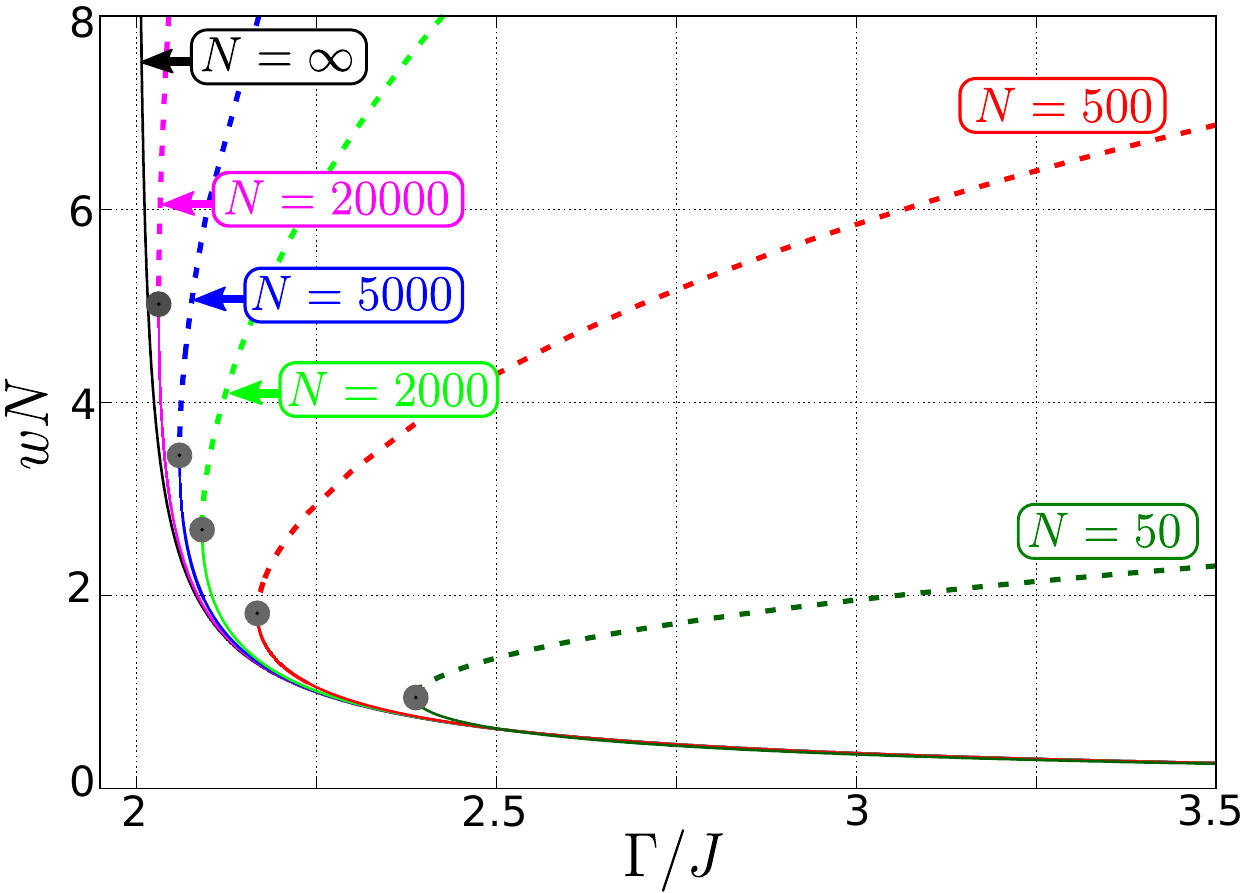}
\caption{Numerical solutions of $w$ for the ferromagnetic model $H_{\mathrm{ferro}}$ in the paramagnetic regime $(\Gamma\ge 2J)$, where the solution of $y$ is zero. The results for several values of $N$ are shown, and each curve has been rescaled by $N$. The solution has an upper (dashed) and a lower (solid) branch, and their meeting point is indicated by a solid circle. Only the lower branch is physically valid, and the lower branches of different $N$ collapse together after rescaling. For finite $N$, $w$ has no valid solution after the meeting point. The curve labeled $N=\infty$ (black) is the analytical solution in the thermodynamic limit, given by Eq. (\ref{eq.app.w soln.analytical.N infinity}).}
\label{fig.app.ferro model.wsoln.para phase}
\end{center}
\end{figure}

The singles equation is obtained by substituting Eqs. (\ref{eq.app.list.ccsd.1flip.Jx}) and (\ref{eq.app.list.ccsd.1flip.Jz2}) into $\langle 1|e^{-T^{\mathrm{SD}}} H_{\mathrm{ferro}}e^{T^{\mathrm{SD}}} |0\rangle$, and the doubles equation Eqs. (\ref{eq.app.list.ccsd.2flip.Jx}) and (\ref{eq.app.list.ccsd.2flip.Jz2}) into $\langle 12|e^{-T^{\mathrm{SD}}} H_{\mathrm{ferro}}e^{T^{\mathrm{SD}}} |0\rangle$. The equations are solved numerically to obtain $y$ and $w$. In the paramagnetic regime $\Gamma\ge 2J$, one has the solution $y=0$, and Fig. \ref{fig.app.ferro model.wsoln.para phase} shows the solution of $w$ for several values of $N$. The solution consists of an upper (dashed) and a lower (solid) branch, meeting at the point $(\Gamma_N, w_NN)$ indicated by a solid circle (grey). Only the lower branch is physically valid. The lower branches of different $N$ collapse together upon rescaling (note the vertical axis). For finite $N$, there is no valid solution of $w$ after the meeting point. Numerically, we found that $\Gamma_N$ approaches the critical point $\Gamma_{\rm c}$ asymptotically as $\Gamma_N-\Gamma_{\mathrm{c}}\sim 3.3 \times N^{-0.47}$, while $w_N\sim 0.32\times N^{-0.73}$. 

\begin{figure}[h]
\begin{center}
\includegraphics[scale=0.7]{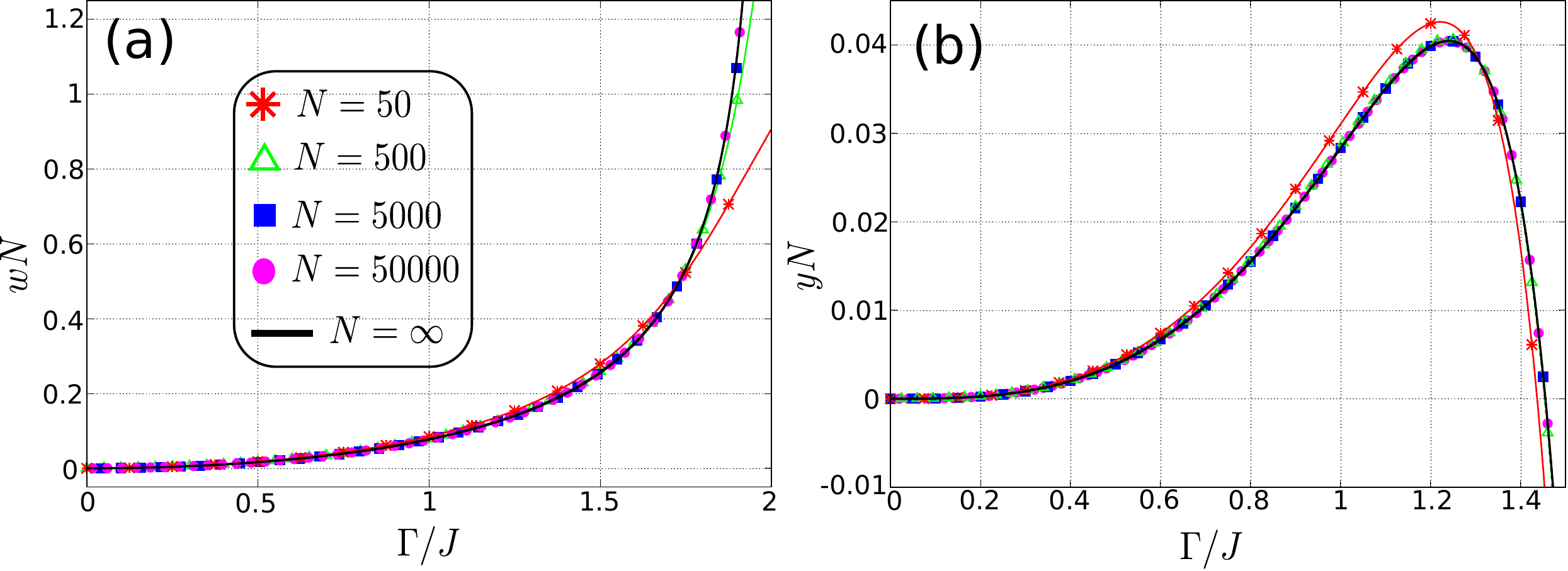}
\caption{Numerical solutions of (a) $w$ and (b) $y$ for the model $H_{\mathrm{ferro}}$ in the ferromagnetic regime $(\Gamma<2J)$. The results for several values of $N$ are shown. Symbols have the same meaning in both panels. The vertical axes have been rescaled, to highlight that the solutions of different $N$ collapse together upon rescaling. The curves labeled $N=\infty$ (black) are the analytical solutions in the thermodynamic limit, and are given by Eq. (\ref{eq.app.w soln.analytical.N infinity}) [for (a)] and Eq. (\ref{eq.app.y soln.analytical.N infinity}) [for (b)].}
\label{fig.app.ferro model.ywsoln.ferro phase}
\end{center}
\end{figure}

Figure \ref{fig.app.ferro model.ywsoln.ferro phase} shows the solutions for $y$ and $w$ in the ferromagnetic regime $\Gamma<2J$. Once again, the solutions of different $N$ collapse together after rescaling.

\subsection{CCSD energy}
\label{}

\begin{figure}[h]
\begin{center}
\includegraphics[scale=0.7]{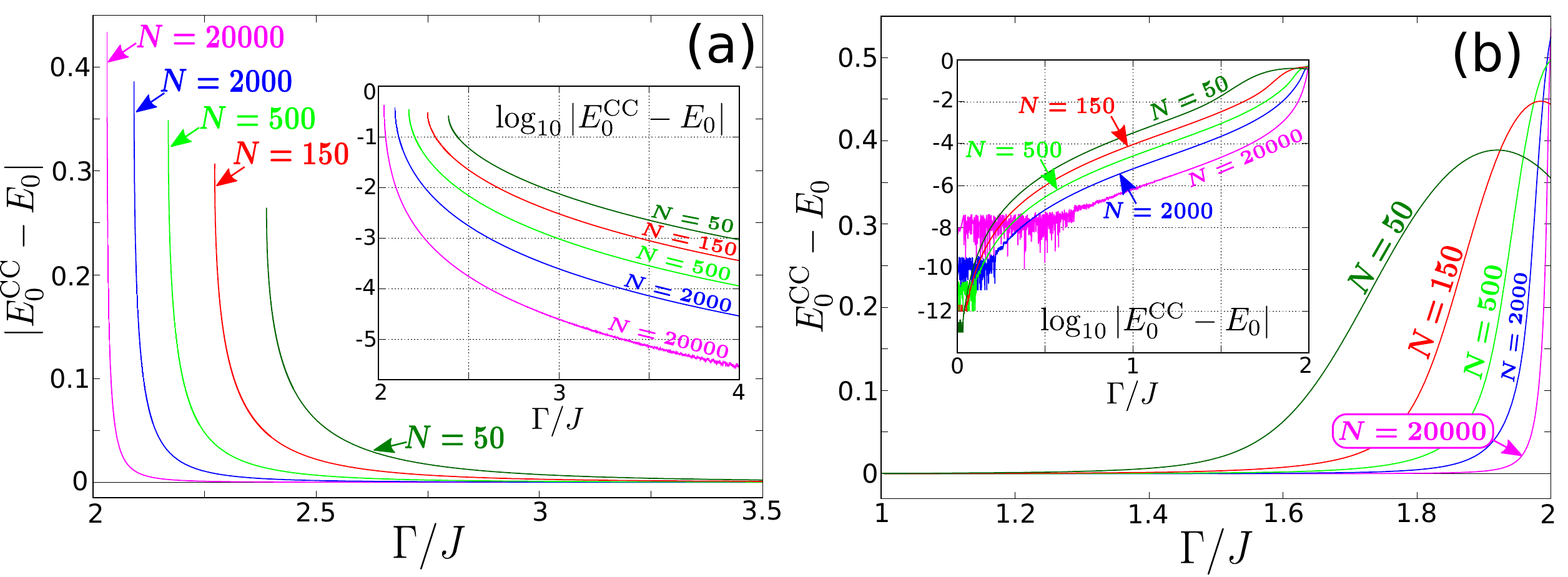}
\caption{Error between the CCSD energy $E_0^{\rm CC}(H_{\rm ferro})$ and the numerically exact ground state energy $E_0(H_{\rm ferro})$ of the model $H_{\rm ferro}$, in the (a) paramagnetic and (b) ferromagnetic regimes. Results for several values of $N$ are shown. One sees that the error decreases as $N$ increases. Insets: The errors replotted on logarithmic scale.}
\label{fig.app.ferro model.error bet Ecc and E0}
\end{center}
\end{figure}

The CCSD energy $E_0^{\rm CC}(H_{\rm ferro})$ is obtained by inserting the solutions of $y$ and $w$ into Eqs. (\ref{eq.app.list.ccsd.energy.Jx}) and (\ref{eq.app.list.ccsd.energy.Jz2}), which are in turn substituted into $\langle 0|e^{-T^{\mathrm{SD}}} H_{\mathrm{ferro}}e^{T^{\mathrm{SD}}} |0\rangle$. We consider the difference between $E_0^{\rm CC}(H_{\rm ferro})$ and the exact ground state energy $E_0(H_{\rm ferro})$, obtained numerically via diagonalization [as done for Eq. (\ref{eq.cubic w antiferro.hamiltonian definition})]. Panels (a) and (b) of Fig. \ref{fig.app.ferro model.error bet Ecc and E0} show the error $E_0^{\rm CC}(H_{\rm ferro})-E_0(H_{\rm ferro})$ in the para and ferromagnetic regimes, respectively. The results for several values of $N$ are shown. It is seen that the agreement between $E_0^{\rm CC}(H_{\rm ferro})$ and $E_0(H_{\rm ferro})$ improves as $N$ increases, with error occurring mainly around the critical point. To highlight the close agreement between the two energies, the insets show the errors replotted on logarithmic scale.

\begin{figure}[h]
\begin{center}
\includegraphics[scale=0.7]{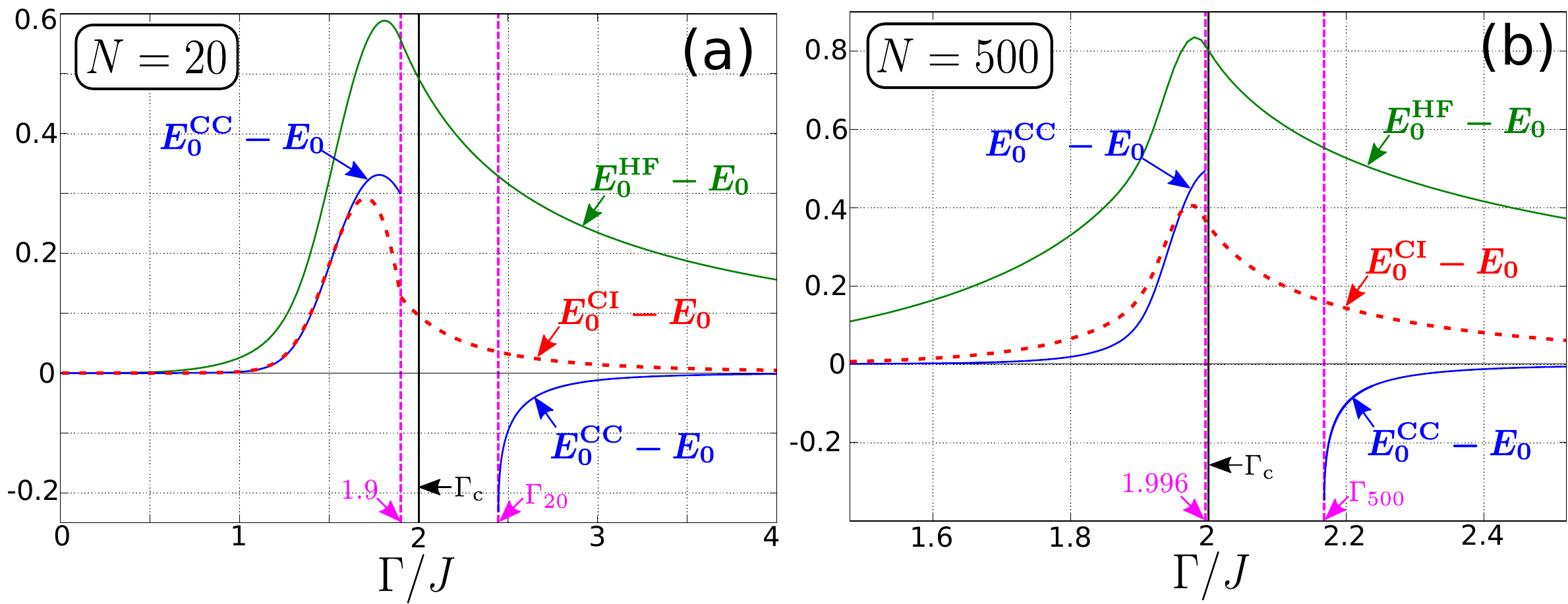}
\caption{Comparing the accuracy of the approximations $E_0^{\rm HF}(H_{\rm ferro})$, $E_0^{\rm CC}(H_{\rm ferro})$, and $E_0^{\rm CI}(H_{\rm ferro})$ for the model $H_{\rm ferro}$. The exact energy $E_0(H_{\rm ferro})$ has been subtracted from all three. The solid vertical line (black) indicates the critical point $\Gamma_{\rm c}$. There is no solution for $E_0^{\rm CC}(H_{\rm ferro})$ between the two dashed vertical lines (magenta), where the left line marks the transition point of $E^{\rm HF}_0(H_{\rm ferro})$ [cf. Eq. (\ref{eq.app.EHF.ferro.two phases})] and the right one marks the vanishing of the solution of $w$ [cf. Fig. \ref{fig.app.ferro model.wsoln.para phase}]. (a) For $N=20$. (b) For $N=500$.}
\label{fig.app.comparing E0 by CC CI HF.ferro}
\end{center}
\end{figure}

It is instructive to see how CCSD fare compared to Hartree-Fock approximation and CISD. For general aspects of Configuration Interaction expansion, the reader is referred to Refs. \cite{Jensen07,Bartlett07}. The CISD approximation of $H_{\rm ferro}$ was studied in Ref. \cite{Koh16}. Figure \ref{fig.app.comparing E0 by CC CI HF.ferro} shows $E_0^{\rm HF}(H_{\rm ferro})$, $E_0^{\rm CC}(H_{\rm ferro})$, and the CISD energy $E_0^{\rm CI}(H_{\rm ferro})$ as a function of $\Gamma$, with the exact energy $E_0(H_{\rm ferro})$ subtracted from all three approximations. Panels (a) and (b) show the results for $N=20$ (small system size) and 500 (large), respectively. In general, CCSD shows evident improvement over CISD in the paramagnetic regime, both for small and large system sizes. In the ferromagnetic regime, however, the advantage of CCSD over CISD only manifests itself when $N$ is large. This observation is in line with the fact that Coupled-Cluster theory accounts for extensivity better than Configuration Interaction expansion. Another notable difference between CISD and CCSD is that for the former the energy $E^{\rm CI}_0(H_{\rm ferro})$ always exists, whereas for the latter there is a small region around the critical point where $E^{\rm CC}_0(H_{\rm ferro})$ disappears. The reason for the vanishing of the CCSD energy is not immediately clear. One possible explanation is that third and higher-order effects begin to dominate the correlation energy around the transition point. In that case, it might be possible to recover the Coupled-Cluster energy by incorporating, say, three- and four-body excitations into the operator $T$.

\subsection{Analytical solutions of $w$, $y$, and $E_0^{\mathrm{CC}}(H_{\mathrm{ferro}})$ in the limit $N\rightarrow\infty$}
\label{}

In the limit $N\rightarrow\infty$, it is possible to obtain analytical solutions of $w$ and $y$. In the CCSD equations, make a change of variable $x=e^{2w}$ followed by $x=1+\delta$; expanding to second order in $\delta$, one solves a quadratic equation in $\delta$ to obtain
\begin{equation}
w
=
\left\{
\begin{array}{ccc}
\frac{1}{2N} \left(\sqrt{\frac{\Gamma}{\Gamma-2J}}-1\right)   \,\, + \,\, O(N^{-2})   \,\, &\mathrm{when}&\,\, \Gamma \ge 2J, \\ [10pt]
\frac{1}{2N} \left(\sqrt{\frac{4J^2}{4J^2-\Gamma^2}}-1\right) \,\, + \,\, O(N^{-2}) \,\, &\mathrm{when}&\,\, \Gamma  < 2J. \\
\end{array}
\right.
\\[10pt]
\label{eq.app.w soln.analytical.N infinity}
\end{equation}
and 
\begin{equation}
y
=
\left\{
\begin{array}{ccc}
0   \,\, &\mathrm{when}&\,\, \Gamma \ge 2J, \\ [10pt]
\frac{\Gamma}{4N}
\left(
\frac{\Gamma^2-16J^2}{(4J^2-\Gamma^2)^{\frac{3}{2}}}
+
\frac{12J}{4J^2-\Gamma^2}
-
\frac{1}{J}
\right)
\,\, + \,\, O(N^{-2}) \,\, &\mathrm{when}&\,\, \Gamma  < 2J. \\
\end{array}
\right.
\label{eq.app.y soln.analytical.N infinity}
\end{equation}
Equations (\ref{eq.app.w soln.analytical.N infinity}) and (\ref{eq.app.y soln.analytical.N infinity}) are plotted in Figs. \ref{fig.app.ferro model.wsoln.para phase} and \ref{fig.app.ferro model.ywsoln.ferro phase} (labeled $N\rightarrow\infty$). It is seen that they agree quite well with the numerical results of large $N$.

\begin{figure}[h]
\begin{center}
\includegraphics[scale=0.7]{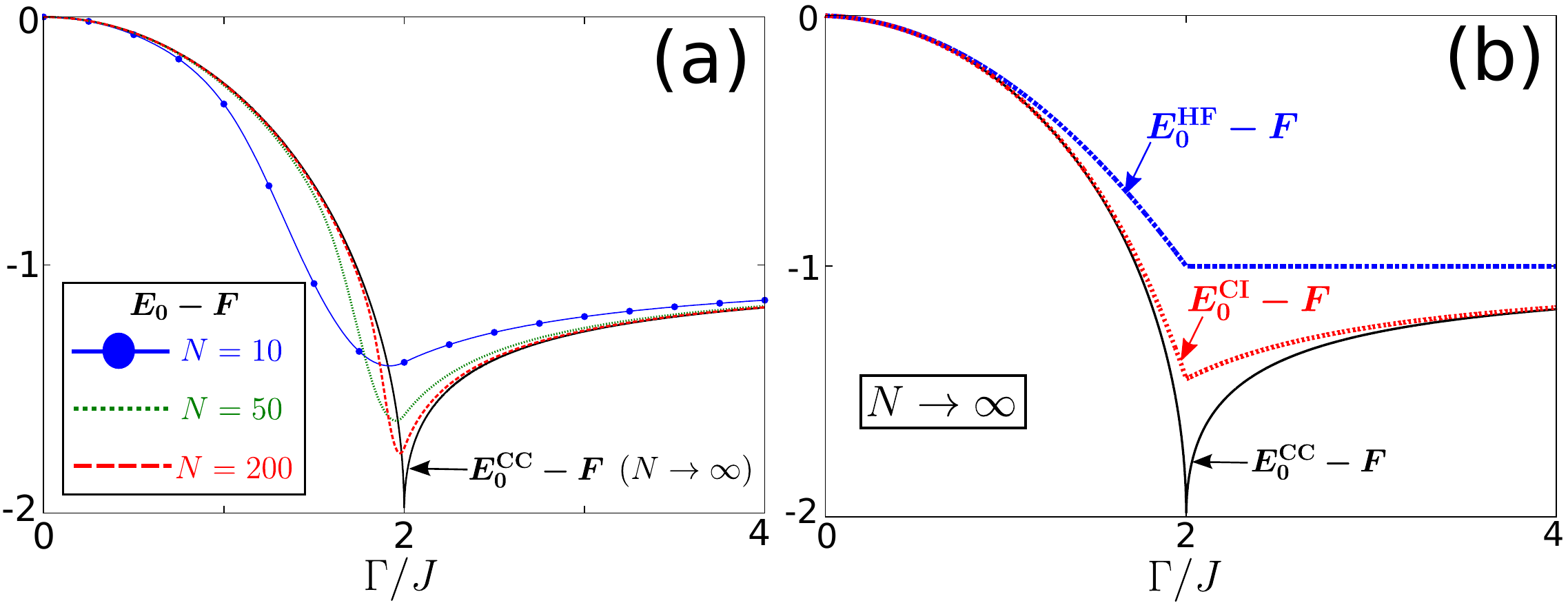}
\caption{(a) Convergence of the exact ground state energy $E_0(H_{\rm ferro})$ (colored) towards the thermodynamic CCSD energy $E_0^{\rm CC}(H_{\rm ferro})$ [Eq. (\ref{eq.app.CCSD energy.analytical.N infinity})] as $N$ increases. The free energy $F$ is defined in the text. (b) Comparison between the three energies $E_0^{\rm HF}(H_{\rm ferro})$, $E_0^{\rm CI}(H_{\rm ferro})$, and $E_0^{\rm CC}(H_{\rm ferro})$ in the limit $N\rightarrow\infty$.}
\label{fig.app.ferro.N infinity limit}
\end{center}
\end{figure}

Inserting the solutions Eqs. (\ref{eq.app.w soln.analytical.N infinity}) and (\ref{eq.app.y soln.analytical.N infinity}) into $E_0^{\rm CC}(H_{\rm ferro})$ and expanding in powers of $N^{-1}$, one obtains for the CCSD energy in the thermodynamic limit
\begin{equation}
E_0^{\mathrm{CC}}(H_{\rm ferro})
=
\left\{
\begin{array}{ccc}
 -N\Gamma + \sqrt{\Gamma(\Gamma-2J)} - \Gamma \,\, + \,\, O(N^{-1})   \,\, &\mathrm{when}&\,\, \Gamma \ge 2J, \\ [10pt]
 -N\left(\frac{4J^2+\Gamma^2}{4J}\right) + \sqrt{4J^2-\Gamma^2} - 2J  \,\, + \,\, O(N^{-1}) \,\, &\mathrm{when}&\,\, \Gamma  < 2J. \\
\end{array}
\right.
\\[10pt]
\label{eq.app.CCSD energy.analytical.N infinity}
\end{equation}
On the right side of Eq. (\ref{eq.app.CCSD energy.analytical.N infinity}), the leading term proportional to $N$ is also known as the free energy, which we denote as $F$. Figure \ref{fig.app.ferro.N infinity limit}(a) shows that the ground state energy $E_0(H_{\rm ferro})$ converges towards Eq. (\ref{eq.app.CCSD energy.analytical.N infinity}) as $N$ increases. In other words, CCSD gives the exact ground state energy in the thermodynamic limit. Panel (b) compares the three approximations $E_0^{\rm HF}(H_{\rm ferro})$ [cf. Eq. (\ref{eq.app.EHF.ferro.two phases})], $E_0^{\rm CI}(H_{\rm ferro})$ \cite{footnote.concerning exact E0 CI expression in Koh16}, and $E_0^{\rm CC}(H_{\rm ferro})$ in the limit $N\rightarrow\infty$. One sees that CISD fails to account for a significant part of the correlation energy around the critical point.

It is interesting to point out that the $N\rightarrow\infty$ solution Eq. (\ref{eq.app.CCSD energy.analytical.N infinity}) can also be obtained via the Holstein-Primakoff transform \cite{Dusuel05,Koh16}. Between the two approaches, the derivation based on Holstein-Primakoff transform is slightly easier compared to CCSD. On the other hand, one corollary from the CCSD approach is that we know that in the thermodynamic limit the ground state of $H_{\rm ferro}$ consists of only singles and doubles excitations, with the wave function taking the simple and compact form $e^{T^{\rm SD}}|0\rangle$. This is an insight from Coupled-Cluster theory which is not immediately apparent from the Holstein-Primakoff approach.

%
%

\section{Wave functions of various approximations in $J_z$ representation}
\label{app.wavefunctions}

The numerical diagonalization of the Hamiltonians Eqs. (\ref{eq.cubic w antiferro.hamiltonian definition}) and (\ref{eq.app.ferro model.definition}) were performed in the representation where the operator $J_z$ is diagonal. In this appendix, we discuss the representations of Hartree-Fock, CISD, and CCSD wave functions in this basis. This allows us to compare these approximations with the energy eigenfunctions that are obtained numerically.

Denote $|m_z\rangle$ as the eigenvector of $J_z$, where $m_z=-N,-N+2,\cdots,N$ is the projection quantum number in the $z$-direction. Expressing $|m_z\rangle$ in the $\sigma_i^z$ basis, one has
\begin{equation}
|m_z\rangle
=
{N \choose n_d}^{-\frac{1}{2}}
\sum_{\{i_1,\cdots,i_{n_d}\}\in \cal{M}}
|i_1 \cdots i_{n_d} )
\label{}
\end{equation}
where $n_d=(N-m_z)/2$ is the number of spins pointing down in $|m_z\rangle$, and $|i_1\cdots i_{n_d})$ denotes a Hartree-Fock state in which the spins with indices $i_1,\cdots,i_{n_d}$ are in the spinor state ${0 \choose 1}$ while the rest are in ${1 \choose 0}$. The symbol $\cal M$ denotes all possible ways to choose $n_d$ indices from $N$, and the binomial coefficient ${N \choose n_d}$ is to ensure normalization of $|m_z\rangle$. In the following, the expansion coefficients of a wave function $|\Psi\rangle$ in the basis $|m_z\rangle$ is obtained by computing the inner product $\langle m_z|\Psi\rangle$.

\subsection{Hartree-Fock approximation}
\label{}

For the Hartree-Fock state $|0\rangle={\alpha\choose \beta}^N$, one has 
\begin{equation}
\langle m_z |0 \rangle= {N \choose n_d}^{\frac{1}{2}} \,\, \alpha^{N-n_d} \,\, \beta^{n_d}
\label{eq.app.HF projected onto mz}
\end{equation}

\subsection{CISD approximation}
\label{}

The CISD wave function is given by \cite{Koh16}
\begin{equation}
|\mathrm{CI}\rangle
=
\left(
\hat{x}
+
\hat{y}
\sum_i \sigma_i^y
+
\hat{w}
\sum_{i>j}\sigma_i^y\sigma_j^y
\right)
|0\rangle
\label{eq.app.CISD wavefunction.definition}
\end{equation}
where $|0\rangle$ is as in Eq. (\ref{eq.app.HF projected onto mz}), and $\hat{x}$, $\hat{y}$, $\hat{w}$ are determined by minimizing $\langle \mathrm{CI}|H| \mathrm{CI}\rangle$ subjected to $\langle \mathrm{CI}|\mathrm{CI}\rangle=1$. The inner product is
\begin{eqnarray}
\langle m_z |\mathrm{CI} \rangle
&=&
\mathcal{N}_{\rm CI}
\left\{
\,\,\,
\hat{x} 
\,
\alpha^{N-n_d} \beta^{n_d} 
\right.
\nonumber
\\[5pt]
&&
\,\,\,\,\,\,\,\,\,\,\,\,\,\,\,\,\,
+
\,
\hat{y}
\left[
n_d 
\,
\alpha^{N-(n_d-1)} \beta^{n_d - 1} 
-
(N-n_d)
\,
\alpha^{N-(n_d+1)} \beta^{n_d+1} 
\,
\right]
\nonumber
\\[10pt]
&&
\,\,\,\,\,\,\,\,\,\,\,\,\,\,\,\,\,\,
+
\frac{\hat{w}}{2}
\left[
n_d(n_d-1)
\alpha^{N-(n_d-2)}
\beta^{n_d-2}
+
(N-n_d)
(N-n_d-1)
\alpha^{N-(n_d+2)}
\beta^{n_d+2}
\right.
\nonumber
\\[10pt]
&&
\left.
\left.
\,\,\,\,\,\,\,\,\,\,\,\,\,\,\,\,\,\,\,\,\,\,\,\,\,\,\,\,\,\,
-
2
\,
n_d
(N-n_d)
\alpha^{N-n_d}
\beta^{n_d}
\right]
\,\,\,
\right\}
\label{eq.app.CISD wavefunction projected onto mz}
\end{eqnarray}
where $\mathcal{N}_{\rm CI}=\left[ {N\choose n_d}^{-1} \left(\hat{x}^2 + N \hat{y}^2 + \frac{N(N-1)}{2}\hat{w}^2\right)\right]^{-\frac{1}{2}}$. To avoid overflows when evaluating Eq. (\ref{eq.app.CISD wavefunction projected onto mz}) numerically, it is advisable to first group the ${N\choose n_d}$ in $\mathcal{N}_{\rm CI}$ together with the powers of $\alpha$ and $\beta$, and then sum up their exponents.

\subsection{CCSD approximation}
\label{}
 
We write the normalized CCSD wave function as
\begin{equation}
|\mathrm{CC}\rangle
=
\mathcal{N}_{\rm CC} \, e^{-Nw} \, e^{T^{\rm SD}}|0\rangle
\label{eq.app.CCSD Normalized wavefunction.definition}
\end{equation}
where $|0\rangle$ is as before, and $\mathcal{N}_{\rm CC}=\left[ 2^{-N}\sum_{i=0}^N {N \choose i}e^{-w(N-2i)^2} \right]^{-\frac{1}{2}}$. 

\vspace{0.25cm}

When $n_d$ is even, one has
\begin{equation}
\langle m_z |\mathrm{CC} \rangle
=
\nu
\,\,
\mathcal{N}'
\sum_{q=0}^{N-n_d}
\sum_{r=0}^{n_d}
(-1)^r
\cos[(\theta+y)(N-2(q+r))]
{N-n_d\choose q}
{n_d\choose r}
e^{-\frac{w}{2}[N-2(q+r)]^2}
\label{eq.app.CCSD wavefunction projected onto mz.even}
\end{equation}
where $\mathcal{N}'=2^{-N} {N \choose n_d}^{\frac{1}{2}} \mathcal{N}_{\rm CC}$, the angle $\theta$ is given by $e^{i\theta}=\alpha + i\beta$, and 
\begin{equation}
\nu=
\left\{
\begin{array}{ccl}
+1 &\,\,\, \rm when \,\,\, & n_d=0,4,8,\cdots, \\
-1 &\,\,\, \rm when \,\,\, & n_d=2,6,10,\cdots. \\
\end{array}
\right.
\label{eq.app.CCSD.v when nd even}
\end{equation}

\vspace{0.25cm}

When $n_d$ is odd, replace the cosine in Eq. (\ref{eq.app.CCSD wavefunction projected onto mz.even}) by sine, and $\nu$ by
\begin{equation}
\tilde{\nu}=
\left\{
\begin{array}{ccl}
+1 &\,\,\, \rm when \,\,\, & n_d=1,5,9,\cdots, \\
-1 &\,\,\, \rm when \,\,\, & n_d=3,7,11,\cdots. \\
\end{array}
\right.
\label{eq.app.CCSD.v when nd odd}
\end{equation}

\vspace{0.25cm}

Although Eq. (\ref{eq.app.CCSD wavefunction projected onto mz.even}) is exact, it is difficult to evaluate accurately when $N$ is large due to numerical overflow. To avoid that, note that before integrating out the random field $m$, one has
\begin{equation}
\langle m_z |\mathrm{CC} \rangle
=
\mathcal{N}_{\rm CC}
\left[
{N \choose n_d}
\left(\frac{N}{2\pi}\right)\right]^{\frac{1}{2}} 
\int 
dm 
\,\, 
\exp
\left[
-Nf(m)
\right]
\label{eq.app.CCSD wave function.before integrating out m}
\end{equation}
where
\begin{equation}
f(m)
=
\frac{1}{2}m^2
-
\left(1-\frac{n_d}{N}\right)
\ln
\left(
\alpha \cos t
-
\beta \sin t
\right)
-
\left(
\frac{n_d}{N}
\right)
\ln
\left(
\beta \cos t
+
\alpha \sin t
\right)
\label{}
\end{equation}
and $t=y+m\sqrt{N w}$. When $N$ is large, the integral Eq. (\ref{eq.app.CCSD wave function.before integrating out m}) can be  approximated by the method of steepest descent
\begin{equation}
\langle m_z |\mathrm{CC} \rangle\approx\mathcal{N}_{\rm CC}\,{N \choose n_d}^{\frac{1}{2}}\,|f''(m_0)|^{-\frac{1}{2}}\,\,e^{-Nf(m_0)}
\label{eq.app.CCSD wave function.steepest descent}
\end{equation}
where $m_0$ is the solution of the saddle-point equation $f'(m_0)=0$.

\subsection{Application to the ferromagnetic model}
\label{}

\begin{figure}[h]
\begin{center}
\includegraphics[scale=0.6]{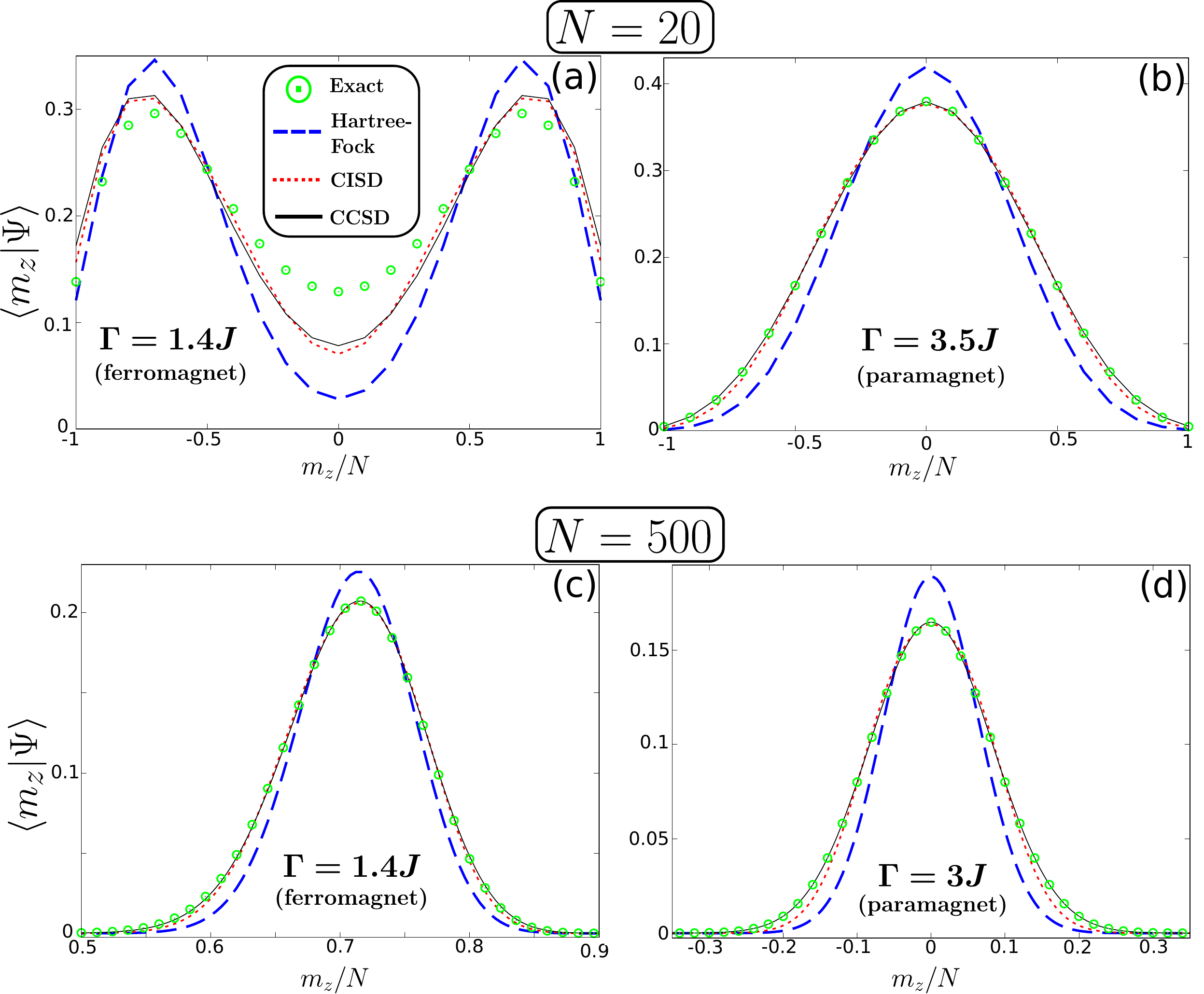}
\caption{Comparing the exact and approximated ground state wave functions of the ferromagnetic model $H_{\rm ferro}$. The wave functions are represented using the eigenvectors $|m_z\rangle$ of the operator $J_z$. Panels (a) and (b): $N=20$. Panels (c) and (d): $N=500$. Panels (a) and (c): Ferromagnetic regime. Panels (b) and (d): Paramagnetic regime. The symbols shown in (a) applies to all panels. The CCSD results for $N=20$ are obtained using Eqs. (\ref{eq.app.CCSD wavefunction projected onto mz.even}) to (\ref{eq.app.CCSD.v when nd odd}), while those for $N=500$ using Eq. (\ref{eq.app.CCSD wave function.steepest descent}). Overall, CCSD (black solid curve) gives the closest approximation to the exact ground state (green circles).}
\label{fig.app.compare wavefunctions.ferro model.}
\end{center}
\end{figure}

Let us apply the above results to the ferromagnetic model Eq. (\ref{eq.app.ferro model.definition}). Each panel in Fig. \ref{fig.app.compare wavefunctions.ferro model.} compares the Hartree-Fock (blue dashed), CISD (red dotted), and CCSD (black solid) wave functions with the exact ground state obtained via diagonalization (green circles). Panels (a) and (b) show the results for a small system size of $N=20$,  while panels (c) and (d) show for a larger one of $N=500$. The left panels (a) and (c) show the situation in the ferromagnetic regime, while the right ones (b) and (d) that in the paramagnetic one. Overall, one sees that both CISD and CCSD are quite accurate, although at the visual level CCSD performs slightly better than CISD. The sole exception is panel (a) ($N=20$, ferromagnetic regime) where all three approximations deviate rather significantly from the exact result. This highlights the difficulty of attaining accurate approximations in the ordered phase when the system size is small.

%
%



\begin{thebibliography}{99} 


\bibitem{Zhang21} F. Zhang, N. Gomes, Y. Yao, P.P. Orth, and T. Iadecola, Phys. Rev. B \textbf{104}, 075159 (2021).

\bibitem{Seki21} Y. Seki, Y. Matsuzaki, and S. Kawabata, J. Phys. Soc. Jpn. \textbf{90}, 054002 (2021).

\bibitem{Albash18} T. Albash and D.A. Lidar, Rev. Mod. Phys. \textbf{90}, 015002 (2018).

\bibitem{Tekur18} S.H. Tekur, U.T. Bhosale, and M.S. Santhanam, Phys. Rev. B \textbf{98}, 104305 (2018).

\bibitem{Beugeling18} W. Beugeling, A. B\"{a}cker, R. Moessner, and M. Haque, Phys. Rev. E \textbf{98}, 022204 (2018).






\bibitem{Dusuel05} S. Dusuel and J. Vidal, Phys. Rev. B \textbf{71}, 224420 (2005).

\bibitem{Koh16} Y.W. Koh, Phys. Rev. B \textbf{93}, 134202 (2016).

\bibitem{footnote.on meanfield vs classical} Technically speaking, the term `classical energy landscape' should more appropriately be refered to as `mean-field energy landscape' or `semiclassical energy landscape'. However, as the former nomenclature sounds more intuitive and is not as heavy as the latter two, we chose to use the former.



\bibitem{Bray80} A.J. Bray and M.A. Moore, J. Phys. C: Solid State Phys. \textbf{13}, L655 (1980).


\bibitem{Yamamoto87} T. Yamamoto and H. Ishii, J. Phys. C: Solid State Phys. \textbf{20}, 6053 (1987).

\bibitem{Thirumalai89} D. Thirumalai, Q. Li, and T.R. Kirkpatrick, J. Phys. A: Math. Gen. \textbf{22}, 3339 (1989).

\bibitem{Ray89} P. Ray, B.K. Chakrabarti, and A. Chakrabarti, Phys. Rev. B \textbf{39}, 11 828 (1989).

\bibitem{Buttner90} G. B\"{u}ttner and K.D. Usadel, Phys. Rev. B \textbf{41}, 428 (1990).

\bibitem{Goldschmidt90} Y.Y. Goldschmidt and P.Y. Lai, Phys. Rev. Lett. \textbf{64}, 2467 (1990).

\bibitem{Usadel91} K.D. Usadel, G. B\"{u}ttner, and T.K. Kope\'{c}, Phys. Rev. B \textbf{44}, 12 583 (1991).



\bibitem{Kopec94} T.K. Kope\'{c}, Phys. Rev. B \textbf{50}, 9963 (1994).

\bibitem{Read95} N. Read, S. Sachdev, and J. Ye, Phys. Rev. B \textbf{52}, 384 (1995).


\bibitem{Dobrosavljevic90} V. Dobrosavljevi\'{c} and D. Thirumalai, J. Phys. A: Math. Gen. \textbf{23}, L767 (1990).

\bibitem{Goldschmidt90R} Y.Y. Goldschmidt, Phys. Rev. B \textbf{41}, 4858(R) (1990).

\bibitem{Cesare96}L. De Cesare, K. Lukierska-Walasek, I. Rabuffo, and K. Walasek, J. Phys. A: Math. Gen. \textbf{29}, 1605 (1996).

\bibitem{Obuchi07} T. Obuchi, H. Nishimori, and D. Sherrington, J. Phys. Soc. Jpn. \textbf{76}, 054002 (2007).



\bibitem{Miller93} J. Miller and D.A. Huse, Phys. Rev. Lett. \textbf{70}, 3147 (1993).

\bibitem{Grempel98} D.R. Grempel and M.J. Rozenberg, Phys. Rev. Lett. \textbf{80}, 389 (1998).

\bibitem{Rozenberg98} D.R. Grempel and M.J. Rozenberg, Phys. Rev. Lett. \textbf{81}, 2550 (1998).

\bibitem{Arrachea01} L. Arrachea and M.J. Rozenberg, Phys. Rev. Lett. \textbf{86}, 5172 (2001).


\bibitem{Ishii85} H. Ishii and T. Yamamoto, J. Phys. C: Solid State Phys. \textbf{18}, 6225 (1985).

\bibitem{Cesare92} L. De Cesare, K. Lukierska-Walasek, and K. Walasek,  Phys. Rev. B \textbf{45}, 8127 (1992).

\bibitem{Biroli01} G. Biroli and L.F. Cugliandolo, Phys. Rev. B \textbf{64}, 014206 (2001).



\bibitem{Gubin12} A. Gubin and L.F. Santos, Am. J. Phys. \textbf{80}, 246 (2012).

\bibitem{Zakrzewski23} J. Zakrzewski, Entropy \textbf{25}, 491 (2023).

\bibitem{Brown08} W.G. Brown, L.F. Santos, D.J. Starling, and L. Viola, Phys. Rev. E \textbf{77}, 021106 (2008).

\bibitem{Georgeot00a} B. Georgeot and D.L. Shepelyansky, Phys. Rev. E \textbf{62}, 3504 (2000).

\bibitem{Georgeot00b} B. Georgeot and D.L. Shepelyansky, Phys. Rev. E \textbf{62}, 6366 (2000).

\bibitem{Georgeot98} B. Georgeot and D.L. Shepelyansky, Phys. Rev. Lett. \textbf{81}, 5129 (1998).

\bibitem{Bera22} S. Bera, K.Y. Venkata Lokesh, and S. Banerjee, Phys. Rev. Lett. \textbf{128}, 115302 (2022).

\bibitem{Winer22} M. Winer, R. Barney, C.L. Baldwin, V. Galitski, and B. Swingle, J. High Energ. Phys. \textbf{2022}, 32 (2022).




\bibitem{Laumann14} C.R. Laumann, A. Pal, and A. Scardicchio, Phys. Rev. Lett. \textbf{113}, 200405 (2014).

\bibitem{Baldwin17} C.L. Baldwin, C.R. Laumann, A. Pal, and A. Scardicchio, Phys. Rev. Lett. \textbf{118}, 127201 (2017).

\bibitem{Mukherjee18} S. Mukherjee, S. Nag, and A. Garg, Phys. Rev. B \textbf{97}, 144202 (2018).

\bibitem{Imbrie16} J. Z. Imbrie, Phys. Rev. Lett. \textbf{117}, 027201 (2016).

\bibitem{Kjall14} J.A. Kj\"{a}ll, J.H. Bardarson, and F. Pollmann, Phys. Rev. Lett. \textbf{113}, 107204 (2014).

\bibitem{Lee17} M. Lee, T.R. Look, D.N. Sheng, and S.P. Lim, Phys. Rev. B \textbf{96}, 075146 (2017).

\bibitem{Biroli20} G. Biroli and M. Tarzia, Phys. Rev. B \textbf{102}, 064211 (2020).

\bibitem{Tarzia20} M. Tarzia, Phys. Rev. B \textbf{102}, 014208 (2020).

\bibitem{Herviou21} L. Herviou, J.H. Bardarson, and N. Regnault, Phys. Rev. B \textbf{103}, 134207 (2021).



\bibitem{Amin15} M.H. Amin, Phys. Rev. A \textbf{92}, 052323 (2015).

\bibitem{Boixo16} S. Boixo, V.N. Smelyanskiy, A. Shabani, S.V. Isakov, M. Dykman, V.S. Denchev, M.H. Amin, A.Y. Smirnov, M. Mohseni, and H. Neven, Nature Comm. \textbf{7}, 10327 (2016).

\bibitem{Denchev16} V.S. Denchev, S. Boixo, S.V. Isakov, N. Ding, R. Babbush, V.N. Smelyanskiy, J. M. Martinis, and H. Neven, Phys. Rev. X \textbf{6}, 031015 (2016).

\bibitem{Jensen07} F. Jensen, \emph{Introduction to Computational Chemistry} (Wiley, Chichester, UK, 2007).

\bibitem{Bartlett07} R.J. Bartlett and M. Musial, Rev. Mod. Phys. \textbf{79}, 291 (2007).

\bibitem{footnote.two sense of the word excitation} The term `excitation', when used in the context of Coupled-Cluster theory, should be understood in terms of correlation effects, and not as the energy difference between two energy eigenstates.

\bibitem{LeBlanc15} J.P.F. LeBlanc, A.E. Antipov, F. Becca, I.W. Bulik, G. K-L Chan, C-M Chung, Y. Deng, M. Ferrero, T.M. Henderson, C.A. Jim\'{e}nez-Hoyos et al., Phys. Rev. X \textbf{5}, 041041 (2015).




\bibitem{Roger90} M. Roger and J.H. Hetherington, Phys. Rev. B \textbf{41}, 200 (1990). 

\bibitem{Bishop} R.F. Bishop, J.B. Parkinson, and Y. Xian, Phys. Rev. B \textbf{44}, 9425 (1991); Phys. Rev. B \textbf{46}, 880 (1992).

\bibitem{Miguel96} B. Miguel and J.-P. Malrieu, Phys. Rev. B \textbf{54}, 1652 (1996).

\bibitem{Farnell01} D.J.J. Farnell, K.A. Gernoth, and R.F. Bishop, Phys. Rev. B \textbf{64}, 172409 (2001).

\bibitem{Kruger06} S.E. Kr\"{u}ger, R. Darradi, J. Richter, and D.J.J. Farnell, Phys. Rev. B \textbf{73}, 094404 (2006).

\bibitem{Darradi08} R. Darradi, O. Derzhko, R. Zinke, J. Schulenberg, S.E. Kr\"{u}ger, and J. Richter, Phys. Rev. B \textbf{78}, 214415 (2008).

\bibitem{Richter10} J. Richter, R. Darradi, J. Schulenberg, D.J.J. Farnell, and H. Rosner, Phys. Rev. B \textbf{81}, 174429 (2010). 

\bibitem{Gotze11} O. G\"{o}tze, D.J.J. Farnell, R.F. Bishop, P.H.Y. Li, and J. Richter, Phys. Rev. B \textbf{84}, 224428 (2011). 

\bibitem{Gotze15} O. G\"{o}tze and J. Richter, Phys. Rev. B \textbf{91}, 104402 (2015). 


\bibitem{Hirata03} S. Hirata, J. Chem. Phys. A \textbf{107}, 9887 (2003).


\bibitem{Seki12} Y. Seki and H. Nishimori, Phys. Rev. E \textbf{85}, 051112 (2012).

\bibitem{footnote.concerning using HF psi over CCSD psi in Hcubic} The more accurate CCSD wave function Eq. (\ref{eq.app.CCSD wave function.steepest descent}) gives a closer correspondence with the energy eigenfunction. The Hartree-Fock wave function, however, is simpler and sufficient to illustrate our point.  


\bibitem{Ma.and.Gong.93} Y.Q. Ma and C.D. Gong, Phys. Rev. B \textbf{48}, 12778 (1993).

\bibitem{Ma.etal.93} Y.Q. Ma, Y.M. Zhang, Y.G. Ma, and C.D. Gong, Phys. Rev. E \textbf{47}, 3985 (1993).

\bibitem{Nishimori96} H. Nishimori and Y. Nonomura, J. Phys. Soc. Jpn. \textbf{65}, 3780 (1996).

\bibitem{AmitBOOK} D.J. Amit, \emph{Modeling Brain Function: The World of Attractor Neural Networks} (Cambridge University Press, Cambridge, 1989).

\bibitem{Rebentrost18} P. Rebentrost, T.R. Bromley, C. Weedbrook, and S. Lloyd, Phys. Rev. A \textbf{98}, 042308 (2018).

\bibitem{Liu20} G. Liu, W.P. Ma, H. Cao, and L.D. Lyu, Laser Phys. Lett. \textbf{17}, 045201 (2020).

\bibitem{Miller21} N.E. Miller and S. Mukhopadhyay, Sci. Rep. \textbf{11}, 23391 (2021).


\bibitem{footnote.the patterns being used in Hhop} The patterns are: $\bm{\xi^{1}}$=(1,1,1,1,1,1,1,1,1,1,-1,-1,-1,1,1,1), $\bm{\xi^{2}}$=(1,1,1,1,1,1,-1,-1,-1,-1,1,1,1,1,1,-1), $\bm{\xi^{3}}$=(1,1,1,-1,-1,-1,1,1,1,1,1,1,1,-1,-1,1), and $\bm{\xi^{4}}$=(1,-1,-1,1,1,1,1,1,1,1,1,1,1,-1,-1,-1).


\bibitem{footnote.omega and pole vectors} We used $\bm{\omega}$=(1,-1,-1,1,1,1,1,1,1,1,0,1,1,-1,-1,-1) and $\bm{p}$=(0,0,-1,-1,2,2,-1,-1,-1,-1,-1,-1,1,1,1,-1).


\bibitem{footnote.concerning Spectra and convergence} Y. Qiu, Spectra, https://spectralib.org. Using this package, we outputted the lowest 1400 eigenvalues of $H_{\rm hop}$, and checked that those that we need, $E_0$ to $E_{450}$, are converged.

\bibitem{Rowe68} D.J. Rowe, Rev. Mod. Phys. \textbf{40}, 153 (1968).


\bibitem{footnote.concerning exact E0 CI expression in Koh16} The analytical expression for $E_0^{\rm CI}(H_{\rm ferro})$ in the limit $N\rightarrow\infty$ is given by Eq. (A6) in Ref. \cite{Koh16}.



\end{thebibliography}
\end{document}